\documentclass[aps,prc,superscriptaddress,reprint]{revtex4-2}
\usepackage{graphicx}
\usepackage{multirow}
\usepackage[dvipsnames]{xcolor}
\usepackage{amssymb}
\usepackage{amsmath}
\usepackage{bm}
\usepackage{lineno}
\usepackage{booktabs}
\usepackage[colorlinks=true, breaklinks=true, linkcolor=blue, citecolor=violet, urlcolor=cyan]{hyperref}
\usepackage{placeins}

\begin{document}

\title{Inclusive breakup of three-body projectiles: \texorpdfstring{\\}{ }
A unified four-body framework for pair-detection and single-particle observables}

\author{Jin Lei}
\email[]{jinl@tongji.edu.cn}
\affiliation{School of Physics Science and Engineering, Tongji University, Shanghai 200092, China.}
\affiliation{Southern Center for Nuclear-Science Theory (SCNT), Institute of Modern Physics, Chinese Academy of Sciences, Huizhou 516000, Guangdong Province, China.}

\date{\today}

\begin{abstract}
Inclusive breakup measurements detect only part of a projectile while
the remaining fragments interact with the target.  For a projectile
with three constituents, these measurements can probe correlations
that are lost when the projectile is treated as two preformed clusters.
I derive cross-section sum rules within the distorted-wave Born
approximation for detecting either a pair or a single constituent,
retaining the full three-body projectile wave function.
For pair detection, the detected state is projected from this wave
function.  Comparing the resulting reaction source with its factorized
limit tests, at the amplitude level, what the preformed-cluster
approximation omits.  For single-particle detection, both unobserved
constituents can interact with the target, and their absorption includes
a contribution mediated by target excitation that is absent from an
independent-fragment description.  The optical reduction recovers the
Carlson-Frederico-Hussein absorption kernel
[Phys. Lett. B \textbf{767}, 53 (2017)]; explicit coupling of the
detected particle to target excitation adds further contributions.
The ${}^6\mathrm{Li}=\alpha+n+p$ example shows how neutron-proton
states beyond the bound deuteron enter the reaction source.  I give a route for calculating
the single-particle spectrum using inelastic target transition form
factors and three-body propagation.
\end{abstract}

\maketitle

\section{Introduction}
\label{sec:intro}

A deuteron can break up in a collision with a nucleus, leaving an
outgoing proton while the neutron scatters from the target, excites it,
or is captured.  If only the proton is measured, its spectrum contains
contributions from all these neutron-target final states.  This is an
inclusive breakup measurement~\cite{Moro2025}.  It gives access to
reactions of the unobserved neutron through the detected proton.
The theoretical task is to calculate this spectrum by summing over the
unobserved states, while retaining the interactions that determine
their contributions.

Early closed-form treatments used unitarity and
surface approximations~\cite{Budzanowski1978,Baur1980,Shyam1980,Baur1983}.
For a two-body projectile $a=b+x$, the unobserved $x+A$ channels can be
summed by completeness, which expresses the inclusive nonelastic
breakup (NEB) cross section through the imaginary part of an optical
Green's function acting on a breakup source.  This is the basis of
the Ichimura-Austern-Vincent (IAV) sum rule and its prior-form
Udagawa-Tamura (UT) and Hussein-McVoy decomposition
~\cite{Ichimura1985,Austern1981,Udagawa1981,Li1984,Hussein1985,GomezRamos2021}.
The resulting theory is now used routinely for $(d,p)$ and
$({}^6\mathrm{Li},\alpha)$ reactions
~\cite{Potel2015,Potel2017,Lei2015,Lei2015b,Lei2017,Carlson2016,Lei2019,Lei2019b,Lei2016review}.
The prior form is convenient for numerical calculations because the
post-form transition integral lacks a natural cutoff at large distances
and converges slowly~\cite{Ichimura1985,Lei2025post}.  Eikonal and Faddeev
formulations provide complementary treatments of the same two-body
projectile problem~\cite{Neoh2016,Deltuva2025}.

Many weakly bound nuclei require three active projectile
constituents.  Examples include the Borromean halos
${}^6\mathrm{He}=\alpha+n+n$ and
${}^{11}\mathrm{Li}={}^9\mathrm{Li}+n+n$, the
$\alpha+n+p$ structure of ${}^6\mathrm{Li}$ when deuteron breakup
matters, and ${}^9\mathrm{Be}=\alpha+\alpha+n$
~\cite{Canto2006,Keeley2007,Lei2017,Lehman1982,Villanueva2024,Descouvemont2015}.
Reducing these projectiles to two clusters hides the effect of their
three-body structure and offers no internal test of that reduction.

For $a=i+j+k$, two different inclusive measurements must be
distinguished.  Detecting a correlated pair $b=(ij)$ leaves a
two-body $k+A$ system unresolved,
$a+A\to b+(k\!+\!A)^*$.  Detecting only $i$ leaves the three-body
$jk+A$ system unresolved,
$a+A\to i+(jk\!+\!A)^*$.  The first channel is close to the usual
IAV problem once its source has been constructed; the second requires
a three-body propagator.  For ${}^6\mathrm{Li}$, inclusive deuteron detection leaves
$\alpha+A$, whereas proton detection leaves $n+\alpha+A$.

Carlson, Frederico, and Hussein (CFH) treated the single-particle
channel by reducing the target-ground-state part of the $jk+A$
propagator to a three-body optical form with absorptive terms
$W_j+W_k+W_{3B}$~\cite{CFH2017,Hussein2017,Hussein2020}.  The
cross term $W_{3B}$ has no counterpart for a two-body projectile.
Existing applications nevertheless use two-body cluster models
~\cite{Souza2021}.  The complementary pair-detection channel has not
been formulated as an IAV sum rule with the detected pair projected
from a full three-body projectile.  Treating the pair as preformed,
as in reduced descriptions of
${}^{197}\mathrm{Au}({}^9\mathrm{Be},{}^8\mathrm{Be})$
~\cite{Villanueva2024}, removes precisely the pair mixing that such a
measurement can test.

I derive both channels from one four-body Hamiltonian, retaining the
full three-body projectile wave function.  For pair detection, the
source contains the interactions of the pair with both the third
particle and the target.  Projection onto the measured pair state gives
a coincidence cross section and allows the preformed-cluster
approximation to be tested by factorizing the projectile wave function.
That factorization recovers the detected-composite formalism of
Ref.~\cite{Lei2026nsp}.  For single-particle detection, the
target-ground-state optical reduction recovers CFH.  Retaining
$V_{iA}-U_{iA}$ in the source adds target-excited contributions, whose
direct part can be reduced with a diagonal-intermediate-target-state
approximation.  The interference terms remain in Feshbach form.

The sum rules use the distorted-wave Born approximation (DWBA) for
the exit-channel residual interaction.  The unresolved propagators
are retained before the optical reductions are made.  In the
two-body-projectile limit, the results can be compared with the exact
Faddeev treatment of Ref.~\cite{Deltuva2025}; an exact treatment of
a three-body projectile would require four-body Yakubovsky equations.
The derivations below separate the sum rules from the approximations
needed to evaluate them.  The ${}^6\mathrm{Li}$ example shows what
is lost in a two-cluster description, and the discussion gives a
numerical route for each observable and the limits of the target
reduction.

\section{Four-body Hamiltonian}
\label{sec:hamiltonian}

I consider the inclusive breakup reaction induced by a three-body
projectile $a = i + j + k$ on a target nucleus $A$.
The full Hamiltonian for the four-body $a + A$ system is
\begin{align}
H &= H_A(\xi) + h_i + h_j + h_k + K_i + K_j + K_k
\notag \\
&\quad + V_{ij} + V_{ik} + V_{jk}
+ V_{iA} + V_{jA} + V_{kA} \,,
\label{eq:H_full}
\end{align}
where $H_A(\xi)$ is the internal Hamiltonian of the target nucleus
with ground state $H_A \Phi_A = E_A \Phi_A$ and internal coordinates
$\xi$; $h_i$, $h_j$, and $h_k$ are the internal Hamiltonians of the
three projectile constituents when they are composite; $K_i$, $K_j$,
and $K_k$ are their center-of-mass kinetic energies; $V_{ij}$,
$V_{ik}$, and $V_{jk}$ are the mutual interactions among the
constituents; and $V_{iA}$, $V_{jA}$, and $V_{kA}$ are their
interactions with the target.
For a composite constituent, $h_i$ acts on its internal coordinates
$\zeta_i$, and likewise for $j$ and $k$.  The pairwise interactions
may depend on these coordinates as well as on the relative positions.

The projectile $a$ is a bound state of its three constituents.
To describe its internal structure, I introduce two sets of Jacobi
coordinates adapted to the two exit-channel partitions.

For the pair-detection channel, in which the pair $b \equiv (ij)$ is
detected and $k$ is unresolved, the natural Jacobi coordinates are
\begin{align}
\boldsymbol\zeta &= \mathbf{r}_i - \mathbf{r}_j \,,
\label{eq:jacobi_A_zeta} \\
\mathbf{y} &= \mathbf{r}_k
  - \frac{m_i \mathbf{r}_i + m_j \mathbf{r}_j}{m_i + m_j} \,,
\label{eq:jacobi_A_y}
\end{align}
where $\boldsymbol\zeta$ is the internal coordinate of the pair and
$\mathbf{y}$ is the relative coordinate of $k$ with respect to the
pair's center of mass.
The kinetic energies transform as
$K_i + K_j + K_k = K_a + K_\zeta + K_y$, where $K_a$ is the
center-of-mass kinetic energy of $a$, $K_\zeta$ is the kinetic
energy conjugate to $\boldsymbol\zeta$ with reduced mass
$\mu_\zeta = m_i m_j/(m_i + m_j)$, and $K_y$ is the kinetic energy
conjugate to $\mathbf{y}$ with reduced mass
$\mu_y = m_k(m_i + m_j)/(m_i + m_j + m_k)$.
The internal Hamiltonian of the pair is
$h_b = h_i + h_j + K_\zeta + V_{ij}$, with eigenstates
$h_b |\phi_\alpha\rangle = \varepsilon_\alpha |\phi_\alpha\rangle$.
The bound states of $b$ (if any) correspond to discrete values of
$\alpha$, while continuum states are labeled by the relative momentum
$\mathbf{q}$ of the pair's constituents.
Here and below, the symbol $\sum_\alpha$ is used as a compact
notation for the sum over discrete bound states plus the integral
over continuum states.
These eigenstates form a complete set in $b$'s internal Hilbert space,
$\sum_\alpha |\phi_\alpha\rangle\langle\phi_\alpha| = \mathbf{1}_\zeta$.

For the single-particle channel, in which particle $i$ is detected
and the pair $(jk)$ is unresolved, the natural Jacobi coordinates are
\begin{align}
\boldsymbol\zeta' &= \mathbf{r}_j - \mathbf{r}_k \,,
\label{eq:jacobi_B_zeta} \\
\mathbf{y}' &= \mathbf{r}_i
  - \frac{m_j \mathbf{r}_j + m_k \mathbf{r}_k}{m_j + m_k} \,.
\label{eq:jacobi_B_y}
\end{align}
Expressing the constituent positions in either Jacobi set gives
\begin{equation}
\begin{aligned}
\begin{pmatrix}
\boldsymbol\zeta' \\[2pt] \mathbf y'
\end{pmatrix}
&= \mathcal J \begin{pmatrix}
\boldsymbol\zeta \\[2pt] \mathbf y
\end{pmatrix}, \\[4pt]
\mathcal J &= \begin{pmatrix}
-\dfrac{m_i}{m_i+m_j} & -1 \\[8pt]
\dfrac{m_j\,M_a}{(m_i+m_j)(m_j+m_k)}
& -\dfrac{m_k}{m_j+m_k}
\end{pmatrix},
\end{aligned}
\label{eq:jacobi_transformation}
\end{equation}
where $M_a = m_i+m_j+m_k$.
The determinant of the matrix in Eq.~(\ref{eq:jacobi_transformation})
is unity,
$\det\mathcal J = m_i m_k/[(m_i+m_j)(m_j+m_k)]
+m_jM_a/[(m_i+m_j)(m_j+m_k)] = 1$,
so the transformation preserves the internal volume element.
The unscaled coordinates are not related by an orthogonal rotation.
For the ${}^{6}\mathrm{Li}=\alpha+n+p$ specialization with
$m_n \approx m_p$ and
$m_i=m_n, m_j=m_p, m_k=m_\alpha$, Eq.~(\ref{eq:jacobi_transformation})
gives
$\boldsymbol\zeta' = -\tfrac{1}{2}\boldsymbol\zeta - \mathbf y$ and
$\mathbf y' = \tfrac{M_a}{2(m_p+m_\alpha)}\boldsymbol\zeta
 - \tfrac{m_\alpha}{m_p+m_\alpha}\mathbf y$,
so both coordinates mix with coefficients of order unity.
The same projectile state has equal normalization in the two sets,
$\Phi_a(\boldsymbol\zeta,\mathbf y) =
\Phi_a(\boldsymbol\zeta',\mathbf y')$, but a product
$\Phi_a\approx\phi_0(\boldsymbol\zeta)f(\mathbf y)$ generally
ceases to factorize after this transformation.  A cluster approximation
suited to pair detection therefore need not hold for single-particle
detection.

The projectile bound-state wave function
$\Phi_a(\boldsymbol\zeta, \mathbf{y})$, or equivalently
$\Phi_a(\boldsymbol\zeta', \mathbf{y}')$ after a Jacobi
transformation, satisfies
\begin{equation}
(K_\zeta + K_y + V_{ij} + V_{ik} + V_{jk}
  + h_i + h_j + h_k)\,\Phi_a
= -\epsilon_a\,\Phi_a \,,
\label{eq:bound_state}
\end{equation}
where $\epsilon_a > 0$ is the projectile binding energy and the
internal Hamiltonians $h_i$, $h_j$, $h_k$ act on the respective
internal coordinates.
For structureless constituents, $h_i = h_j = h_k = 0$ and
$\Phi_a$ depends only on $\boldsymbol\zeta$ and $\mathbf{y}$.
The total energy is $E = E_a - \epsilon_a + E_A$, where
$E_a = \hbar^2 k_a^2/(2\mu_a)$ is the entrance-channel kinetic
energy with reduced mass $\mu_a = m_a m_A/(m_a + m_A)$.
Throughout this paper I take the target ground-state energy as
the energy origin, $E_A = 0$, so that $E = E_a - \epsilon_a$ in
all subsequent formulas and the kinematic energies
$E_b, E_i, E_{k,\alpha}, E_{jkA}$ defined below are referred to
this origin.
Restoring $E_A\neq 0$ amounts to a uniform additive shift of
$\mathcal H_{PP}$ in Eq.~(\ref{eq:GPP_inverse}) and of all
target-excited energies $\omega_{A'}$ in
Eqs.~(\ref{eq:HjkAprime})--(\ref{eq:Qsector_reduction}) by the
same constant.
The entrance-channel distorted wave
$\chi_a^{(+)}(\mathbf R)$ is the solution of
$(K_a + U_a - E_a)\chi_a^{(+)} = 0$ with outgoing asymptotic
boundary conditions, where $\mathbf R$ is the projectile-target
relative coordinate and $U_a$ is the entrance-channel optical
potential describing elastic $a+A$ scattering.
Analogously, the exit-channel distorted waves
$\chi_b^{(-)}$ and $\chi_i^{(-)}$ satisfy the corresponding
exit-channel optical-model equations with incoming boundary
conditions and kinetic energies $E_b, E_i$ respectively.
I retain the finite target mass and use relative Jacobi coordinates
in the overall center-of-mass frame.
The natural exit-channel Jacobi coordinates are
$\mathbf{r}_b$, the detected-pair center of mass (CoM) relative to the CoM of the
residual $k+A$ system in Partition A, and $\mathbf{r}_i$, the
detected-particle position relative to the CoM of the residual
$jk+A$ system in Partition B.
They are conjugate to the exit-channel kinetic operators $K_b$ and
$K_i$ and are the arguments of the distorted waves
$\chi_b^{(-)}$, $\chi_i^{(-)}$ and of the auxiliary potentials
$U_b$, $U_i$ that generate them.
The microscopic interactions $V_{iA}, V_{jA}, V_{kA}$, by contrast,
are local in the fragment-target separations, which differ from the
Jacobi coordinates by the recoil of the residual system,
\begin{align}
\mathbf{r}_{bA}
&= \mathbf{r}_b + \frac{m_k}{m_k+m_A}\,\mathbf{r}_{kA},
\notag \\
\mathbf{r}_{iA}
&= \mathbf{r}_i + \frac{m_j+m_k}{m_j+m_k+m_A}\,\mathbf{r}_{(jk)A},
\label{eq:fragtarget}
\end{align}
where $\mathbf{r}_{bA}$ ($\mathbf{r}_{iA}$) is the position of the
pair CoM (of particle $i$) relative to the target,
$\mathbf{r}_{kA}=\mathbf{r}_k-\mathbf{r}_A$ is the unresolved
$k$-$A$ relative coordinate in Partition A, and $\mathbf{r}_{(jk)A}$
is the corresponding $(jk)$-CoM--$A$ coordinate in Partition B.
The fragment-target and exit Jacobi coordinates coincide only as
$m_A\to\infty$.  I use Eq.~(\ref{eq:fragtarget}) in the source
integrals [Eqs.~(\ref{eq:ViA_VjA}), (\ref{eq:rho_A_coup_coord}),
(\ref{eq:source_6Li_ref}), and (\ref{eq:source_6Li_B})] and multipole
expansions [Eqs.~(\ref{eq:tidal_6Li}), (\ref{eq:ViA_multipole})].
The recoil shift adds a dependence on the residual coordinate
$\mathbf r_{kA}$ or $\mathbf r_{(jk)A}$ to the interaction inside
the source integral.  These are already coordinates of the unresolved
system: in Partition A, $\mathbf r_{kA}$ is the output coordinate
$\mathbf r_k$ of $\rho_\alpha^{(A)}(\mathbf r_k,\xi)$.
Recoil thus introduces no additional integration variable.

The entrance-channel arguments follow from the same exact geometry:
in Partition A,
\begin{equation}
\mathbf{y} = \mathbf{r}_{kA} - \mathbf{r}_{bA} \,,
\qquad
\mathbf{R} = \frac{m_b \mathbf{r}_{bA} + m_k \mathbf{r}_{kA}}{M_a}
\,,
\label{eq:exit_to_entrance}
\end{equation}
with $m_b = m_i + m_j$, and analogously
$\mathbf{y}' = \mathbf{r}_{iA} - \mathbf{r}_{(jk)A}$ and
$\mathbf{R} = [m_i \mathbf{r}_{iA}
+ (m_j+m_k)\,\mathbf{r}_{(jk)A}]/M_a$ in Partition B.
These linear transformations have unit Jacobian.

For kinetic operators, the symbols $K_i, K_j, K_k$ in
Eq.~(\ref{eq:H_full}) are lab-frame single-particle kinetic
energies; in any subsystem Hamiltonian they are replaced by
reduced-mass kinetic operators conjugate to the appropriate
Jacobi coordinates.
Specifically, in $H_{\mathrm{exit}}^{(A)}$
[Eq.~(\ref{eq:Hexit_A})] $K_b$ has reduced mass
$\mu_b = m_b m_{kA}/(m_b+m_{kA})$ with $m_b=m_i+m_j$,
$m_{kA}=m_k+m_A$;
in $H_{kA}$ the operator $K_k$ has reduced mass
$\mu_{kA} = m_k m_A/(m_k+m_A)$;
in $H_{\mathrm{exit}}^{(B)}$ [Eq.~(\ref{eq:Hexit_B})] $K_i$ has
reduced mass $\mu_i = m_i m_{jkA}/(m_i+m_{jkA})$ with
$m_{jkA}=m_j+m_k+m_A$;
and in $H_{jkA}$ [Eq.~(\ref{eq:HjkA})] $K_j+K_k$ stand for the
two reduced-mass kinetic operators conjugate to the $jk$ internal
and $(jk)$-$A$ relative coordinates.

For later use, I define the composite interactions between the pair
$b = (ij)$ and the remaining particle $k$ and target $A$:
\begin{align}
V_{bk} &= V_{ik} + V_{jk} \,,
\label{eq:Vbk} \\
V_{bA} &= V_{iA} + V_{jA} \,.
\label{eq:VbA}
\end{align}
These interactions depend on the pair's internal coordinate
$\boldsymbol\zeta$ because the positions of $i$ and $j$ relative
to the target or to $k$ depend on their separation within the pair.

\section{Pair-detection inclusive breakup}
\label{sec:pair}

I first consider the pair-detection channel
$a + A \to b + (k\!+\!A)^*$, in which the correlated pair
$b = (ij)$ is detected in a definite internal state $\phi_\alpha$
while the remaining particle $k$ and target $A$ are unresolved.
The pair is selected by projecting the full projectile wave function
onto $\phi_\alpha$; it need not be a preformed entrance-channel
cluster.  Summing over the unobserved $k+A$ states then follows the
usual IAV construction.

The exit-channel Hamiltonian for this partition is
\begin{equation}
H_{\mathrm{exit}}^{(A)} = h_b + K_b + U_b + H_{kA} \,,
\label{eq:Hexit_A}
\end{equation}
where $h_b = h_i + h_j + K_\zeta + V_{ij}$ is the internal
Hamiltonian of the pair, $K_b$ is the center-of-mass kinetic energy
of $b$, $U_b$ is an auxiliary optical potential that generates the
distorted wave $\chi_b^{(-)}$ for the pair, and
$H_{kA} = H_A + h_k + K_k + V_{kA}$ is the full $k + A$ Hamiltonian.
This exit-channel Hamiltonian is separable in $b$'s degrees of
freedom and the $k + A$ system, so the exit-channel eigenstates
factorize as
$|\chi_b^{(-)}(\mathbf{k}_b)\rangle\,|\phi_\alpha\rangle\,
|\Psi_{kA}^c\rangle$.
The post-form residual interaction is
\begin{equation}
V_{\mathrm{post}}^{(A)} = H - H_{\mathrm{exit}}^{(A)}
= V_{bk} + V_{bA} - U_b \,,
\label{eq:Vpost_A}
\end{equation}
Both $V_{bk}$ and $V_{bA}$ retain their dependence on
$\boldsymbol\zeta$ through the positions of $i$ and $j$.

The DWBA transition amplitude for detecting $b$ in internal state
$\phi_\alpha$ while the $k + A$ system is in eigenstate $c$ is
\begin{equation}
T_{\alpha,c}^{(A)}
= \langle \chi_b^{(-)} \phi_\alpha\, \Psi_{kA}^c |\,
V_{\mathrm{post}}^{(A)} \,| \chi_a^{(+)} \Phi_a \Phi_A \rangle \,.
\label{eq:T_A}
\end{equation}
The state-resolved doubly differential inclusive cross section, summed
over all unobserved final states of the $k + A$ system, is
\begin{equation}
\frac{d^2\sigma_\alpha^{(A)}}{dE_b\, d\Omega_b}
= \frac{(2\pi)^4}{v_a}\,
\sum_c |T_{\alpha,c}^{(A)}|^2\, \delta(E_{k,\alpha} - E^c) \,,
\label{eq:xsec_A}
\end{equation}
where $v_a$ is the $a$-$A$ relative velocity and
$E_{k,\alpha} = E - E_b - \varepsilon_\alpha$ is the energy available
to the $k + A$ system when the pair is in internal state $\phi_\alpha$
with center-of-mass kinetic energy $E_b$.
The factor $(2\pi)^4/v_a$ corresponds to the standard IAV
normalization in which momentum eigenstates are normalized as
$\langle \mathbf{k}|\mathbf{k}'\rangle = \delta(\mathbf{k}-\mathbf{k}')$
(plane-wave $\delta$-normalization) and the detected-fragment
phase-space factor is included in the chosen differential
$T$-matrix convention.
Throughout, the entrance and exit distorted waves
$\chi_a^{(+)}(\mathbf R), \chi_b^{(-)}(\mathbf r_b),
\chi_i^{(-)}(\mathbf r_i)$ are momentum-normalized in the same
convention, with asymptotic plane-wave amplitudes of unit
strength; the corresponding $\delta(\mathbf k-\mathbf k')$
normalization on the asymptotic momentum carries through to the
on-shell $T$-matrix and to the spectral identity for the
unresolved-subsystem resolvent.
Using a different scattering-state normalization would multiply all
sum-rule expressions below by the same kinematic factor and would not
alter any source or kernel identity.

I define the source function by projecting onto the pair's internal
state and the distorted wave,
\begin{equation}
|\rho_\alpha^{(A)}\rangle
= \langle \phi_\alpha\, \chi_b^{(-)} |\,
V_{\mathrm{post}}^{(A)} \,| \chi_a^{(+)} \Phi_a \Phi_A \rangle \,,
\label{eq:source_A}
\end{equation}
which is a state in the $k + A$ Hilbert space depending on
$\mathbf{r}_k$ and $\xi$; here and below, $\mathbf{r}_k$ denotes
the $k$-$A$ relative coordinate $\mathbf{r}_{kA}$ of
Eq.~(\ref{eq:fragtarget}), with the subscript $A$ dropped on this
variable for brevity.
Substitution into Eq.~(\ref{eq:xsec_A}) gives
\begin{equation}
\frac{d^2\sigma_\alpha^{(A)}}{dE_b\, d\Omega_b}
= \frac{(2\pi)^4}{v_a}
\sum_c \langle \rho_\alpha^{(A)}|\Psi_{kA}^c\rangle
\langle \Psi_{kA}^c|\rho_\alpha^{(A)}\rangle\,
\delta(E_{k,\alpha} - E^c),
\end{equation}
so that the summation over unobserved final states
$\{|\Psi_{kA}^c\rangle\}$ is an on-shell projection.
Using the spectral representation of the resolvent
$G_{kA}^{\mathrm{full}}(z) = \sum_c |\Psi_{kA}^c\rangle(z-E^c)^{-1}
\langle\Psi_{kA}^c|$ and the Sokhotski-Plemelj identity
$(E - H_{kA} + i0^+)^{-1} - (E - H_{kA} - i0^+)^{-1}
= -2\pi i \delta(E-H_{kA})$, one has
$\sum_c |\Psi_{kA}^c\rangle\delta(E-E^c)\langle\Psi_{kA}^c|
= -\pi^{-1}\mathrm{Im}\,G_{kA}^{\mathrm{full}}(E+i0^+)$, which
converts the on-shell sum into the imaginary part of the resolvent
and yields the master sum rule
\begin{equation}
\frac{d^2\sigma_\alpha^{(A)}}{dE_b\, d\Omega_b}
= -\frac{(2\pi)^4}{\pi v_a}\,
\mathrm{Im}\,
\langle \rho_\alpha^{(A)} |\, G_{kA}^{\mathrm{full}} \,|
\rho_\alpha^{(A)} \rangle \,.
\label{eq:master_A}
\end{equation}
where $G_{kA}^{\mathrm{full}} = (E_{k,\alpha}^+ - H_{kA})^{-1}$ is
the full $k + A$ resolvent.
This sum rule is exact once the DWBA transition amplitude
[Eq.~(\ref{eq:T_A})] is adopted.
The resolvent $G_{kA}^{\mathrm{full}}$ describes \emph{two-body}
relative motion of $k$ and $A$ on top of the internal target
dynamics carried by $H_A$.
The fully two-body optical propagator $G_k$ used in the practical
formulas below arises only after Feshbach projection onto the target
ground state.
All the physics of the three-body projectile and the pair's internal
structure enters through the source $\rho_\alpha^{(A)}$.

For later comparison with optical-model calculations, I introduce a
reference interaction $U_{bA}$ that is local in the exact
pair-target separation $\mathbf{r}_{bA}$ [Eq.~(\ref{eq:fragtarget})]
and is diagonal in the target ground-state space.
It need not be identical to the distorting potential $U_b$, which is
a function of the Jacobi coordinate $\mathbf{r}_b$, although
choosing the two with the same functional form is often convenient
in practical calculations.
Because
$V_{\mathrm{post}}^{(A)} = (V_{bk} + U_{bA} - U_b) + (V_{bA} - U_{bA})$,
the source decomposes into a reference-channel part and an explicit
pair-target coupling part,
$|\rho_\alpha^{(A)}\rangle = |\rho_\alpha^{(A,\mathrm{ref})}\rangle
+ |\rho_\alpha^{(A,\mathrm{coup})}\rangle$, where
\begin{align}
|\rho_\alpha^{(A,\mathrm{ref})}\rangle
&= \langle \phi_\alpha\, \chi_b^{(-)} |\,
(V_{bk} + U_{bA} - U_b) \,| \chi_a^{(+)} \Phi_a \Phi_A \rangle \,,
\label{eq:rho_A_ref} \\
|\rho_\alpha^{(A,\mathrm{coup})}\rangle
&= \langle \phi_\alpha\, \chi_b^{(-)} |\,
(V_{bA} - U_{bA}) \,| \chi_a^{(+)} \Phi_a \Phi_A \rangle \,.
\label{eq:rho_A_coup}
\end{align}
None of the terms in $V_{bk}+U_{bA}-U_b$ excites the target.
The interaction $V_{bk}=V_{ik}+V_{jk}$ depends on
$\boldsymbol\zeta$ and couples different pair states.  The local
term $U_{bA}-U_b$ leaves the target unchanged.  If these potentials
have the same functional form, their difference comes only from
the recoil shift between $\mathbf r_{bA}$ and $\mathbf r_b$
[Eq.~(\ref{eq:fragtarget})] and vanishes as $m_A\to\infty$.
The remaining interaction $V_{bA}-U_{bA}$ depends on both
$\boldsymbol\zeta$ and the target coordinates $\xi$.  Its source is
\begin{align}
&\rho_\alpha^{(A,\mathrm{coup})}(\mathbf{r}_k, \xi)
= \int d\mathbf{r}_b\, d\boldsymbol\zeta\;
\chi_b^{(-)*}\,\phi_\alpha^*
\notag \\
&\qquad \times
[V_{bA}(\mathbf{r}_{bA}, \boldsymbol\zeta, \xi)
- U_{bA}(\mathbf{r}_{bA})]\,
\chi_a^{(+)}\,\Phi_a\,\Phi_A(\xi) \,,
\label{eq:rho_A_coup_coord}
\end{align}
where the integration runs over the Jacobi coordinate
$\mathbf{r}_b$ and the pair internal coordinate $\boldsymbol\zeta$,
while $V_{bA}$ and $U_{bA}$ are evaluated at the exact pair-target
separation
$\mathbf{r}_{bA} = \mathbf{r}_b + \frac{m_k}{m_k+m_A}\,\mathbf{r}_k$
via the recoil relation [Eq.~(\ref{eq:fragtarget})], with
$\mathbf{r}_k$ the output variable of the source.
The arguments of $\chi_a^{(+)}$ and $\Phi_a$ are likewise the exact
combinations $\mathbf{R}$ and $\mathbf{y}$ of
Eq.~(\ref{eq:exit_to_entrance}).
For the explicit pair-target interaction, $V_{bA} = V_{iA} + V_{jA}$
with
\begin{align}
V_{iA} &= V_{iA}(\mathbf{r}_{bA} + \tfrac{m_j}{m_i+m_j}
  \boldsymbol\zeta,\, \xi) \,,
\notag \\
V_{jA} &= V_{jA}(\mathbf{r}_{bA} - \tfrac{m_i}{m_i+m_j}
  \boldsymbol\zeta,\, \xi) \,,
\label{eq:ViA_VjA}
\end{align}
The arguments are the constituent-target separations, including the
recoil shift in $\mathbf r_{bA}$ from Eq.~(\ref{eq:fragtarget}).

The cross section separates into three terms,
\begin{equation}
\frac{d^2\sigma_\alpha^{(A)}}{dE_b\, d\Omega_b}
= \sigma_\alpha^{(\mathrm{ref})}
+ \sigma_\alpha^{(\mathrm{int})}
+ \sigma_\alpha^{(\mathrm{coup})} \,,
\label{eq:three_terms_A}
\end{equation}
Equation~(\ref{eq:three_terms_A}) follows by expanding the quadratic
form in the two source terms; no approximation is made.
To evaluate these terms, I split the sources with the target-ground-state
projector $P = |\Phi_A\rangle\langle\Phi_A|$ and its complement
$Q = \mathbf{1} - P$, both acting in target Hilbert space only.
Because the reference-channel operator $V_{bk} + U_{bA} - U_b$
contains no explicit target-excitation piece, the reference source lives
entirely in $P$-space,
$P|\rho_\alpha^{(A,\mathrm{ref})}\rangle = |\rho_\alpha^{(A,\mathrm{ref})}\rangle$.
The explicit coupling operator $V_{bA} - U_{bA}$, on the other hand,
has both a component within $P$ and transitions from $P$ to $Q$, so
\begin{equation}
|\rho_\alpha^{(A,\mathrm{coup})}\rangle
= |\rho_\alpha^{(A,\mathrm{coup}),P}\rangle
+ |\rho_\alpha^{(A,\mathrm{coup}),Q}\rangle \,,
\label{eq:rho_PQ}
\end{equation}
with the $P$- and $Q$-components defined by acting with $P$ or $Q$
on the target side of $V_{bA}\,\Phi_A$ before contracting with
$\langle\phi_\alpha\,\chi_b^{(-)}|$,
\begin{align}
|\rho_\alpha^{(A,\mathrm{coup}),P}\rangle
&= |\Phi_A\rangle\,\langle\phi_\alpha\,\chi_b^{(-)}|
\bigl(\langle\Phi_A| V_{bA} |\Phi_A\rangle - U_{bA}\bigr)
\notag \\
&\qquad \times
| \chi_a^{(+)}\,\Phi_a\rangle,
\notag \\
|\rho_\alpha^{(A,\mathrm{coup}),Q}\rangle
&= \langle\phi_\alpha\,\chi_b^{(-)}|\,
\bigl[Q\,V_{bA}\,|\Phi_A\rangle\bigr]\,
| \chi_a^{(+)}\,\Phi_a\rangle.
\label{eq:rho_coup_PQ_defs}
\end{align}
The $P$-diagonal piece vanishes identically when $U_{bA}$ is chosen
to coincide with the ground-state folding potential
$\langle\Phi_A|V_{bA}|\Phi_A\rangle$; for any other choice it
is the difference between the folding and reference potentials
acting in the source, as in Ref.~\cite{Lei2026nsp}.
The $Q$ component describes target excitation by $V_{bA}$, which
is absent from ordinary closure-based IAV.

To decompose $G_{kA}^{\mathrm{full}}$, first define the resolvent
restricted to the excited-target space,
\begin{equation}
g_Q^{(A)} = (E_{k,\alpha}^+ - QH_{kA}Q)^{-1}.
\label{eq:gQ_A}
\end{equation}
Then
\begin{align}
G_{PP}^{\mathrm{exact}}
&= P G_{kA}^{\mathrm{full}} P,
\notag\\
G_{PQ}
&= G_{PP}^{\mathrm{exact}}\,P V_{kA}Q\,g_Q^{(A)},
\notag\\
G_{QP}
&= g_Q^{(A)}\,Q V_{kA}P\,G_{PP}^{\mathrm{exact}},
\notag\\
G_{QQ}^{\mathrm{full}}
&= g_Q^{(A)}
 + g_Q^{(A)} Q V_{kA}P\,G_{PP}^{\mathrm{exact}}\,
   P V_{kA}Q\,g_Q^{(A)} .
\label{eq:Feshbach_A_blocks}
\end{align}
The full $Q$-$Q$ block contains
the direct propagation in the target-excited space plus
$Q\to P\to Q$ rescattering through the target-ground-state block.
I abbreviate the prefactor and sources as
\begin{align*}
C_a &= \frac{(2\pi)^4}{\pi v_a},\\
|\rho_r\rangle &= |\rho_\alpha^{(A,\mathrm{ref})}\rangle,\\
|\rho_P\rangle &= |\rho_\alpha^{(A,\mathrm{coup}),P}\rangle,\\
|\rho_Q\rangle &= |\rho_\alpha^{(A,\mathrm{coup}),Q}\rangle .
\end{align*}
In the following expressions I write the $P$-sector block as
$G_k$ when it is represented by the usual elastic optical propagator,
with the dynamic polarization from eliminated target excitations
absorbed into $U_k$.
This replacement is the same optical reduction used in ordinary IAV;
it is not part of the unreduced DWBA identity in
Eq.~(\ref{eq:master_A}).
The off-diagonal blocks $G_{PQ}$, $G_{QP}$ and the full block
$G_{QQ}^{\mathrm{full}}$ are kept as Feshbach-block notation unless
an additional reduction is stated.
\begin{widetext}
\begin{align}
\sigma_\alpha^{(\mathrm{ref})}
&= -C_a\,\mathrm{Im}\,
\langle \rho_r |G_k| \rho_r \rangle \,,
\label{eq:sigma_ref_A}\\
\sigma_\alpha^{(\mathrm{int})}
&= -C_a\,\mathrm{Im}\,
\Bigl[
\langle \rho_r |G_k| \rho_P \rangle
+ \langle \rho_P |G_k| \rho_r \rangle
+ \langle \rho_r |G_{PQ}| \rho_Q \rangle
+ \langle \rho_Q |G_{QP}| \rho_r \rangle
\Bigr] \,,
\label{eq:sigma_int_A}\\
\sigma_\alpha^{(\mathrm{coup})}
&= -C_a\,\mathrm{Im}\,
\Bigl[
\langle \rho_P |G_k| \rho_P \rangle
+ \langle \rho_P |G_{PQ}| \rho_Q \rangle
+ \langle \rho_Q |G_{QP}| \rho_P \rangle
+ \langle \rho_Q |G_{QQ}^{\mathrm{full}}| \rho_Q \rangle
\Bigr] \,.
\label{eq:sigma_coup_A}
\end{align}
\end{widetext}
Equation~(\ref{eq:sigma_coup_A}) contains propagation in the target
ground state, interference between ground and excited target states,
and propagation between excited target states.  Changing $U_{bA}$
redistributes strength between $\rho_r$ and $\rho_P$, so the
separated contributions are not observable.  Their sum is fixed by
$\rho_\alpha^{(A)}$.
The last term cannot be absorbed into a choice of the elastic
reference interaction $U_{bA}$.
The interference terms have been kept as explicit $G_{PQ}$ and
$G_{QP}$ matrix elements because the outgoing resolvent is not a
Hermitian operator; only after a further reduction to an absorptive
kernel can they be recast as real interference terms.
For the usual elastic-breakup (EBU) and NEB separation, I first
restrict the source to the target ground state.  The excited-target piece
$\rho_\alpha^{(A,\mathrm{coup}),Q}$ of
Eq.~(\ref{eq:rho_coup_PQ_defs}) is dropped, leaving the
single-channel total source
$\tilde\rho_\alpha^{(A,\mathrm{tot})}
=\tilde\rho_\alpha^{(A,\mathrm{ref})}
+\tilde\rho_\alpha^{(A,\mathrm{coup}),P}$
entirely in the target-ground-state $P$-sector.
I then apply the two-potential
(or Kawai) identity~\cite{Ichimura1985,Lei2018prior},
which represents the imaginary part of the full $k+A$ resolvent in the
optical-model approximation as
\begin{equation}
-\pi^{-1}\mathrm{Im}\,G_{kA}^{\mathrm{full}}
\;\approx\;
|\chi_k^{(+)}\rangle\langle\chi_k^{(+)}|_{\rm on\text{-}shell}
- \pi^{-1}\,G_k^{\dagger}\,W_k\,G_k \,,
\label{eq:two-potential-A}
\end{equation}
where $G_k = (E_{k,\alpha}^+ - K_k - U_k)^{-1}$ is the
elastic-channel optical Green's function for the $k+A$ system,
$U_k = V_k + iW_k$ with $W_k\le 0$, and $|\chi_k^{(+)}\rangle$ is
the outgoing optical distorted wave at energy $E_{k,\alpha}$
(with $E_A=0$ from Sec.~\ref{sec:hamiltonian}; restoring
$E_A\neq 0$ adds $-E_A$ to the denominator).
The first term is the elastic-channel on-shell
density of $G_{kA}^{\mathrm{full}}$, identified at the optical
level with $|\chi_k^{(+)}\rangle\langle\chi_k^{(+)}|$, and is
not extractable from $\mathrm{Im}\,G_k$ alone (the optical $G_k$
has no real-axis pole); the second is the inelastic spectral
density of $G_{kA}^{\mathrm{full}}$ approximated by the absorptive
part of $G_k$, which follows from the algebraic identity
$G_k - G_k^\dagger = 2i\,G_k^\dagger W_k G_k$ and represents the
non-elastic flux loss in the optical sense.
Both pieces contribute to the master formula
[Eq.~(\ref{eq:master_A})] when the source is restricted to its
$P$-sector, yielding the decomposition into NEB and EBU
\begin{align}
\sigma_\alpha^{(A,\mathrm{NEB})}
&= -\frac{(2\pi)^4}{\pi v_a}
\langle \tilde\rho_\alpha^{(A,\mathrm{tot})}|
G_k^\dagger W_k G_k
|\tilde\rho_\alpha^{(A,\mathrm{tot})}\rangle,
\label{eq:NEB_A}\\
\sigma_\alpha^{(A,\mathrm{EBU})}
&= \frac{(2\pi)^4}{v_a}
\bigl|\langle\chi_k^{(+)}|
\tilde\rho_\alpha^{(A,\mathrm{tot})}\rangle\bigr|^2 \,,
\label{eq:EBU_A}\\
\Bigl[\frac{d^2\sigma_\alpha^{(A)}}{dE_b\, d\Omega_b}
\Bigr]_{\!\rm 1\text{-}ch}
&= \sigma_\alpha^{(A,\mathrm{NEB})}+\sigma_\alpha^{(A,\mathrm{EBU})}.
\label{eq:master_A_practical}
\end{align}
The subscript ``$1$-ch'' denotes the restriction of the source to
$P$.  Equation~(\ref{eq:master_A_practical}) omits all matrix
elements containing $\rho_Q$: the $PQ$ and $QP$ interferences in
Eqs.~(\ref{eq:sigma_int_A})--(\ref{eq:sigma_coup_A}) and
$\langle\rho_Q|G_{QQ}^{\mathrm{full}}|\rho_Q\rangle$.
These terms remain in the unreduced sum rule.

Equation~(\ref{eq:NEB_A}) has the IAV NEB form, with $G_k$ in
place of $G_x$ and a pair-projected source.  In the EBU term
[Eq.~(\ref{eq:EBU_A})], $k$ scatters elastically from the target
and the pair exits in state $\phi_\alpha$.  The two contributions
are separate reaction channels within this reduction.  For a detected
deuteron, $\alpha=0$ and $\phi_\alpha=\phi_d$, the cluster limit
recovers the usual DWBA elastic-breakup amplitude.  Both NEB and EBU
sample the projectile structure through the source; the propagator
remains that of $k+A$.

Choosing $\alpha = 0$ (the pair's ground state, if bound) gives the
cross section for detecting an intact pair, while
$\alpha = (\mathbf{q})$ (a continuum state with relative momentum
$\mathbf{q}$) gives the inclusive coincidence cross section for
detecting the pair's breakup products at specified relative energy
and angle.
The total inclusive cross section, summed over all internal states of
the pair, is
\begin{equation}
\frac{d^2\sigma^{(A)}}{dE_b\, d\Omega_b}
= \sum_\alpha
\frac{d^2\sigma_\alpha^{(A)}}{dE_b\, d\Omega_b} \,,
\label{eq:total_A}
\end{equation}
where the shorthand $\sum_\alpha$ stands for a discrete sum over
bound pair states plus a continuum integral.
For the discrete part each
$d^2\sigma_\alpha^{(A)}/(dE_b\,d\Omega_b)$ is added directly; the
continuum part is added either as a phase-space integral over the
density $d^5\sigma/(dE_b\,d\Omega_b\,d^3\mathbf q)$ in the
continuum normalization, or as a sum over discrete-bin cross
sections in the bin normalization, using the conventions specified
below.
Only in the additional approximation that the $\alpha$ dependence of
the unresolved propagator is weak over the populated pair states,
so that
$G_{k,\alpha}^{\mathrm{full}}
= (E - E_b - \varepsilon_\alpha - H_{kA})^{-1}$
may be replaced by a common reduced propagator
$G_k(\bar E_k)$ at some representative energy
$\bar E_k = E - E_b - \bar\varepsilon$, with
$\bar\varepsilon$ a representative pair-state energy chosen
within the populated $\alpha$-window (e.g., a centroid value),
does closure give
\begin{equation}
\frac{d^2\sigma^{(A)}}{dE_b\, d\Omega_b}
\approx -\frac{(2\pi)^4}{\pi v_a}\,
\mathrm{Im}\,
\langle \mathcal{S}^{(A)} |\,
\mathbf{1}_\zeta \otimes G_k(\bar E_k)\,
| \mathcal{S}^{(A)} \rangle \,,
\label{eq:closure_A}
\end{equation}
where $\mathcal{S}^{(A)}(\mathbf{r}_k, \boldsymbol\zeta) =
\langle\chi_b^{(-)}|V_{\mathrm{post}}^{(A)}\,\chi_a^{(+)}
\Phi_a\rangle_{\mathbf{r}_b}$ is the unprojected source and
$\mathbf{1}_\zeta = \sum_\alpha|\phi_\alpha\rangle\langle\phi_\alpha|$
is the pair-Hilbert completeness relation.
The replacement $G_{k,\alpha}\to G_k(\bar E_k)$ is the only step
beyond exactness in Eq.~(\ref{eq:closure_A}).
The operator-level error has the formal form
\begin{equation}
G_{k,\alpha} - G_k(\bar E_k)
= G_{k,\alpha}\,(\varepsilon_\alpha-\bar\varepsilon)\,
G_k(\bar E_k),
\label{eq:closure_bound}
\end{equation}
which is a resolvent identity, not a uniform norm bound:
both $G_{k,\alpha}$ and $G_k(\bar E_k)$ are unbounded operators
on the continuous spectrum, so a strict $\|\cdot\|_{\rm op}$
estimate would require restricting to states sufficiently close
to the on-shell pole, where the absorptive width $|W_k|$
regulates the resolvent.
The schematic dimensional scale relevant for that on-shell region
is $\Delta\varepsilon_{\mathrm{pair}}/|W_k|$, where
$\Delta\varepsilon_{\mathrm{pair}}\equiv
\sup_{\alpha\in\text{populated}}|\varepsilon_\alpha-\bar\varepsilon|$
is the energy spread of the populated pair states relative to
the chosen $\bar\varepsilon$, and $|W_k|$ stands for the magnitude
of the on-shell absorptive optical potential of the $kA$ system,
$\langle\chi_k^{(+)}||W_k|\,|\chi_k^{(+)}\rangle$ (treated as a
positive scalar scale here).
The corresponding on-shell width
$\Gamma_{kA}\equiv -2\langle\chi_k^{(+)}|W_k|\chi_k^{(+)}\rangle
= 2|W_k|$ (in the same on-shell sense) provides the natural
absorptive scale.
Equation~(\ref{eq:closure_A}) is therefore reliable
when $\Delta\varepsilon_\mathrm{pair}\ll\Gamma_{kA}$,
trivial when only a single bound pair state is retained, and
degrades once the populated continuum window approaches the $kA$
optical imaginary scale, in agreement with the qualitative
discussion of Ref.~\cite{Lei2026nsp}.
The state-resolved sum rule [Eqs.~(\ref{eq:master_A_practical})
and~(\ref{eq:total_A})] bypasses this replacement entirely.

For the coincidence observable, the pair's internal state label
$\alpha$ is the relative momentum $\mathbf{q}$ of the constituents.
Two normalization conventions are used in practice.
The continuum-normalized pair state
$|\phi_{\mathbf q}\rangle$ satisfies
$\langle\phi_{\mathbf q}|\phi_{\mathbf q'}\rangle_\zeta
= \delta^3(\mathbf q - \mathbf q')$, for which the state-resolved
cross section
$d^2\sigma_{\mathbf q}^{(A)}/(dE_b\,d\Omega_b)$ inherits an extra
phase-space factor per unit $\mathbf q$ and is more appropriately
written as a quintuple differential
$d^5\sigma/(dE_b\,d\Omega_b\,d^3\mathbf q)$.
The bin state
$|\tilde\phi_n\rangle = \mathcal N_n^{-1}
\int_{q_n}^{q_{n+1}} dq\,q^2
\int d\Omega_{\mathbf q}\,g_n(\hat{\mathbf q})|\phi_{\mathbf q}\rangle$,
with weight $g_n$ and normalization
$\mathcal N_n^2 = \int_{q_n}^{q_{n+1}}dq\,q^2
\int d\Omega_{\mathbf q}|g_n|^2$, satisfies
$\langle\tilde\phi_n|\tilde\phi_m\rangle = \delta_{nm}$, so that
$d^2\sigma_{\tilde\phi_n}^{(A)}/(dE_b\,d\Omega_b)$ retains the
dimensions of an ordinary doubly differential cross section and
can be summed as in Eq.~(\ref{eq:total_A}) with discrete $\alpha$.
The bin cross section is the coherent square of a bin amplitude,
not a literal $|g_n|^2$-weighted integral of the continuum density.
Equivalently,
\begin{equation}
T_{\tilde\phi_n}^{(A)}
=\mathcal N_n^{-1}\int_{q_n}^{q_{n+1}}dq\,q^2
\int d\Omega_{\mathbf q}\,g_n(\hat{\mathbf q})T_{\mathbf q}^{(A)} ,
\end{equation}
and
$d^2\sigma_{\tilde\phi_n}^{(A)}/(dE_b\,d\Omega_b)
=(2\pi)^4v_a^{-1}|T_{\tilde\phi_n}^{(A)}|^2$.
Only in the narrow-bin limit, where $T_{\mathbf q}^{(A)}$ varies
slowly across the bin, does this coherent expression reduce to the
integrated continuum density (for the usual top-hat bin weight).
Bin and continuum formulations carry the same physical content
when the bin widths are small compared with the variation scale of
the source and the $kA$ resolvent.
One may then transform from the pair center-of-mass variables
$(E_b, \Omega_b, \mathbf{q})$ to the laboratory variables of the
individual particles $(E_i, \Omega_i, E_j, \Omega_j)$ via the
appropriate Jacobian to obtain the corresponding pair-resolved
inclusive distribution,
\begin{equation}
\frac{d^6\sigma}{dE_i\, d\Omega_i\, dE_j\, d\Omega_j}
= J(\mathbf{q}; E_i, \Omega_i, E_j, \Omega_j)\;
\frac{d^5\sigma}{dE_b\, d\Omega_b\, d^3\mathbf q} \,,
\label{eq:coincidence}
\end{equation}
where $J$ is the kinematic Jacobian and the right-hand side is
evaluated at the $(E_b, \Omega_b, \mathbf{q})$ values corresponding
to the specified laboratory kinematics through momentum and energy
conservation.
If a discretized bin state is used instead, Eq.~(\ref{eq:coincidence})
is first applied to the continuum density and then integrated over
the bin with the weight appearing in Eq.~(\ref{eq:total_A}); the bin
cross section itself should not be read as a density in
$d^3\mathbf q$.
This observable is semi-inclusive: it is exclusive with respect to the measured kinematics
of particles $i$ and $j$, but remains inclusive with respect to all
final states of the unresolved $k + A$ subsystem.
Once particles $i$ and $j$ are kinematically measured, four-body energy-momentum
conservation together with a ground-state target leaves only a
discrete set of kinematic configurations for $k$; under that
specialization the present observable reduces to an exclusive
three-body breakup measurement of the type
$A(a,ij)k$~\cite{Keeley2007,Villanueva2024}.
The inclusive sum in Eq.~(\ref{eq:coincidence}) runs over target
internal states and those
$k$ kinematics that are not already fixed by the measured
$(\mathbf{k}_i,\mathbf{k}_j)$, i.e.\ target-excitation channels
together with compound-nucleus formation in $k+A$.

The practical Jacobian $J$ is the standard kinematic transformation
from the pair center-of-mass variables $(E_b,\Omega_b,\mathbf q)$
to the individual-particle laboratory variables, determined by
momentum conservation inside the pair and independent of the
dynamical content of the sum rule.
Equation~(\ref{eq:coincidence}) is the laboratory-frame representation
of the pair-resolved sum rule; it does not supply an exact three-body
asymptotic breakup amplitude.

\section{Single-particle inclusive breakup}
\label{sec:single}

I now consider the single-particle channel
$a + A \to i + (jk\!+\!A)^*$, in which particle $i$ is detected
while the subsystem $j + k + A$ is unresolved.
I treat $i$ as structureless below.  For composite $i$, one retains
$h_i$ in the exit Hamiltonian and projects onto its internal state,
as in Ref.~\cite{Lei2026nsp}.

The exit-channel Hamiltonian for this partition is
\begin{equation}
H_{\mathrm{exit}}^{(B)} = h_i + K_i + U_i + H_{jkA} \,,
\label{eq:Hexit_B}
\end{equation}
where $U_i$ is an auxiliary optical potential generating the distorted
wave $\chi_i^{(-)}$ for particle $i$, and
\begin{equation}
H_{jkA} = H_A + h_j + h_k + K_j + K_k + V_{jk} + V_{jA} + V_{kA} \,,
\label{eq:HjkA}
\end{equation}
is the full three-body $jk + A$ Hamiltonian.
Within this unresolved subsystem I use $\mathbf r_j$ and
$\mathbf r_k$ to denote the exact $j$-$A$ and $k$-$A$ relative
coordinates, the natural variables of the folded and absorptive
kernels below; the $(jk)$-CoM--$A$ coordinate of
Eq.~(\ref{eq:fragtarget}) is then
$\mathbf r_{(jk)A} = (m_j \mathbf r_j + m_k \mathbf r_k)
/(m_j+m_k)$, and the kinetic operators $K_j + K_k$ in
Eq.~(\ref{eq:HjkA}) remain the reduced-mass Jacobi operators
specified in Sec.~\ref{sec:hamiltonian}.
The post-form residual interaction is
\begin{equation}
V_{\mathrm{post}}^{(B)} = H - H_{\mathrm{exit}}^{(B)}
= V_{ij} + V_{ik} + V_{iA} - U_i \,.
\label{eq:Vpost_B}
\end{equation}
The residual contains the interactions of the detected particle with
$j$, $k$, and $A$.

The DWBA transition amplitude for detecting $i$ while the $jk + A$
system is in eigenstate $c$ is
\begin{equation}
T_{c}^{(B)}
= \langle \chi_i^{(-)} \Psi_{jkA}^c |\,
V_{\mathrm{post}}^{(B)} \,| \chi_a^{(+)} \Phi_a \Phi_A \rangle \,,
\label{eq:T_B}
\end{equation}
where $H_{jkA}|\Psi_{jkA}^c\rangle = E^c |\Psi_{jkA}^c\rangle$.
Defining the source
\begin{equation}
|\rho_i^{(B)}\rangle
= \langle \chi_i^{(-)} |\,
V_{\mathrm{post}}^{(B)} \,| \chi_a^{(+)} \Phi_a \Phi_A \rangle \,,
\label{eq:source_B}
\end{equation}
and using the spectral identity for the $jk + A$ system, the
doubly differential inclusive cross section becomes
\begin{equation}
\frac{d^2\sigma^{(B)}}{dE_i\, d\Omega_i}
= -\frac{(2\pi)^4}{\pi v_a}\,
\mathrm{Im}\,
\langle \rho_i^{(B)} |\, G_{jkA} \,| \rho_i^{(B)} \rangle \,.
\label{eq:master_B}
\end{equation}
where $G_{jkA} = (E_{jkA}^+ - H_{jkA})^{-1}$ is the full three-body
$jk + A$ resolvent and
$E_{jkA} = E - E_i$ is the energy available to the $jk + A$ system.

Equation~(\ref{eq:master_B}) follows exactly from the adopted DWBA
amplitude [Eq.~(\ref{eq:T_B})].  At this stage $V_{iA}$ enters
only through the source.  Eliminating target-excited sectors can
also generate effective couplings in the reduced kernel.

Table~\ref{tab:comparison} compares the two measurements and their
post-form interactions.

\begin{table*}[t]
\caption{Comparison of the two exit-channel partitions for inclusive
breakup of a three-body projectile $a = i + j + k$.}
\label{tab:comparison}
\begin{ruledtabular}
\begin{tabular}{lcc}
 & Partition A & Partition B \\
\hline
Detected & pair $b = (ij)$ & particle $i$ \\
Unresolved & $k + A$ & $jk + A$ \\
Propagator & $G_{kA}$ (2-body) & $G_{jkA}$ (3-body) \\
$V_{\mathrm{post}}$ & $V_{bk} + V_{bA} - U_b$
  & $V_{ij} + V_{ik} + V_{iA} - U_i$ \\
Explicit target coupling & $V_{bA} - U_{bA}$
  & $V_{iA} - U_{iA}$ \\
\end{tabular}
\end{ruledtabular}
\end{table*}

Introducing a reference interaction $U_{iA}$, analogous to $U_{bA}$
in Partition A and likewise local in the exact $i$-$A$ separation
$\mathbf{r}_{iA}$ [Eq.~(\ref{eq:fragtarget})], the source
decomposes as
$|\rho_i^{(B)}\rangle = |\rho_i^{(B,\mathrm{ref})}\rangle
+ |\rho_i^{(B,\mathrm{coup})}\rangle$.
In the post form [Eq.~(\ref{eq:Vpost_B})], $V_{iA}$ enters the
source, while $V_{jA}$ and $V_{kA}$ enter $H_{jkA}$ and its
propagator.
In the prior form, the residual $V_{\mathrm{prior}} = V_{iA}+V_{jA}+V_{kA}-U_a$
[Eq.~(\ref{eq:Vprior})] contains all three fragment-target
couplings: only the detected-particle coupling $V_{iA}-U_{iA}$ is
assigned to the source, whereas $V_{jA}, V_{kA}$ remain inside
the unresolved-system Hamiltonian $H_{jkA}$ and feed the Feshbach
kernel of $G_{jkA}$ through $\Delta\mathcal H_{PP}$
[Eq.~(\ref{eq:DeltaHPP})] and the associated off-diagonal and
$Q$-space blocks.
Treating $QV_{jA}P$ or $QV_{kA}P$ simultaneously as prior-form
source corrections and as Feshbach kernel couplings would
double-count the same target-excitation amplitude.
The prior-form reduction in Sec.~\ref{sec:prior} uses this same
assignment.

The reference-channel source is
\begin{equation}
|\rho_i^{(B,\mathrm{ref})}\rangle
= \langle \chi_i^{(-)} |\,
(V_{ij} + V_{ik} + U_{iA} - U_i) \,| \chi_a^{(+)} \Phi_a \Phi_A \rangle \,,
\label{eq:rho_B_ref}
\end{equation}
built from interactions that do not excite the target, and the
explicit target-coupling source
\begin{equation}
|\rho_i^{(B,\mathrm{coup})}\rangle
= \langle \chi_i^{(-)} |\,
(V_{iA} - U_{iA}) \,| \chi_a^{(+)} \Phi_a \Phi_A \rangle \,,
\label{eq:rho_B_coup}
\end{equation}
is the part of the detected-particle--target interaction that is
not represented by the elastic reference interaction.
If $i$ is structureless, $V_{iA}$ depends only on the exact
$i$-$A$ separation $\mathbf{r}_{iA}$ [Eq.~(\ref{eq:fragtarget})]
and $\xi$, and $V_{iA} - U_{iA}$ represents the difference between
the true $i$-$A$ interaction and the elastic optical model.
If $i$ is composite, this operator acquires the same
$\boldsymbol\zeta_i$-dependence discussed in Ref.~\cite{Lei2026nsp}.

The cross section again separates into reference, interference, and
explicit-coupling terms.
When $V_{iA}$ is replaced by its target-elastic reference
$U_{iA}$, the explicit coupling source vanishes, the source reduces
to $|\rho_i^{(B,\mathrm{ref})}\rangle$, and the result becomes
\begin{equation}
\frac{d^2\sigma^{(B,\mathrm{ref})}}{dE_i\, d\Omega_i}
= -\frac{(2\pi)^4}{\pi v_a}\,
\mathrm{Im}\,
\langle \rho_i^{(B,\mathrm{ref})} |\, G_{jkA} \,|
\rho_i^{(B,\mathrm{ref})} \rangle \,,
\label{eq:sigma_B_ref}
\end{equation}
which is the DWBA sum-rule starting point of the CFH
formalism~\cite{CFH2017}.
I reproduce Eqs.~(19)--(23) of Ref.~\cite{CFH2017} in the
present notation because the explicit-coupling terms require the same
$P$-$Q$ decomposition of $G_{jkA}$.
The resulting absorptive kernel is the CFH result, included here for
self-containment.

Apply $P = |\Phi_A\rangle\langle\Phi_A|$, $Q = \mathbf 1 - P$ to
$G_{jkA} = (E_{jkA}^+ - H_{jkA})^{-1}$.
With the global $E_A = 0$ convention adopted in
Sec.~\ref{sec:hamiltonian}, $E_{jkA} = E - E_i$ is the energy
available to the $jk$ relative motion in the target-ground-state
sector.
The Feshbach projection of the full target-ground-state block,
denoted $G_{PP}^{\rm exact}\equiv P\,G_{jkA}\,P$, satisfies
\begin{equation}
(G_{PP}^{\rm exact})^{-1}
= E_{jkA}^+ - \mathcal H_{PP} - \Delta \mathcal H_{PP},
\label{eq:GPP_inverse}
\end{equation}
with
\begin{align}
\mathcal H_{PP} &= h_j + h_k + K_j + K_k + V_{jk}
 + U_j^{\mathrm{fold}} + U_k^{\mathrm{fold}},
\notag \\
U_j^{\mathrm{fold}}(\mathbf r_j) &= \langle\Phi_A|V_{jA}|\Phi_A\rangle,
\quad
U_k^{\mathrm{fold}}(\mathbf r_k) = \langle\Phi_A|V_{kA}|\Phi_A\rangle.
\label{eq:folding_defs}
\end{align}
Equivalent expressions retaining $E_A$ explicitly amount to
adding $-E_A$ to $\mathcal H_{PP}$ in
Eq.~(\ref{eq:GPP_inverse}) and shifting all target-excited
energies $\omega_{A'}$ in
Eqs.~(\ref{eq:HjkAprime})--(\ref{eq:Qsector_reduction}) by the
same constant.
The reduced $Q$-space resolvent entering the Feshbach kernel is
\begin{equation}
g_Q \equiv (E_{jkA}^+ - QH_{jkA}Q)^{-1}.
\label{eq:gQ_def}
\end{equation}
The polarization kernel is
\begin{equation}
\Delta \mathcal H_{PP}
= P(V_{jA}+V_{kA}) Q\, g_Q\,
Q(V_{jA}+V_{kA}) P \,.
\label{eq:DeltaHPP}
\end{equation}
The operator $g_Q$ in Eq.~(\ref{eq:DeltaHPP}) still contains both
fragments and all target-excited channels; the usual optical-model
step replaces its one-fragment subblocks by the phenomenological
dynamic-polarization contributions that are already included in
elastic $j+A$ and $k+A$ optical potentials.
The kernel $\Delta\mathcal H_{PP}$ is non-Hermitian because $g_Q$
is the retarded resolvent ($+i0^+$ prescription), so its imaginary
part is well defined as
$\mathrm{Im}\,\Delta\mathcal H_{PP}
= -\pi\, P(V_{jA}+V_{kA}) Q\,
\delta(E_{jkA}-QH_{jkA}Q)\,Q(V_{jA}+V_{kA}) P$, a negative
semidefinite operator on $P$-space.
Expanding the square in Eq.~(\ref{eq:DeltaHPP}) produces three
distinct contributions,
\begin{equation}
\Delta \mathcal H_{PP} = \delta U_j + \delta U_k + \mathcal V_{3B} \,,
\label{eq:DeltaHPP_decomp}
\end{equation}
with
\begin{align}
\delta U_j
&= PV_{jA} Q\, g_Q\, Q V_{jA} P,
\notag \\
\delta U_k
&= PV_{kA} Q\, g_Q\, Q V_{kA} P,
\notag \\
\mathcal V_{3B}
&= PV_{jA} Q\, g_Q\, Q V_{kA} P
 + PV_{kA} Q\, g_Q\, Q V_{jA} P \,.
\label{eq:V3B_def}
\end{align}
The cross term $\mathcal V_{3B}$ has $V_{jA}$ at one end of
the propagator and $V_{kA}$ at the other.
Its absorptive part, defined symmetrically as
\begin{equation}
W_{3B} \equiv \mathrm{Im}\,\mathcal V_{3B}
\equiv \frac{1}{2i}\bigl(\mathcal V_{3B}-\mathcal V_{3B}^\dagger\bigr),
\label{eq:W3B_def}
\end{equation}
is the genuine four-body absorption identified in
Ref.~\cite{CFH2017}, supported where both $j$ and $k$
simultaneously sit near the surface and the inelastic couplings
are strong.
With the convention $U=V+iW$, $W\le 0$, one has
$\mathrm{Im}\,\delta U_j = (PV_{jA}Q)\,
\mathrm{Im}\,g_Q\,(QV_{jA}P)\le 0$ as a quadratic form, and
analogously for $\mathrm{Im}\,\delta U_k$;
$\mathrm{Im}(\Delta\mathcal H_{PP})
= P(V_{jA}+V_{kA})Q\,\mathrm{Im}\,g_Q\,Q(V_{jA}+V_{kA})P
\le 0$ is the imaginary part of the full polarization kernel.
The cross piece $W_{3B}$ alone is \emph{not} sign-definite; only
the microscopic combination $W_j^{\mathrm{micro}}+W_k^{\mathrm{micro}}
+W_{3B}$, identified with
$\mathrm{Im}(U_j^{\mathrm{fold}}+\delta U_j+U_k^{\mathrm{fold}}
+\delta U_k+\mathcal V_{3B})$, inherits the negative
semi-definiteness of the full polarization kernel.
The phenomenological $U_j,U_k$ obtained
from $j+A$ and $k+A$ elastic data already incorporate dynamic
polarization not separable into the diagonal Feshbach pieces of
Eq.~(\ref{eq:V3B_def}); their imaginary parts $W_j,W_k$ are
chosen to be absorptive but need not equal $\mathrm{Im}(U_j^{\mathrm{fold}}+\delta
U_j)$ etc.
The negative semi-definiteness of $W_j+W_k+W_{3B}$ as a sum
holds rigorously only at the microscopic Feshbach level; once
$U_j$ and $U_k$ are replaced by phenomenological optical
potentials, the strict inheritance of negative semi-definiteness
relies on the residual $W_{3B}$ being a small perturbation around
phenomenologically absorptive $W_j+W_k$, an assumption common to
the CFH optical reduction.
A sufficient practical condition is that
$|\langle\psi|W_{3B}|\psi\rangle|\ll
 |\langle\psi|(W_j+W_k)|\psi\rangle|$ pointwise in the
quadratic-form sense on states $|\psi\rangle$ in the
source-supported subspace, which can be checked a posteriori from
the microscopic decomposition of the polarization kernel
[Eq.~(\ref{eq:V3B_def})] before $W_{3B}$ is treated as an
absorptive operator alongside $W_j$ and $W_k$.
Under this standard optical reduction, the parts of $g_Q$ in
which only $j$ (or only $k$) couples to target excitations dress
the static folding potentials into phenomenological $U_j$ and
$U_k$, with the other constituent remaining inert in that
single-coupling piece, while the cross piece $\mathcal V_{3B}$ is
retained as a genuine three-body kernel.
The CFH-optical effective Hamiltonian on the $P$-sector is
therefore
\begin{equation}
\mathcal H_{PP}^{\mathrm{opt}}
\equiv h_j+h_k+K_j+K_k+V_{jk}+U_j+U_k+\mathcal V_{3B},
\label{eq:HPP_opt}
\end{equation}
and the corresponding optical propagator is
\begin{equation}
G_{PP}^{\mathrm{opt}}
\equiv \bigl(E_{jkA}^+ - \mathcal H_{PP}^{\mathrm{opt}}\bigr)^{-1},
\label{eq:GPP_opt}
\end{equation}
where the optical Hamiltonian replaces the energy-dependent Feshbach
polarization in Eq.~(\ref{eq:GPP_inverse}) by an energy-independent
representation.  The diagonal terms are absorbed into $U_j$ and
$U_k$, while $\mathcal V_{3B}$ is retained.  Its imaginary part
$W_{3B}$ contributes alongside $W_j$ and $W_k$.

From this point on, $G_{PP}$ refers to the CFH-optical operator
$G_{PP}^{\mathrm{opt}}$ unless stated otherwise; the exact
Feshbach block is denoted $G_{PP}^{\rm exact}$ where the
distinction matters.
The two-potential identity applied to $G_{PP}^{\mathrm{opt}}$
gives
\begin{align}
-\pi^{-1}\mathrm{Im}\,G_{PP}^{\mathrm{opt}}
&\;\approx\; |\chi_{jk}^{(+)}\rangle\langle\chi_{jk}^{(+)}|_{\rm on\text{-}shell}
\notag \\
&\quad
- \pi^{-1}\,G_{PP}^{\mathrm{opt}\dagger}\bigl[W_j + W_k
+ W_{3B}\bigr]G_{PP}^{\mathrm{opt}},
\label{eq:ImGPP_CFH}
\end{align}
as in Eq.~(\ref{eq:two-potential-A}).  The two terms follow from
$G_{PP}^{\mathrm{opt}}-G_{PP}^{\mathrm{opt}\dagger}
=2i\,G_{PP}^{\mathrm{opt}\dagger}[W_j+W_k+W_{3B}]
G_{PP}^{\mathrm{opt}}$ together with isolation of the elastic-channel
on-shell pole, with
$W_j = \mathrm{Im}\,U_j$ and $W_k = \mathrm{Im}\,U_k$ in the
phenomenological representation.
In the microscopic Feshbach limit these reduce to
$\mathrm{Im}(U_j^{\mathrm{fold}}+\delta U_j)$ and
$\mathrm{Im}(U_k^{\mathrm{fold}}+\delta U_k)$, respectively.
Here $|\chi_{jk}^{(+)}\rangle$ is the three-body
$jk$ elastic scattering wave function in the ground-state target
field.
The first term gives the EBU density.  It comes from isolating the
elastic-channel pole in $\mathrm{Im}(G_{jkA}-G_{PP}^{\mathrm{opt}})$,
as in Eq.~(\ref{eq:two-potential-A}).  The second gives the NEB
density when $W_j+W_k+W_{3B}$ is absorptive.  For phenomenological
potentials, this sign condition must be checked as discussed below
Eq.~(\ref{eq:V3B_def}).

The CFH-reduced NEB cross section therefore reads
\begin{align}
\frac{d^2\sigma^{(B,\mathrm{NEB})}}{dE_i\, d\Omega_i}
&= -\frac{(2\pi)^4}{\pi v_a}
\notag \\
&\quad \times
\langle \hat\rho_{jk}|
G_{PP}^{\mathrm{opt}\dagger}\bigl[W_j+W_k+W_{3B}\bigr]
G_{PP}^{\mathrm{opt}}
|\hat\rho_{jk}\rangle,
\label{eq:CFH_NEB}
\end{align}
where $\hat\rho_{jk} = \langle\Phi_A|\rho_i^{(B,\mathrm{ref})}\rangle$
is the reference-channel source after projection onto the target
ground state.
Equations~(\ref{eq:GPP_inverse})--(\ref{eq:CFH_NEB}) are the CFH
reference-channel absorptive kernel in the present notation.
Retaining instead the on-shell elastic projector of
Eq.~(\ref{eq:ImGPP_CFH}) in the master sum rule gives the EBU
counterpart
\begin{equation}
\frac{d^2\sigma^{(B,\mathrm{EBU})}}{dE_i\, d\Omega_i}
= \frac{(2\pi)^4}{v_a}
\bigl|\langle\chi_{jk}^{(+)}|\hat\rho_{jk}\rangle\bigr|^2.
\label{eq:EBU_B}
\end{equation}
Computing $|\chi_{jk}^{(+)}\rangle$ requires three-body scattering
methods, such as those of Ref.~\cite{Deltuva2025}.
Section~\ref{sec:discussion} discusses how to calculate this state
and the action of $G_{PP}^{\mathrm{opt}}$ on a source.

To include $V_{iA}-U_{iA}$, I project both sources onto
$P=|\Phi_A\rangle\langle\Phi_A|$ and $Q=\mathbf1-P$.
The reference source lies entirely in $P$, since
$V_{ij}+V_{ik}+U_{iA}-U_i$ does not excite the target.
The explicit source has two parts: the difference
$\langle\Phi_A|V_{iA}|\Phi_A\rangle-U_{iA}$ within $P$, and
transitions to excited target states through $QV_{iA}P$.

Writing
$|\rho_i^{(B,\mathrm{coup})}\rangle
= |\rho_i^{(B,\mathrm{coup}),P}\rangle
+ |\rho_i^{(B,\mathrm{coup}),Q}\rangle$ with $P$ and $Q$ acting on
the target side of $V_{iA}\Phi_A$,
\begin{align}
|\rho_i^{(B,\mathrm{coup}),P}\rangle
&= |\Phi_A\rangle\,
\langle\chi_i^{(-)}|\bigl(\langle\Phi_A|V_{iA}|\Phi_A\rangle
 - U_{iA}\bigr)
\notag \\
&\qquad \times
|\chi_a^{(+)}\Phi_a\rangle,
\notag \\
|\rho_i^{(B,\mathrm{coup}),Q}\rangle
&= \langle\chi_i^{(-)}|\,
\bigl[Q\,V_{iA}\,|\Phi_A\rangle\bigr]\,
|\chi_a^{(+)}\Phi_a\rangle \,,
\label{eq:rho_B_coup_PQ}
\end{align}
which live in the $P$-sector and $Q$-sector of target Hilbert space
respectively, tensored with the $jk$ relative Hilbert space.
The Feshbach decomposition
$G_{jkA} = G_{PP}^{\rm exact} + G_{PQ} + G_{QP}
+ G_{QQ}^{\mathrm{full}}$ has blocks
\begin{align}
(G_{PP}^{\rm exact})^{-1}
&= E_{jkA} - P H_{jkA} P
  - P H_{jkA} Q\,g_Q\,Q H_{jkA} P,
\notag \\
G_{PQ}
&= G_{PP}^{\rm exact}\, P(V_{jA}+V_{kA}) Q\, g_Q,
\notag \\
G_{QP}
&= g_Q\,Q(V_{jA}+V_{kA})P\,G_{PP}^{\rm exact},
\notag \\
G_{QQ}^{\mathrm{full}}
&= g_Q
\notag\\
&\quad
 + g_Q\,Q(V_{jA}+V_{kA})P\,G_{PP}^{\rm exact}\,
\notag\\
&\qquad \times
   P(V_{jA}+V_{kA})Q\,g_Q,
\label{eq:Feshbach_B}
\end{align}
where $g_Q$ is defined in Eq.~(\ref{eq:gQ_def}).  These relations
are exact before optical reduction.  I then replace $G_{PP}^{\rm exact}$ by
$G_{PP}^{\rm opt}$ [Eq.~(\ref{eq:GPP_opt})], using the
energy-independent Hamiltonian of Eq.~(\ref{eq:HPP_opt}).
The off-diagonal blocks and $G_{QQ}^{\mathrm{full}}$ retain their
Feshbach form.

With this convention the cross section takes the form
\begin{widetext}
\begin{align}
\sigma^{(B)}
&= -\frac{(2\pi)^4}{\pi v_a}\,\mathrm{Im}\,
\Bigl\{
\mathcal I_{\mathrm{ref}}+\mathcal I_P+\mathcal I_{PQ}
+\mathcal I_{\mathrm{coup},P}+\mathcal I_{\mathrm{coup},PQ}
+\mathcal I_Q
\Bigr\},
\label{eq:target_excited_CFH_B}\\
\mathcal I_{\mathrm{ref}}
&= \langle \rho_i^{(B,\mathrm{ref})} |
G_{PP}^{\mathrm{opt}} | \rho_i^{(B,\mathrm{ref})} \rangle,
\notag \\
\mathcal I_P
&= \langle \rho_i^{(B,\mathrm{ref})} |
G_{PP}^{\mathrm{opt}} | \rho_i^{(B,\mathrm{coup}),P} \rangle
+ \langle \rho_i^{(B,\mathrm{coup}),P} |
G_{PP}^{\mathrm{opt}} | \rho_i^{(B,\mathrm{ref})} \rangle,
\notag \\
\mathcal I_{PQ}
&= \langle \rho_i^{(B,\mathrm{ref})} |
G_{PQ} | \rho_i^{(B,\mathrm{coup}),Q} \rangle
+ \langle \rho_i^{(B,\mathrm{coup}),Q} |
G_{QP} | \rho_i^{(B,\mathrm{ref})} \rangle,
\notag \\
\mathcal I_{\mathrm{coup},P}
&= \langle \rho_i^{(B,\mathrm{coup}),P} |
G_{PP}^{\mathrm{opt}} | \rho_i^{(B,\mathrm{coup}),P} \rangle,
\notag \\
\mathcal I_{\mathrm{coup},PQ}
&= \langle \rho_i^{(B,\mathrm{coup}),P} |
G_{PQ} | \rho_i^{(B,\mathrm{coup}),Q} \rangle
+ \langle \rho_i^{(B,\mathrm{coup}),Q} |
G_{QP} | \rho_i^{(B,\mathrm{coup}),P} \rangle,
\notag \\
\mathcal I_Q
&= \langle \rho_i^{(B,\mathrm{coup}),Q} |
G_{QQ}^{\mathrm{full}} | \rho_i^{(B,\mathrm{coup}),Q} \rangle .
\notag
\end{align}
\end{widetext}
The sources on either side of each propagator select its target-space
block.  The reference source and $\rho_i^{(B,\mathrm{coup}),P}$
lie in $P$, while $\rho_i^{(B,\mathrm{coup}),Q}$ lies in $Q$.

The term $\mathcal I_{\mathrm{ref}}$ is the CFH reference-channel
contribution whose absorptive reduction gives the
$W_j + W_k + W_{3B}$ form of Eq.~(\ref{eq:CFH_NEB}); the remaining
$\mathcal I$ terms are generated by the explicit target-coupling
source.
The intra-$P$ term $\mathcal I_P$ vanishes identically when
$U_{iA}$ is chosen as the elastic folding potential, and in general
represents the renormalization freedom already discussed in
Ref.~\cite{Lei2026nsp}.
The terms involving $G_{PQ}$ and $G_{QP}$ are interferences between the
elastic $jk+A$ channel space and target-excited $jk+A^*$ channels,
coupled through $V_{jA}+V_{kA}$ at the propagator level and through
$Q V_{iA}$ at the source level.
As in Eq.~(\ref{eq:sigma_coup_A}), these interferences are written
as explicit $PQ$ and $QP$ pairs before the final imaginary part is
taken; replacing them by $2\mathrm{Re}$ would only be justified after
a Hermitian absorptive-kernel reduction.
The term $\mathcal I_Q$ is the additional target-excited
contribution: absorption into
$jk+A^*$ continua driven by the explicit target-coupling source.
Because it contains $G_{QQ}^{\mathrm{full}}$, it includes both
direct propagation in the $Q$ space and the induced
$Q\to P\to Q$ rescattering shown in Eq.~(\ref{eq:Feshbach_B}).
The separation between $\rho_i^{(B,\mathrm{ref})}$ and
$\rho_i^{(B,\mathrm{coup}),P}$ depends on the reference potential
$U_{iA}$, so changes of $U_{iA}$ reshuffle strength among the
$P$-sector terms.
Before optical reduction, the sum of the source-kernel matrix elements
recovers Eq.~(\ref{eq:master_B}).  With $G_{PP}^{\mathrm{exact}}$
replaced by $G_{PP}^{\mathrm{opt}}$, the sum recovers the DWBA
expression within that optical approximation.
The $Q$-sector assignment is fixed here by the double-counting rule:
the detected-particle transition $QV_{iA}P$ is a source term, while
the unresolved-fragment transitions $QV_{jA}P$ and $QV_{kA}P$ remain
inside the Feshbach kernels.

For the direct $g_Q$ contribution, I expand the target space as
$Q=\sum_{A'\neq0}|\Phi_{A'}\rangle\langle\Phi_{A'}|$.

Here and below the $|\Phi_{A'}\rangle$ are the
particle-number-conserving eigenstates of the target internal
Hamiltonian $H_A$; the consequence of this choice for
pairing-correlated targets is stated in
Sec.~\ref{sec:discussion}.
For a given excited target state $|\Phi_{A'}\rangle$, the
restricted projector $Q_{A'} = |\Phi_{A'}\rangle\langle\Phi_{A'}|$
satisfies $Q_{A'} g_Q Q_{A'} \equiv \tilde G_{jk}^{(A')}$, where
$\tilde G_{jk}^{(A')}$ is a three-body $jk$ relative-motion
resolvent at the shifted energy
$E_{jk}^{(A')} = E_{jkA} - \omega_{A'}$, with
$\omega_{A'}\equiv E_{A'}-E_A$ the excitation energy of the target
state $|\Phi_{A'}\rangle$ above the ground state (which equals
$E_{A'}$ under the global $E_A=0$ convention of
Sec.~\ref{sec:hamiltonian}).
The resolvent $\tilde G_{jk}^{(A')}$ is built from the
target-excited effective Hamiltonian
\begin{align}
\tilde{\mathcal H}_{jk}^{(A')}
&= h_j + h_k + K_j + K_k + V_{jk}
\notag \\
&\quad
 + \tilde U_j^{(A')}(\mathbf r_j)
 + \tilde U_k^{(A')}(\mathbf r_k)
 + \tilde{\mathcal V}_{3B}^{(A')},
\label{eq:HjkAprime}
\end{align}
where the optical interactions are now defined relative to
$|\Phi_{A'}\rangle$.  A second Feshbach projection onto
$Q_\perp^{(A')}=Q-|\Phi_{A'}\rangle\langle\Phi_{A'}|$ gives
\begin{align}
\tilde U_j^{(A')}(\mathbf r_j)
&= \langle\Phi_{A'}|V_{jA}|\Phi_{A'}\rangle
\notag\\
&\quad
+ \langle\Phi_{A'}|V_{jA}\,Q_\perp^{(A')}\,
\bar G_{Q_\perp}^{(A')}\,Q_\perp^{(A')}\,V_{jA}|\Phi_{A'}\rangle,
\notag\\
\tilde{\mathcal V}_{3B}^{(A')}
&= \langle\Phi_{A'}|V_{jA}\,Q_\perp^{(A')}\,
\bar G_{Q_\perp}^{(A')}\,Q_\perp^{(A')}\,V_{kA}|\Phi_{A'}\rangle
\notag\\
&\quad
+ (j\leftrightarrow k),
\label{eq:Qperp_decomp}
\end{align}
and analogously for $\tilde U_k^{(A')}$, where
$\bar G_{Q_\perp}^{(A')}$ is the resolvent on the
$Q_\perp^{(A')}$ subspace, i.e.\ states whose target component is
neither $|\Phi_A\rangle$ nor $|\Phi_{A'}\rangle$ but any other
target eigenstate (whether energetically above or below
$|\Phi_{A'}\rangle$).
Excluding both $|\Phi_A\rangle$ and $|\Phi_{A'}\rangle$ from
$Q_\perp^{(A')}$ separates these polarization terms from those
of the ground-state block.

I now drop the matrix elements
$\langle\Phi_{A''}|g_Q|\Phi_{A'}\rangle$ with $A''\neq A'$.
This approximation removes propagation between different target labels
at the two ends of $g_Q$.  Each retained diagonal block still
contains excursions through other excited states and back to $A'$.
The interaction matrix elements
$\langle\Phi_{A'}|V_{jA}|\Phi_{A''}\rangle$ and
$\langle\Phi_{A'}|V_{kA}|\Phi_{A''}\rangle$ ($A''\neq A,A'$)
therefore remain in the polarization terms of
Eq.~(\ref{eq:Qperp_decomp}).  Keeping the off-diagonal blocks of
$g_Q$ would instead require coupled propagation in the excited-target
basis.
Representing the polarization terms within each retained block by
optical interactions is a further approximation to the nested
Feshbach expressions in Eq.~(\ref{eq:Qperp_decomp}); these optical
interactions are diagonal in the intermediate target label.

Under the diagonal approximation,
\begin{equation}
g_Q \approx \sum_{A'\neq 0}
|\Phi_{A'}\rangle \tilde G_{jk}^{(A')}\langle\Phi_{A'}|,
\label{eq:gQ_diag}
\end{equation}
and
\begin{align}
-\pi^{-1}\mathrm{Im}\,\tilde G_{jk}^{(A')}
&\;\approx\; |\tilde\chi_{jk}^{(A',+)}\rangle
\langle\tilde\chi_{jk}^{(A',+)}|_{\rm on\text{-}shell}
\notag \\
&\quad
- \pi^{-1}\,\tilde G_{jk}^{(A')\dagger}\,
\tilde W_{\mathrm{CFH}}^{(A')}\,
\tilde G_{jk}^{(A')},
\label{eq:ImGjkAprime}
\end{align}
with $\tilde W_{\mathrm{CFH}}^{(A')}\equiv
\tilde W_j^{(A')}+\tilde W_k^{(A')}+\tilde W_{3B}^{(A')}$,
in the same two-potential form as Eqs.~(\ref{eq:two-potential-A})
and (\ref{eq:ImGPP_CFH}); here
$|\tilde\chi_{jk}^{(A',+)}\rangle$ is the three-body $jk$ elastic
scattering wave function in the target field of the excited state
$|\Phi_{A'}\rangle$ and energy $E_{jk}^{(A')}$, and
$\tilde W_{j,k}^{(A')}, \tilde W_{3B}^{(A')}$ are the target-excited
analogs of the CFH absorption operators, constructed from
$\tilde U_j^{(A')}, \tilde U_k^{(A')}$ and from the cross term
$\tilde{\mathcal V}_{3B}^{(A')}$ with propagation in
$Q_\perp^{(A')}$.
Substituting Eq.~(\ref{eq:gQ_diag}) into the direct $g_Q$ part of
the last line of Eq.~(\ref{eq:target_excited_CFH_B}) gives, in the
diagonal-intermediate-target-states approximation,
\begin{align}
\langle \rho_i^{(B,\mathrm{coup}),Q}|g_Q|
\rho_i^{(B,\mathrm{coup}),Q}\rangle
&\approx \sum_{A'\neq 0}
\langle\hat\rho_{jk}^{(A')}|
\tilde G_{jk}^{(A')}|\hat\rho_{jk}^{(A')}\rangle,
\notag \\
\hat\rho_{jk}^{(A')}
&= \langle\Phi_{A'}|\rho_i^{(B,\mathrm{coup}),Q}\rangle.
\label{eq:Qsector_reduction}
\end{align}
To obtain the direct-$g_Q$ contribution to the cross section, multiply
the imaginary part of Eq.~(\ref{eq:Qsector_reduction}) by
$-(2\pi)^4(\pi v_a)^{-1}$.  Applying the two-potential identity
Eq.~(\ref{eq:ImGjkAprime}) to each $\tilde G_{jk}^{(A')}$ then
splits this into target-excited NEB and EBU densities,
\begin{align}
&-\pi^{-1}\sum_{A'\neq 0}\mathrm{Im}\,
\langle\hat\rho_{jk}^{(A')}|\tilde G_{jk}^{(A')}|
\hat\rho_{jk}^{(A')}\rangle
\notag\\
&\quad
\approx \sum_{A'\neq 0}\Bigl\{
\bigl|\langle\tilde\chi_{jk}^{(A',+)}|
\hat\rho_{jk}^{(A')}\rangle\bigr|^2
\notag\\
&\qquad
- \pi^{-1}\langle\hat\rho_{jk}^{(A')}|
\tilde G_{jk}^{(A')\dagger}\,\tilde W_{\mathrm{CFH}}^{(A')}\,
\tilde G_{jk}^{(A')}|
\hat\rho_{jk}^{(A')}\rangle\Bigr\},
\label{eq:Qsector_NEBEBU}
\end{align}
so that the direct target-excited absorption contributes a sum of
CFH-like NEB structures indexed by the populated target excitation
$A'$, each carrying its own
$\tilde W_j^{(A')}+\tilde W_k^{(A')}+\tilde W_{3B}^{(A')}$
decomposition, accompanied by an EBU contribution into the
$jk$ elastic continuum on top of $|\Phi_{A'}\rangle$.
For targets with strong low-lying collective modes, leading
contributions are expected from surface dipole and quadrupole
excitations reached by $QV_{iA}P$ acting on $|\Phi_A\rangle$.
The $\tilde U_j^{(A')}, \tilde U_k^{(A')}$ are the shifted-energy
analogs of nucleon-nucleus optical interactions in the
target-excited sector, and must be distinguished from the
ground-state optical potentials entering the CFH reference term to
avoid double counting the same target-polarization strength.
In the CFH reference limit $V_{iA} \to U_{iA}$, only
$\mathcal I_{\mathrm{ref}}$ survives, recovering
Eq.~(\ref{eq:CFH_NEB}).

For a structureless detected particle $i$, the operator
$V_{iA} - U_{iA}$ coincides with the familiar non-elastic
(target-excitation) correction of ordinary DWBA; the
$\rho_i^{(B,\mathrm{coup}),Q}$ term in
Eq.~(\ref{eq:target_excited_CFH_B}) then corresponds to the conventional
target-polarization correction, and the extent to which
Eq.~(\ref{eq:target_excited_CFH_B}) contains physics beyond existing DWBA
practice is confined to how this correction interferes with the
$W_{3B}$ channel of CFH.
For a composite detected particle ($i = d$, $\alpha$, or
${}^8\mathrm{Be}$), the $\boldsymbol\zeta_i$-dependence of
$V_{iA}$ brings in the multipole structure of
Ref.~\cite{Lei2026nsp} and Eq.~(\ref{eq:tidal_6Li}) below, which
can be substantial.
The reduction in Eq.~(\ref{eq:Qsector_reduction}) applies only to
the direct $g_Q$ contribution.  Off-diagonal propagation between
excited target labels and the associated source nonorthogonality
require the full coupled-channel treatment; the nested polarization
inside each diagonal block remains as in Eq.~(\ref{eq:Qperp_decomp}).

\section{Prior forms and post-prior relations}
\label{sec:prior}

The entrance-channel Hamiltonian for the four-body system is
\begin{equation}
H_{\mathrm{ent}} = H_A + K_a + U_a + h_a \,,
\label{eq:Hent}
\end{equation}
where $K_a$ is the $a$-$A$ center-of-mass kinetic energy, $U_a$ is
the entrance-channel optical potential, and $h_a$ is the internal
Hamiltonian of the projectile,
$h_a = h_i + h_j + h_k + K_\zeta + K_y + V_{ij} + V_{ik} + V_{jk}$.
The prior-form residual interaction is common to both partitions,
\begin{equation}
V_{\mathrm{prior}} = H - H_{\mathrm{ent}}
= V_{iA} + V_{jA} + V_{kA} - U_a \,,
\label{eq:Vprior}
\end{equation}
The same residual interaction therefore generates both prior sources.
The post-prior identity is established below for pair detection and,
for single-particle detection, within the CFH target-ground-state
optical reduction.  The target-excited terms are retained through the
Feshbach decomposition; no separate post-prior identity is proved for
them.  These prior forms provide the starting point for calculations
where the corresponding IAV and UT convergence properties apply.

For Partition A, the prior-form T-matrix is
$T_{\alpha,c}^{\prime(A)}
= \langle \chi_b^{(-)} \phi_\alpha\, \Psi_{kA}^c |\,
V_{\mathrm{prior}} \,| \chi_a^{(+)} \Phi_a \Phi_A \rangle$,
and the prior source is
\begin{equation}
|\rho_\alpha^{\prime(A)}\rangle
= \langle \phi_\alpha\, \chi_b^{(-)} |\,
V_{\mathrm{prior}} \,| \chi_a^{(+)} \Phi_a \Phi_A \rangle \,.
\label{eq:source_prior_A}
\end{equation}
In the reduced single-channel limit, the prior source decomposes as
$\tilde\rho_\alpha^{\prime(A)} = \tilde\rho_\alpha^{\prime(A,\mathrm{UT})}
+ \tilde\rho_\alpha^{\prime(A,\mathrm{coup})}$, where the UT source
involves $U_k + U_{bA} - U_a$ and the explicit coupling source involves
$V_{bA} - U_{bA}$, the same operator as in the post form.
The UT NEB cross section in prior form is
\begin{equation}
\sigma_\alpha^{(A,\mathrm{UT})}
= -\frac{(2\pi)^4}{\pi v_a}\,
\langle \tilde\rho_\alpha^{\prime(A,\mathrm{UT})} |
G_k^\dagger\, W_k\, G_k
| \tilde\rho_\alpha^{\prime(A,\mathrm{UT})} \rangle \,,
\label{eq:sigma_UT_A}
\end{equation}
and the nonorthogonality (NO) correction, i.e., the Hussein-McVoy term,
and the interference (IN) correction take the standard
form~\cite{Hussein1985,Lei2018prior}.
The nonorthogonality overlap, projected onto the pair's internal state,
is
\begin{equation}
\psi_\alpha^{(A,\mathrm{NO})}(\mathbf{r}_k)
= \int d\mathbf{r}_b\, d\boldsymbol\zeta\;
\chi_b^{(-)*}\,\phi_\alpha^*\,
\chi_a^{(+)}\,\Phi_a \,,
\label{eq:psi_NO_A}
\end{equation}
which does not involve any interaction and is independent of the
choice of reference interaction.

For Partition B, the prior source is
\begin{equation}
|\rho_i^{\prime(B)}\rangle
= \langle \chi_i^{(-)} |\,
V_{\mathrm{prior}} \,| \chi_a^{(+)} \Phi_a \Phi_A \rangle \,,
\label{eq:source_prior_B}
\end{equation}
which decomposes into UT and explicit-coupling parts with the UT source
built from $U_j + U_k + U_{iA} - U_a$ and the explicit coupling source
from $V_{iA} - U_{iA}$.
As in the CFH reduction of Sec.~\ref{sec:single}, the target couplings
of the unresolved
constituents $j$ and $k$ are assigned to the three-body propagator
and to its absorptive kernel, rather than to an additional source
term.

The difference between post and prior residual interactions is, for
Partition A,
\begin{equation}
V_{\mathrm{post}}^{(A)} - V_{\mathrm{prior}}
= V_{bk} - V_{kA} + U_a - U_b \,,
\label{eq:postprior_diff_A}
\end{equation}
where $V_{bA}$ cancels exactly, just as in the two-body
case~\cite{Ichimura1985,Lei2026nsp}.
For Partition B,
\begin{equation}
V_{\mathrm{post}}^{(B)} - V_{\mathrm{prior}}
= V_{ij} + V_{ik} - V_{jA} - V_{kA} + U_a - U_i \,,
\label{eq:postprior_diff_B}
\end{equation}
where $V_{iA}$ cancels exactly.
The explicit target coupling of the detected pair in Partition A,
$V_{bA}-U_{bA}$, and of the detected particle in Partition B,
$V_{iA}-U_{iA}$, each enters the post and prior sources identically
and therefore drops out of their difference.
Equations~(\ref{eq:postprior_diff_A}) and~(\ref{eq:postprior_diff_B})
are operator identities valid on any state; the corresponding
matrix-element identity $T^{\rm post} = T^{\rm prior}$ requires in
addition that the entrance and exit channel functions satisfy their
respective Schr\"odinger equations up to surface terms.
For Partition A, the entrance state $\chi_a^{(+)}\Phi_a\Phi_A$ and the
exit state $\chi_b^{(-)}\phi_\alpha\Psi_{kA}^c$ are governed by
two-body asymptotic Schr\"odinger equations.
For charged systems, the long-range Coulomb pieces must first be
absorbed consistently into the channel distortions, so that the
operators entering the post-prior difference are short-ranged.
Under that standard assumption, the surface terms vanish and the
full-space DWBA post-prior relation for Partition A follows by the
same algebra as in the two-body
derivation~\cite{Ichimura1985,Lei2026nsp,Lei2018prior}.
For Partition B the analogous full-space step requires Faddeev
asymptotics for the three-body exit state $\Psi_{jkA}^c$ and is left
out of scope, as discussed below; the reduced identity used in this
paper is established only at the CFH-reduced level.
The post-prior identity connecting the two representations is,
for Partition A in the reduced single-channel limit,
\begin{equation}
G_k\,|\tilde\rho_\alpha^{\prime(A,\mathrm{UT})}\rangle
= G_k\,|\tilde\rho_\alpha^{(A,\mathrm{ref})}\rangle
- |\psi_\alpha^{(A,\mathrm{NO})}\rangle \,,
\label{eq:postprior_identity_A}
\end{equation}
and the explicit coupling sources are identical in both forms,
$\tilde\rho_\alpha^{\prime(A,\mathrm{coup})}
= \tilde\rho_\alpha^{(A,\mathrm{coup})}$.
Substituting Eq.~(\ref{eq:postprior_identity_A}) into the prior-form
NEB cross section produces the Hussein-McVoy decomposition,
extended here to include the explicit coupling source.
Defining the auxiliary kets
$|\Lambda_\alpha^{\mathrm{UT}}\rangle
= G_k|\tilde\rho_\alpha^{\prime(A,\mathrm{UT})}\rangle$,
$|\Lambda_\alpha^{\mathrm{NO}}\rangle
= |\psi_\alpha^{(A,\mathrm{NO})}\rangle$, and
$|\Lambda_\alpha^{\mathrm{coup}}\rangle
= G_k|\tilde\rho_\alpha^{(A,\mathrm{coup})}\rangle$, the
NEB cross section splits into five terms,
\begin{align}
\sigma_\alpha^{(A,\mathrm{NEB})}
&= \sigma_\alpha^{\mathrm{UT}}
 + \sigma_\alpha^{\mathrm{NO}}
 + \sigma_\alpha^{\mathrm{IN}}
 + \sigma_\alpha^{\mathrm{coup,pure}}
 + \sigma_\alpha^{\mathrm{coup,IN}},
\notag \\
\sigma_\alpha^{\mathrm{UT}}
&= -\frac{(2\pi)^4}{\pi v_a}
\langle\Lambda_\alpha^{\mathrm{UT}}|W_k|\Lambda_\alpha^{\mathrm{UT}}\rangle,
\notag \\
\sigma_\alpha^{\mathrm{NO}}
&= -\frac{(2\pi)^4}{\pi v_a}
\langle\Lambda_\alpha^{\mathrm{NO}}|W_k|\Lambda_\alpha^{\mathrm{NO}}\rangle,
\notag \\
\sigma_\alpha^{\mathrm{IN}}
&= -\frac{2(2\pi)^4}{\pi v_a}\,\mathrm{Re}\,
\langle\Lambda_\alpha^{\mathrm{UT}}|W_k|\Lambda_\alpha^{\mathrm{NO}}\rangle,
\notag \\
\sigma_\alpha^{\mathrm{coup,pure}}
&= -\frac{(2\pi)^4}{\pi v_a}
\langle\Lambda_\alpha^{\mathrm{coup}}|W_k|
\Lambda_\alpha^{\mathrm{coup}}\rangle,
\notag \\
\sigma_\alpha^{\mathrm{coup,IN}}
&= -\frac{2(2\pi)^4}{\pi v_a}\,\mathrm{Re}\,
\langle\Lambda_\alpha^{\mathrm{UT}}+\Lambda_\alpha^{\mathrm{NO}}|W_k|
\Lambda_\alpha^{\mathrm{coup}}\rangle.
\label{eq:HM_decomposition}
\end{align}
The UT, NO, and IN terms are the usual Hussein-McVoy
contributions~\cite{Hussein1985,Lei2018prior,GomezRamos2021}, now
projected onto $\phi_\alpha$.  The overlap in
Eq.~(\ref{eq:psi_NO_A}) retains the full $\Phi_a$.
The last two terms give the square of the explicit-coupling amplitude
and its interference with the UT and NO amplitudes.  Both are needed
even though the explicit source is identical in post and prior forms.

With the convention $W_k\le 0$, the quadratic UT and NO terms are
non-negative, while the IN piece can have either sign.
The latter is known to be numerically important near and above the Coulomb
barrier~\cite{Hussein1985,Lei2018prior}.
Substituting Eq.~(\ref{eq:HM_decomposition}) back into the post
form verifies post-prior equivalence for the reduced single-channel
NEB cross section in Partition A, including the reference and
explicit-coupling source contributions at that level.

For Partition B, a full-space post-prior identity at the level of
the unreduced DWBA matrix element would require Faddeev-asymptotic
exit states with Dollard-regularized Coulomb phases on each
two-fragment partition of $jk+A$, together with a controlled
treatment of the connected breakup hyperplane.
This is beyond the scope of the present DWBA-level derivation; the
analogous problem in the two-body-projectile limit is discussed in
Ref.~\cite{Deltuva2025}.
No result below depends on a full-space identity for Partition B.
The target-excited terms retain the Feshbach decomposition and the
source and propagator assignments of Sec.~\ref{sec:single}.

For the target-ground-state sector, I use the energy-independent
CFH optical Hamiltonian to derive the counterpart of
Eq.~(\ref{eq:postprior_identity_A}).

The projected post- and prior-form sources are
$\hat\rho^{(B,\mathrm{ref})}
= \langle\Phi_A|\rho_i^{(B,\mathrm{ref})}\rangle$ and
$\hat\rho^{\prime(B,\mathrm{UT})}
= \langle\Phi_A|\rho_i^{\prime(B,\mathrm{UT})}\rangle$, with
$\rho_i^{\prime(B,\mathrm{UT})}$ built from
$U_j + U_k + U_{iA} - U_a$ in the prior residual.
The reduced exit-channel Hamiltonian on the $P$-sector is
$\bar H_{\mathrm{exit}}^{(B,P)} = h_i + K_i + U_i
+ \mathcal H_{PP}^{\mathrm{opt}}$, with
$\mathcal H_{PP}^{\mathrm{opt}}
= h_j+h_k+K_j+K_k+V_{jk}+U_j+U_k+\mathcal V_{3B}$ from
Eq.~(\ref{eq:HPP_opt}), so that the optical
$G_{PP}^{\mathrm{opt}}$ obeys
$(E - \bar H_{\mathrm{exit}}^{(B,P)})G_{PP}^{\mathrm{opt}}
= \mathbf 1$ in the $P$-sector relative-motion Hilbert space.
The interaction $\mathcal V_{3B}$ is included in this Hamiltonian
and in $G_{PP}^{\mathrm{opt}}$.  As in the optical reduction of
Sec.~\ref{sec:single}, $\Delta\mathcal H_{PP}$ is treated as
energy-independent.  Retaining its energy dependence would introduce
$\partial_E\Delta\mathcal H_{PP}$ terms, which are not treated
in this derivation.

Applying $\langle\Phi_A|$ to the post and prior sources turns the
microscopic operators $V_{jA},V_{kA},V_{iA}$ into their
ground-state expectations, absorbed into the optical
$U_j+U_k+U_{iA}$, while the $V_{ij}+V_{ik}$ part of the post-form
residual is target-trivial.
Acting $G_{PP}^{\mathrm{opt}}$ from the left and using the on-shell
Schr\"odinger identity for the asymptotic exit state
$|\chi_i^{(-)}\rangle\otimes|\chi_{jk}^{(+)}\rangle\,\Phi_A$,
exactly as in the two-body derivation of
Eq.~(\ref{eq:postprior_identity_A}), one obtains
\begin{equation}
G_{PP}^{\mathrm{opt}}\,|\hat\rho^{\prime(B,\mathrm{UT})}\rangle
= G_{PP}^{\mathrm{opt}}\,|\hat\rho^{(B,\mathrm{ref})}\rangle
 - |\hat\psi_{3B}^{(B,\mathrm{NO})}\rangle,
\label{eq:postprior_identity_B_reduced}
\end{equation}
where the three-body analog of the nonorthogonality overlap is
\begin{align}
\hat\psi_{3B}^{(B,\mathrm{NO})}(\mathbf r_j,\mathbf r_k)
&= \int d\mathbf r_i\, d\xi\,
\chi_i^{(-)*}(\mathbf r_i)\,\Phi_A^*(\xi)
\notag\\
&\quad\times
\chi_a^{(+)}\bigl(\mathbf R(\mathbf r_i,\mathbf r_j,\mathbf r_k)\bigr)
\notag\\
&\quad\times
\Phi_a\bigl(\boldsymbol\zeta'(\mathbf r_j,\mathbf r_k),\,
\mathbf y'(\mathbf r_i,\mathbf r_j,\mathbf r_k)\bigr)\,\Phi_A(\xi),
\label{eq:psi_NO_B}
\end{align}
where the dependence of the projectile Jacobi coordinates and of
the entrance distorted-wave argument on
$(\mathbf r_i,\mathbf r_j,\mathbf r_k)$ is given by the kinematic
relations of Sec.~\ref{sec:hamiltonian}.
The $\mathbf r_i$ integration projects out the detected-particle
coordinate; target-state orthonormality
$\langle\Phi_A|\Phi_A\rangle=1$ removes $\xi$.
The result is a function of the unresolved-pair coordinates
$(\mathbf r_j,\mathbf r_k)$ alone, equivalently a function of the
$(jk)$-internal Jacobi coordinate $\boldsymbol\zeta''=\mathbf
r_j-\mathbf r_k$ and the $(jk)$-CoM--target coordinate
$\mathbf r_{(jk)A} = (m_j \mathbf r_j + m_k \mathbf r_k)
/(m_j+m_k)$ of Eq.~(\ref{eq:fragtarget}).
Thus $\hat\psi_{3B}^{(B,\mathrm{NO})}$ is the entrance-exit
overlap after projection onto the target ground state and the detected
particle.  Equation~(\ref{eq:postprior_identity_B_reduced}) has
the form of Eq.~(\ref{eq:postprior_identity_A}), with
$G_k\to G_{PP}^{\mathrm{opt}}$ and
$\psi_\alpha^{(A,\mathrm{NO})}\to\hat\psi_{3B}^{(B,\mathrm{NO})}$.
This is not an exact Feshbach-level identity; it retains the optical
and Coulomb assumptions used above.

Within this ground-state-sector reduction, the detected-particle
explicit source operator is the same in post and prior forms,
$\hat\rho^{\prime(B,\mathrm{coup}),P}
= \hat\rho^{(B,\mathrm{coup}),P}$.
Substituting Eq.~(\ref{eq:postprior_identity_B_reduced}) into the
prior-form reduced cross section produces a Hussein-McVoy-type
decomposition with five explicit terms, analogous to
Eq.~(\ref{eq:HM_decomposition}).
Defining the auxiliary kets
$|\Lambda^{\mathrm{UT,B}}\rangle
= G_{PP}^{\mathrm{opt}}|\hat\rho^{\prime(B,\mathrm{UT})}\rangle$,
$|\Lambda^{\mathrm{NO,B}}\rangle
= |\hat\psi_{3B}^{(B,\mathrm{NO})}\rangle$, and
$|\Lambda^{\mathrm{coup,B}}\rangle
= G_{PP}^{\mathrm{opt}}|\hat\rho^{(B,\mathrm{coup}),P}\rangle$,
and writing $W_{\mathrm{CFH}}\equiv W_j+W_k+W_{3B}$, the
ground-state-sector NEB cross section reads
\begin{align}
\sigma^{(B,\mathrm{NEB})}_{\mathrm{CFH}}
&= \sigma^{\mathrm{UT,B}} + \sigma^{\mathrm{NO,B}}
 + \sigma^{\mathrm{IN,B}}
\notag \\
&\quad
 + \sigma^{\mathrm{coup,pure,B}}
 + \sigma^{\mathrm{coup,IN,B}},
\notag \\
\sigma^{\mathrm{UT,B}}
&= -\frac{(2\pi)^4}{\pi v_a}
\langle\Lambda^{\mathrm{UT,B}}|W_{\mathrm{CFH}}|
\Lambda^{\mathrm{UT,B}}\rangle,
\notag \\
\sigma^{\mathrm{NO,B}}
&= -\frac{(2\pi)^4}{\pi v_a}
\langle\Lambda^{\mathrm{NO,B}}|W_{\mathrm{CFH}}|
\Lambda^{\mathrm{NO,B}}\rangle,
\notag \\
\sigma^{\mathrm{IN,B}}
&= -\frac{2(2\pi)^4}{\pi v_a}\,\mathrm{Re}\,
\notag \\
&\quad\times
\langle\Lambda^{\mathrm{UT,B}}|W_{\mathrm{CFH}}|
\Lambda^{\mathrm{NO,B}}\rangle,
\notag \\
\sigma^{\mathrm{coup,pure,B}}
&= -\frac{(2\pi)^4}{\pi v_a}
\langle\Lambda^{\mathrm{coup,B}}|W_{\mathrm{CFH}}|
\Lambda^{\mathrm{coup,B}}\rangle,
\notag \\
\sigma^{\mathrm{coup,IN,B}}
&= -\frac{2(2\pi)^4}{\pi v_a}\,\mathrm{Re}\,
\notag \\
&\quad\times
\langle\Lambda^{\mathrm{UT,B}}+\Lambda^{\mathrm{NO,B}}|
W_{\mathrm{CFH}}|\Lambda^{\mathrm{coup,B}}\rangle.
\label{eq:HM_decomposition_B}
\end{align}
The two-body $W_k$ in Eq.~(\ref{eq:HM_decomposition}) is replaced
here by $W_{\mathrm{CFH}}=W_j+W_k+W_{3B}$, and the overlap by
$\hat\psi_{3B}^{(B,\mathrm{NO})}$ [Eq.~(\ref{eq:psi_NO_B})].
If $W_{\mathrm{CFH}}$ is negative semidefinite, the UT, NO, and
pure-coupling terms are non-negative, while both interferences may
have either sign.  This sign condition holds for the microscopic
Feshbach kernel and is assumed in its phenomenological replacement
as discussed in Sec.~\ref{sec:single}.

Equation~(\ref{eq:HM_decomposition_B}) holds in the CFH optical
reduction of the target-ground-state sector.  The target-excited terms
[Eqs.~(\ref{eq:target_excited_CFH_B})--(\ref{eq:Qsector_reduction})]
remain separate.  In the microscopic prior residual,
$QV_{\mathrm{prior}}P$ contains $QV_{iA}P$, $QV_{jA}P$, and
$QV_{kA}P$.  The last two already enter the Feshbach kernel of
$jk+A$ and cannot be counted again as source corrections.  The
explicit target-excited source therefore contains $QV_{iA}P$ alone
[Eq.~(\ref{eq:rho_B_coup_PQ})].  A post-prior identity for this
sector would require the full coupled-channel $Q$-space metric,
with the source and kernel varied together.  It is not established
by the diagonal-intermediate-states reduction used here.

\section{Limiting reductions and connections to existing formalisms}
\label{sec:limits}

The two-body IAV sum rule, the detected-cluster result of
Ref.~\cite{Lei2026nsp}, and CFH~\cite{CFH2017} follow by restricting
the projectile wave function or the target coupling as described below.

For Partition A, replacing the full pair-target interaction by the
reference interaction, $V_{bA} \to U_{bA}$, removes the explicit
coupling source and leaves
$\rho_\alpha^{(A,\mathrm{ref})}$.
If the pair $b = (ij)$ is further reduced to a pre-formed
structureless cluster, one recovers the standard two-body IAV with
$a = b + k$ and $x = k$.
Taking $\alpha=0$ and absorbing the internal
$\boldsymbol\zeta$ dependence into cluster interactions turns
$V_{bk}+U_{bA}-U_b$ into $V_{bx}+U_{bA}-U_b$, with $x=k$.
For ${}^6\mathrm{Li}$, $V_{bx}$ is the folded interaction
$\langle\phi_d|V_{n\alpha}+V_{p\alpha}|\phi_d\rangle_\zeta$,
equal to $\mathcal F_{dd}$ in Eq.~(\ref{eq:pair_form_factor}).

Under both operations, the source
\begin{align}
|\rho_\alpha^{(A,\mathrm{ref})}\rangle
&\xrightarrow{\mathrm{cluster}}
\langle\chi_b^{(-)}|(V_{bx} + U_{bA} - U_b)
|\chi_a^{(+)}\varphi_a\rangle
\notag \\
&\equiv |\rho_{\text{2-body IAV}}\rangle,
\label{eq:reduction_to_2body}
\end{align}
and Eq.~(\ref{eq:master_A_practical}) becomes the standard
IAV~\cite{Ichimura1985} master formula with the two-body
projectile wave function
$\varphi_a(\mathbf r_{bk}) = f(\mathbf{y})|_{\text{cluster limit}}$,
where $\mathbf r_{bk} = -\mathbf y$ is the pair-to-$k$ relative
coordinate in the two-body convention of
Refs.~\cite{Ichimura1985,Lei2026nsp}.

The connection to the two-fragment detected-cluster formalism of
Ref.~\cite{Lei2026nsp} is obtained by a different reduction.
If the projectile $a$ happens to have a dominant two-body cluster
structure $a \approx b + k$ with $b$ a bound state of $(ij)$, then
the three-body wave function factorizes approximately as
$\Phi_a(\boldsymbol\zeta, \mathbf{y})
\approx \phi_0(\boldsymbol\zeta)\, f(\mathbf{y})$,
where $f(\mathbf{y})$ describes the $b$-$k$ relative motion.
In this limit, the pair projection
$\langle\phi_0|\Phi_a\rangle_\zeta = f(\mathbf{y})$ reduces to the
two-body wave function, and the four-body source reduces to the
two-fragment source of Ref.~\cite{Lei2026nsp} with
$\varphi_a(\mathbf{r}_{bk}, \boldsymbol\zeta)
= \Phi_a(\boldsymbol\zeta, \mathbf{y})$.
In the full source, the detected pair is projected from the three-body
projectile wave function rather than prescribed as an entrance-channel
cluster, and its internal states are resolved by the
$\boldsymbol\zeta$-dependent operators.
Comparing the source calculated with the full $\Phi_a$ and with
$\phi_0 f$ measures the cluster approximation at the amplitude
level.

For Partition B in the CFH reference limit ($V_{iA} \to U_{iA}$), the
result reduces to Eq.~(\ref{eq:sigma_B_ref}), which is the starting
point of the CFH formalism.
The CFH reference source involves the projectile wave function
$\Phi_a$ projected onto the distorted waves $\chi_i^{(-)}$ and
$\chi_a^{(+)}$, and the three-body resolvent $G_{jkA}$ is treated
within the optical-model framework for $j$ and $k$ separately.
The full four-body source also contains $V_{iA} - U_{iA}$.
For a composite detected particle this term contains the finite-size
tidal coupling of $i$ to the target; for a structureless detected
particle it is the difference between the microscopic and
optical-model $i$-$A$ interactions.
For structureless $i$, the target-excitation part can be expanded in
multipoles:
\begin{equation}
V_{iA}(\mathbf r_{iA}, \xi)
= \sum_{\lambda \mu} v_\lambda(r_{iA})\,
T^\lambda_\mu(\xi) Y^{\lambda*}_\mu(\hat{\mathbf r}_{iA}),
\label{eq:ViA_multipole}
\end{equation}
where $\mathbf r_{iA}$ is the exact $i$-$A$ separation of
Eq.~(\ref{eq:fragtarget}), evaluated inside the Partition-B source
integrals at
$\mathbf r_{iA} = \mathbf r_i + \frac{m_j+m_k}{m_j+m_k+m_A}
\,\mathbf r_{(jk)A}$ with $\mathbf r_{(jk)A}$ determined by the
source output variables while the integration runs over the Jacobi
coordinate $\mathbf r_i$; $T^\lambda_\mu(\xi)$ are target multipole
operators (whose
isoscalar or isovector character depends on the underlying
$NN$-effective interaction and on whether neutron-proton
asymmetries of the target one-body density are resolved; the
even-$\lambda$ pieces typically receive sizeable isoscalar
contributions and the odd-$\lambda$ pieces receive isovector
contributions, but the exact decomposition is target- and
energy-specific) and $v_\lambda(r_{iA})$ are radial form factors
obtained, e.g., by folding a $t$- or $g$-matrix with the target
one-body density.
With these definitions one has
\begin{align}
\langle\Phi_A|V_{iA}|\Phi_A\rangle
&= v_0(r_{iA})\,\langle\Phi_A|T^0_0|\Phi_A\rangle,
\notag \\
\langle\Phi_{A'}|V_{iA}|\Phi_A\rangle
&= \sum_{\lambda\mu} v_\lambda(r_{iA})
Y^{\lambda*}_\mu(\hat{\mathbf r}_{iA})\,
\langle\Phi_{A'}|T^\lambda_\mu|\Phi_A\rangle,
\label{eq:ViA_multipole_matrix}
\end{align}
the first of which equals $U_{iA}$ (up to the polarization
renormalization already absorbed in the phenomenological optical
potential) and the second controls $Q V_{iA} P$.
The target-transition matrix elements
$\langle\Phi_{A'}|T^\lambda_\mu|\Phi_A\rangle$ are exactly the
reduced matrix elements entering inelastic scattering of $i$
from the target; they are known from structure calculations or
from fits to inelastic $(p,p')$, $(\alpha,\alpha')$, etc.\ data
and are quantitatively controlled by low-lying collective
excitations of $A$ (surface dipole, quadrupole, octupole).
The explicit coupling source in the structureless case therefore
inherits the same inelastic form factors that dominate ordinary
inelastic scattering, and the Q-sector reduction of
Eq.~(\ref{eq:Qsector_reduction}) becomes, for each low-lying
$A'$,
\begin{equation}
\hat\rho_{jk}^{(A')}(\mathbf r_j,\mathbf r_k)
= \sum_{\lambda\mu}
\langle\Phi_{A'}|T^\lambda_\mu|\Phi_A\rangle\,
\mathcal R_{\lambda\mu}(\mathbf r_j,\mathbf r_k),
\label{eq:rho_A_prime_multipole}
\end{equation}
with
$\mathcal R_{\lambda\mu}(\mathbf r_j,\mathbf r_k) =
\langle\chi_i^{(-)}|\,v_\lambda(r_{iA})
Y^{\lambda*}_\mu(\hat{\mathbf r}_{iA})|\chi_a^{(+)}\Phi_a\rangle$ a
multipole form factor in the three-body relative-motion space.
The source therefore requires the inelastic form factors $v_\lambda$
and target transition strengths
$B(E\lambda;0\to A')\equiv (2J_0+1)^{-1}|\langle\Phi_{A'}\Vert
T^\lambda\Vert\Phi_A\rangle|^2$ that determine single-particle
inelastic scattering, folded with the three-body projectile
structure.

Table~\ref{tab:limits} collects the cluster and optical reductions.
The target-excited kernels apply to the direct $g_Q$ contribution
under the diagonal approximation; the interference terms remain in
Feshbach form.

\begin{table*}[t]
\caption{Reduction of the present four-body framework to existing
formalisms in various limits.}
\label{tab:limits}
\begin{tabular}{p{0.52\textwidth}p{0.36\textwidth}}
\hline\hline
Limit & Recovered / obtained formalism \\
\hline
A, $V_{bA} \to U_{bA}$, $\Phi_a \to \phi_0 f$
  & Two-body IAV~\cite{Ichimura1985} \\
A, full $V_{bA}$, $\Phi_a \to \phi_0 f$
  & Two-fragment detected-cluster result~\cite{Lei2026nsp} \\
A, full $V_{bA}$, full $\Phi_a$
  & This work, pair detection \\
B, $V_{iA} \to U_{iA}$, reduced $G_{jkA}$
  & CFH~\cite{CFH2017} \\
B, full $V_{iA}$, unreduced $G_{jkA}$
  & This work, single-particle starting point \\
B, full $V_{iA}$, diagonal direct-$g_Q$ reduction
  & Target-excited CFH-like kernel for the direct $g_Q$ part of
    $\mathcal I_Q$;
    $\mathcal I_{PQ},\mathcal I_{\mathrm{coup},PQ}$
    interferences and the induced $Q\to P\to Q$ piece left
    unreduced \\
\hline
\end{tabular}
\end{table*}

\section{Application to \texorpdfstring{${}^6\mathrm{Li}$}{6Li} reactions}
\label{sec:application}

For ${}^6\mathrm{Li}+A\to d+(\alpha+A)^*$, I use the
three-body description ${}^6\mathrm{Li}=\alpha+n+p$, with the
detected pair $b=d=(np)$ and unresolved fragment $k=\alpha$.
Spin labels were suppressed in Secs.~\ref{sec:hamiltonian}--\ref{sec:limits}.
The deuteron has $J^\pi=1^+$ and $L=0,2$ components.  The multipole
expressions below use its $M_J$-averaged density
[Eqs.~(\ref{eq:UdA_fold_def}), (\ref{eq:tidal_6Li})]; the cross
sections require the spin sums and averages appropriate to the
measurement.

The identification of the constituents is $i = n$, $j = p$,
$k = \alpha$, so the pair internal coordinate
$\boldsymbol\zeta = \mathbf{r}_n - \mathbf{r}_p$ is the $n$-$p$
relative coordinate and
$\mathbf{y} = \mathbf{r}_\alpha -
(m_n \mathbf{r}_n + m_p \mathbf{r}_p)/(m_n + m_p)$
is the $\alpha$-$d$ relative coordinate.
The internal Hamiltonian of the pair is
$h_b = K_\zeta + V_{np}$, with the deuteron ground state
$h_b |\phi_d\rangle = -\epsilon_d |\phi_d\rangle$ at binding energy
$\epsilon_d = 2.224$~MeV.
The three-body projectile wave function
$\Phi_a(\boldsymbol\zeta, \mathbf{y})$ satisfies
\begin{equation}
(K_\zeta + K_y + V_{np} + V_{n\alpha} + V_{p\alpha})\,\Phi_a
= -\epsilon_a\,\Phi_a \,,
\label{eq:6Li_bound}
\end{equation}
where $\epsilon_a = 3.70$~MeV is the ${}^6\mathrm{Li} \to
\alpha + n + p$ three-body breakup threshold.

The bound deuteron avoids the continuum-pair normalization while
retaining sensitivity to its internal structure.  I return to an
unbound detected pair, ${}^8\mathrm{Be}$ from ${}^9\mathrm{Be}$,
at the end of this section.

The composite interactions [Eqs.~(\ref{eq:Vbk}) and (\ref{eq:VbA})]
become
\begin{align}
V_{bk} &= V_{n\alpha} + V_{p\alpha} \,,
\label{eq:Vbk_6Li} \\
V_{bA} &= V_{nA} + V_{pA} \,,
\label{eq:VbA_6Li}
\end{align}
where, relative to the pair center of mass, the nucleons sit at
$\pm\boldsymbol\zeta/2$ (for $m_n \approx m_p$), so that their
exact separations from the target are
$\mathbf{r}_{dA} + \boldsymbol\zeta/2$ (neutron) and
$\mathbf{r}_{dA} - \boldsymbol\zeta/2$ (proton), with
$\mathbf{r}_{dA} = \mathbf{r}_b
+ \frac{m_\alpha}{m_\alpha+m_A}\,\mathbf{r}_\alpha$ the exact
deuteron-target separation [Eq.~(\ref{eq:fragtarget}) with
$k=\alpha$] and $\mathbf{r}_\alpha$ the $\alpha$-$A$ relative
coordinate.
The post-form residual interaction [Eq.~(\ref{eq:Vpost_A})] is
$V_{\mathrm{post}}^{(A)} = V_{n\alpha} + V_{p\alpha}
+ V_{nA} + V_{pA} - U_d$.

The reference-channel source for detecting the deuteron
[Eq.~(\ref{eq:rho_A_ref})] has the explicit form
\begin{align}
&\tilde\rho_d^{(\mathrm{ref})}(\mathbf{r}_\alpha)
= \int d\mathbf{r}_b\, d\boldsymbol\zeta\;
\chi_d^{(-)*}\,\phi_d^*
\notag \\
&\qquad \times
(V_{n\alpha} + V_{p\alpha} + U_{dA} - U_d)\,
\chi_a^{(+)}\,\Phi_a(\boldsymbol\zeta, \mathbf{y}) \,,
\label{eq:source_6Li_ref}
\end{align}
where $\mathbf{r}_\alpha$ is the $\alpha$-$A$ relative coordinate
(the variable of the $\alpha + A$ propagator),
$\mathbf{y} = \mathbf{r}_\alpha - \mathbf{r}_{dA}$ and the
entrance distorted-wave argument
$\mathbf{R} = (m_d \mathbf{r}_{dA}
+ m_\alpha \mathbf{r}_\alpha)/M_a$ follow from the exact relations
of Eq.~(\ref{eq:exit_to_entrance}), $U_{dA}$ is evaluated at
$\mathbf{r}_{dA}$, and the integration runs over the deuteron
Jacobi coordinate $\mathbf{r}_b$ and the $n$-$p$ relative
coordinate $\boldsymbol\zeta$.
$V_{n\alpha}$ and $V_{p\alpha}$ depend on
$\boldsymbol\zeta$ through the $\alpha$-nucleon separations, which
equal $\mathbf{y} - \boldsymbol\zeta/2$ ($\alpha$ relative to $n$)
and $\mathbf{y} + \boldsymbol\zeta/2$ ($\alpha$ relative to $p$)
for $m_n \approx m_p$, so the $\boldsymbol\zeta$ integration does
not factorize even for the reference-channel source.

The explicit pair-target coupling involves the tidal operator
$V_{bA} - U_{bA} = V_{nA} + V_{pA} - U_{dA}$.
Expanding in $\boldsymbol\zeta$ separates the pair's response into
dipole, quadrupole, and monopole terms.  These also occur in the
factorized two-fragment limit of Ref.~\cite{Lei2026nsp}, with mass
factors $\eta_{n,p}=m_{n,p}/(m_n+m_p)$.  Here they enter the full
source through $V_{nA}+V_{pA}$.  The pair matrix elements
$\langle\phi_d|\boldsymbol\zeta^n|\phi_\nu\rangle_\zeta$
are weighted by the three-body amplitudes $a_\nu(\mathbf y)$.
The additional dependence on projectile structure appears in the
form factors $\mathcal F_{d\nu}$ [Eq.~(\ref{eq:pair_form_factor})]
and the source difference [Eq.~(\ref{eq:cluster_correction})]
derived below.  The off-diagonal terms $\mathcal F_{d\nu}$ with
$\nu\neq d$ and their contribution to this source difference are
absent from the factorized cluster description of
Ref.~\cite{Lei2026nsp}.

Using the exact nucleon-target separations
$\mathbf r_{nA} = \mathbf r_{dA} + \eta_p\boldsymbol\zeta$,
$\mathbf r_{pA} = \mathbf r_{dA} - \eta_n\boldsymbol\zeta$ with
$\eta_{n,p} = m_{n,p}/(m_n+m_p)$ and expanding each nucleon-target
interaction about the pair-target separation $\mathbf r_{dA}$,
\begin{align}
V_{bA}(\mathbf r_{dA}, \boldsymbol\zeta)
&= V_\Sigma(\mathbf r_{dA})
 + \bigl(\eta_p \nabla V_{nA} - \eta_n \nabla V_{pA}\bigr)
 \cdot\boldsymbol\zeta
\notag \\
&\quad
 + \tfrac{1}{2}\bigl[\eta_p^2\,\partial_a\partial_b V_{nA}
 + \eta_n^2\,\partial_a\partial_b V_{pA}\bigr]
 \zeta^a\zeta^b
\notag \\
&\quad + O(\zeta^3) \,,
\label{eq:VbA_expand}
\end{align}
where $V_\Sigma = V_{nA} + V_{pA}$ is the isoscalar-like sum and
repeated Cartesian indices are summed.
Equation~(\ref{eq:VbA_expand}) is the general-mass expansion;
the form factors retain $\eta_n,\eta_p$ explicitly so the same
expansion applies to any pair $b=(ij)$ with arbitrary mass ratio
(e.g., $\alpha\alpha$ in ${}^9\mathrm{Be}$).
Specializing to $m_n \approx m_p$ so that
$\eta_n = \eta_p = 1/2$, and introducing the isovector combination
$V_\Delta = V_{nA} - V_{pA}$ (dominated by the Lane symmetry term and
by the Coulomb difference on charged targets), the linear term
collapses to $\tfrac{1}{2}\boldsymbol\zeta\cdot\nabla V_\Delta$ and
the quadratic term to
$\tfrac{1}{8}(\boldsymbol\zeta\cdot\nabla)^2 V_\Sigma$.
Averaging the interaction over the deuteron magnetic substates gives
the adiabatic folding potential,
\begin{equation}
U_{dA}^{\mathrm{fold}}(\mathbf r_{dA}) = \int d\boldsymbol\zeta\,
|\overline{\phi_d}(\boldsymbol\zeta)|^2\,
V_{bA}(\mathbf r_{dA}, \boldsymbol\zeta),
\label{eq:UdA_fold_def}
\end{equation}
where $|\overline{\phi_d}|^2 \equiv (2J_d+1)^{-1}
\sum_{M_J}|\phi_d^{M_J}(\boldsymbol\zeta)|^2$ is spherically
symmetric.
With this convention the odd-parity linear term averages to zero
and the spherical part of the quadratic term contributes
$\frac{\langle\zeta^2\rangle_d}{24}\nabla^2 V_\Sigma$, while the
traceless rank-2 part of the quadratic term has zero spherically
averaged expectation and survives as an off-diagonal operator
acting between magnetic substates and between $L=0,2$ components
of the deuteron.
For the equal-mass deuteron specialization,
\begin{align}
V_{bA} - U_{dA}^{\mathrm{fold}}
&=
\underbrace{\tfrac{1}{2}\,\boldsymbol\zeta \cdot \nabla V_\Delta
    (\mathbf r_{dA})}_{\text{E1 (isovector dipole)}}
\notag \\
&\quad
+ \underbrace{\tfrac{1}{8}\,
    \hat Q^{ab}(\boldsymbol\zeta)\,
    \partial_a\partial_b V_\Sigma(\mathbf r_{dA})}_{\text{E2 (tensor
    quadrupole)}}
\notag \\
&\quad
+ \underbrace{\tfrac{1}{24}\bigl(\zeta^2-\langle\zeta^2\rangle_d\bigr)
    \nabla^2 V_\Sigma(\mathbf r_{dA})}_{\text{monopole breathing}}
\notag \\
&\quad + O(\zeta^3) \,,
\label{eq:tidal_6Li}
\end{align}
where
$\hat Q^{ab}(\boldsymbol\zeta) = \zeta^a\zeta^b
- \tfrac{1}{3}\zeta^2\delta^{ab}$ is the traceless rank-2 tensor of
the pair.  Its Cartesian indices distinguish it from the target-space
projector $Q=\mathbf1-P$ of Sec.~\ref{sec:pair}.

The selection rules of the three pieces in
Eq.~(\ref{eq:tidal_6Li}) parallel those established in Sec.~V of
Ref.~\cite{Lei2026nsp} for the two-body deuteron-projectile case
and are recalled briefly here:
the E1 piece connects $|\phi_d\rangle$ only to odd-parity ${}^3P$
continuum states because $\langle\phi_d|\boldsymbol\zeta|\phi_d
\rangle=0$;
the E2 piece has zero spherically-averaged ground-state
expectation by construction of $U_{dA}^{\mathrm{fold}}$
[Eq.~(\ref{eq:UdA_fold_def})] but is operator-active through the
deuteron $L=0$--$L=2$ mixing and on the continuum via the
$^3S_1$--$^3D_1$ coupling;
the monopole breathing piece vanishes identically on
$|\phi_d\rangle$ and enters only through continuum pair states.
With surface estimates $|\nabla V_\Delta|\sim 20$~MeV/fm and
$|\nabla^2 V_\Sigma|\sim 10$~MeV/fm$^2$ on $^{208}$Pb (from a Lane
isovector $|V_p-V_n|/2\sim 5$--$8$~MeV dropping over surface
diffuseness $\sim 0.7$~fm and a Woods-Saxon isoscalar field with
$V_0\sim 50$~MeV, $R\sim 7$~fm, $a\sim 0.65$~fm), and the deuteron
$\zeta_{\mathrm{rms}}\equiv\sqrt{\langle\zeta^2\rangle_d}
\approx 3.9$~fm, the three operator pieces of
Eq.~(\ref{eq:tidal_6Li}) carry the order-of-magnitude scales
\begin{align}
\tfrac{1}{2}\zeta_{\mathrm{rms}}|\nabla V_\Delta|/\sqrt{3}
&\sim 20\text{--}25~\text{MeV},
\notag \\
\tfrac{1}{8}\zeta^2_{\mathrm{rms}}|\nabla^2 V_\Sigma|
&\sim 15\text{--}25~\text{MeV},
\notag \\
\tfrac{1}{24}\zeta^2_{\mathrm{rms}}|\nabla^2 V_\Sigma|
&\sim 5\text{--}8~\text{MeV},
\label{eq:multipole_scales}
\end{align}
These reproduce the estimates of Ref.~\cite{Lei2026nsp}, since
the pair moments and target-field inputs are the same.

The overall sign of the E1 piece reflects the convention
$\boldsymbol\zeta=\mathbf r_n-\mathbf r_p$ adopted here, opposite
to Ref.~\cite{Lei2026nsp}; only $|\nabla V_\Delta|$ enters the
scale estimate.

The scales in Eq.~(\ref{eq:multipole_scales}) do not estimate
cross-section corrections.  For $L_{\mathrm{diff}}\sim0.6$~fm,
$\zeta_{\mathrm{rms}}/L_{\mathrm{diff}}\sim6$; higher powers of
$\boldsymbol\zeta$ are not suppressed.  A calculation must therefore
use the full coordinate-dependent interaction.  The source integral
[Eq.~(\ref{eq:source_6Li_ref})] also contains radial phase-space
weights and cancellations between distorted waves, which cannot be
inferred from surface operator magnitudes.  The multipole expansion
identifies the selection rules, but its truncated terms do not rank
the contributions to the cross section.

In the usual two-body IAV calculation of
${}^6\mathrm{Li}+A\to d+(\alpha+A)^*$, the entrance state is an
$\alpha+d$ cluster wave function.  The three-body source instead
projects $\Phi_a$ onto the detected deuteron after the interaction
acts.  It samples the wave function in the reaction region, so the
norm of the deuteron component alone cannot determine the correction.
To evaluate this difference, expand $\Phi_a$ in pair eigenstates
at fixed $\mathbf y$:
\begin{equation}
\Phi_a(\boldsymbol\zeta, \mathbf{y})
= \sum_\nu a_\nu(\mathbf{y})\, \phi_\nu(\boldsymbol\zeta) \,,
\qquad
a_\nu(\mathbf{y})
= \langle\phi_\nu|\Phi_a\rangle_\zeta \,,
\label{eq:pair_expansion}
\end{equation}
where $\nu$ labels the $np$ pair eigenstate and $\sum_\nu$ includes
a discrete sum over bound pair states
(only the deuteron for $b=(np)$) and a continuum integral over $np$
scattering states.
Substituting into Eq.~(\ref{eq:source_6Li_ref}) and integrating
over $\boldsymbol\zeta$ gives
\begin{widetext}
\begin{align}
\tilde\rho_d^{(\mathrm{ref})}(\mathbf{r}_\alpha)
&= \underbrace{\int d\mathbf r_b\;
\chi_d^{(-)*}(\mathbf r_b)\,
\bigl[U_{dA}(\mathbf r_{dA}) - U_d(\mathbf r_b)\bigr]\,
\chi_a^{(+)}(\mathbf R)\,
f_d(\mathbf y)}_{\boldsymbol\zeta\text{-independent operator: pair
projection selects }\nu=d}
\notag \\
&\quad
+ \underbrace{\sum_\nu\int d\mathbf r_b\;
\chi_d^{(-)*}(\mathbf r_b)\,
\mathcal{F}_{d\nu}(\mathbf y)\,
\chi_a^{(+)}(\mathbf R)\,
a_\nu(\mathbf y)}_{\text{pair form factors generated by }V_{bk}},
\label{eq:rho_factorized}
\end{align}
\end{widetext}
where $\mathbf R$ and $\mathbf y$ retain the arguments of
Eq.~(\ref{eq:exit_to_entrance}).  The pair form factor
\begin{equation}
\mathcal{F}_{d\nu}(\mathbf y)
= \int d\boldsymbol\zeta\,
\phi_d^*(\boldsymbol\zeta)\,
[V_{n\alpha} + V_{p\alpha}](\mathbf y, \boldsymbol\zeta)\,
\phi_\nu(\boldsymbol\zeta)
\label{eq:pair_form_factor}
\end{equation}
couples the pair state $\phi_\nu$ to the deuteron through
$V_{bk}=V_{n\alpha}+V_{p\alpha}$.  For $m_n\approx m_p$, the
$\alpha$-nucleon separations are $\mathbf y-\boldsymbol\zeta/2$
and $\mathbf y+\boldsymbol\zeta/2$.  Expansion about $\mathbf y$
gives
\begin{align}
\mathcal F_{d\nu}(\mathbf y)
&= \langle\phi_d|V_{n\alpha}+V_{p\alpha}|\phi_\nu\rangle_\zeta
\notag \\
&= \bigl[V_{n\alpha}(\mathbf y) + V_{p\alpha}(\mathbf y)\bigr]\,
 \delta_{d\nu}
\notag \\
&\quad
 - \tfrac{1}{2}\bigl[\nabla V_{n\alpha} - \nabla V_{p\alpha}\bigr]
 \cdot
 \langle\phi_d|\boldsymbol\zeta|\phi_\nu\rangle_\zeta
\notag \\
&\quad
 + \tfrac{1}{8}\partial_a\partial_b
 \bigl[V_{n\alpha} + V_{p\alpha}\bigr]
 \langle\phi_d|\zeta^a\zeta^b|\phi_\nu\rangle_\zeta
\notag \\
&\quad + O(\zeta^3),
\label{eq:pair_form_factor_multipole}
\end{align}
The zeroth-order term survives only for $\nu=d$ and gives the
ordinary cluster form factor.  The remaining terms involve the same
pair moments $\langle\phi_d|\boldsymbol\zeta^n|\phi_\nu\rangle_\zeta$
as Eq.~(\ref{eq:tidal_6Li}), with the E1, E2, and monopole
selection rules stated there.  Here the radial fields come from the
$N$-$\alpha$ interactions and are evaluated at $\mathbf y$,
rather than from the target fields at $\mathbf r_{dA}$.

The reference-channel source in Eq.~(\ref{eq:rho_factorized}) contains
the deuteron spectroscopic amplitude
\begin{equation}
f_d(\mathbf{y}) = a_d(\mathbf{y})
= \langle\phi_d | \Phi_a\rangle_\zeta
= \int d\boldsymbol\zeta\;
\phi_d^*(\boldsymbol\zeta)\,
\Phi_a(\boldsymbol\zeta, \mathbf{y}) \,,
\label{eq:spectroscopic_amplitude}
\end{equation}
which multiplies $U_{dA}-U_d$.  The other pair components enter
through $\mathcal F_{d\nu}a_\nu$.  The term $\nu=d$ is diagonal;
$\nu\neq d$ describes mixing with other pair states.

If $\Phi_a$ factorizes as
$\Phi_a(\boldsymbol\zeta,\mathbf y)
= \phi_d(\boldsymbol\zeta)\,f(\mathbf y)$ with the deuteron in its
ground state, then
$a_\nu(\mathbf y) = \langle\phi_\nu|\phi_d\rangle_\zeta
f(\mathbf y) = \delta_{\nu d}\,f(\mathbf y)$ by the orthogonality
of the pair eigenstates, so that $a_d(\mathbf y) = f(\mathbf y)$
and $a_\nu(\mathbf y) = 0$ for all $\nu \neq d$.
The second line of Eq.~(\ref{eq:rho_factorized}) then collapses to
its single term $\mathcal F_{dd}(\mathbf y)\,
f(\mathbf y)$, and $\mathcal F_{dd}$ reduces to the
folding-potential correction
$\langle\phi_d|V_{n\alpha}+V_{p\alpha}|\phi_d\rangle_\zeta$ that
combines with $U_{dA} - U_d$ into the ordinary two-body IAV
residual $V_{bx} + U_{bA} - U_b$ of Ref.~\cite{Lei2026nsp}, with
the identifications $b \to d$ and $x \to \alpha$ in the two-body
framework.
In the full three-body wave function, $a_\nu(\mathbf y) \neq 0$
for $\nu \neq d$, and the second line of
Eq.~(\ref{eq:rho_factorized}) contributes through the off-diagonal
form factors $\mathcal{F}_{d\nu}$; the same structure appears in
the explicit coupling source, with $V_{bA} - U_{bA}$ replacing $V_{bk}$
in $\mathcal{F}_{d\nu}$.
Constructing the cluster source with the same projected amplitude
$f_d=a_d$ isolates this mixing.  The difference between the full
and cluster sources is
\begin{align}
\tilde\rho_d^{(\mathrm{ref})}
- \tilde\rho_d^{(\mathrm{ref, cluster})}
&= \sum_{\nu\neq d}
\int d\mathbf r_b\,\chi_d^{(-)*}\,
\notag \\
&\quad \times
\mathcal F_{d\nu}(\mathbf y)\,
\chi_a^{(+)}\,a_\nu(\mathbf y),
\label{eq:cluster_correction}
\end{align}
so both the non-deuteron amplitudes $a_\nu$ and the matrix elements
of $V_{n\alpha}+V_{p\alpha}$ enter the correction.  The leading
dipole term couples the even-parity deuteron to odd-parity continuum
states, while the quadrupole term couples to even-parity components.
The same pair moments enter the explicit pair-target coupling, with
different radial fields.

The three-body wave function $\Phi_a$ for ${}^6\mathrm{Li}$ can be
obtained from established three-body methods, including the Gaussian
expansion method~\cite{Hiyama2003}, the Lagrange-mesh
method~\cite{Descouvemont2003,Tursunov2006}, and coupled-channel
approaches~\cite{Thompson1988}.
The deuteron spectroscopic factor
$S_d=\int|f_d(\mathbf y)|^2\,d\mathbf y$ measures the norm of
the $\alpha+d$ component.  It is model dependent and is not an
observable.  Nor does $1-S_d$ give the reaction-amplitude correction:
Eq.~(\ref{eq:cluster_correction}) weights the non-deuteron
components by the form factors and distorted waves at each radius.
These off-diagonal contributions are absent from the two-body IAV
cluster model.

Choosing $\alpha=0$ gives the inclusive deuteron spectrum.
Choosing a continuum state $\alpha=(\mathbf q)$ gives the
$np$ coincidence distribution for
${}^6\mathrm{Li}+A\to(np)_{\mathbf q}+(\alpha+A)^*$ at specified
pair relative energy and angle.  Such a measurement would probe
$np$ correlations during breakup.

The same ${}^6\mathrm{Li} = \alpha + n + p$ system also illustrates
Partition B.
Relabeling the constituents so that the detected particle is
$i = p$, with $j = n$ and $k = \alpha$,
the unresolved system is $n + \alpha + A$ and the propagator
$G_{jkA} = G_{n\alpha,A}$ describes their three-body propagation.
Under this relabeling, the Partition-B Jacobi coordinates
defined by Eqs.~(\ref{eq:jacobi_B_zeta}) and (\ref{eq:jacobi_B_y})
become $\boldsymbol\zeta' = \mathbf{r}_n - \mathbf{r}_\alpha$ (the
$n$-$\alpha$ relative coordinate, internal to the unresolved pair)
and, by the Partition-B relation below
Eq.~(\ref{eq:exit_to_entrance}),
$\mathbf{y}' = \mathbf{r}_{pA} - \mathbf{r}_{(n\alpha)A}$ (the
position of the detected proton relative to the $n$-$\alpha$ center
of mass), where $\mathbf{r}_{pA}$ is the exact proton-target
separation,
$\mathbf{r}_{(n\alpha)A} = (m_n \mathbf{r}_n
+ m_\alpha \mathbf{r}_\alpha)/(m_n+m_\alpha)$, and
$\mathbf{r}_n, \mathbf{r}_\alpha$ are the $n$-$A$ and $\alpha$-$A$
relative coordinates of the unresolved subsystem, following the
convention of Sec.~\ref{sec:single}.
These differ from the Partition-A coordinates
$\boldsymbol\zeta = \mathbf{r}_n - \mathbf{r}_p$ and
$\mathbf{y} = \mathbf{r}_\alpha - (m_n\mathbf{r}_n
+ m_p\mathbf{r}_p)/(m_n + m_p)$ used earlier in this section; the
two parametrizations of the same $\Phi_a$ are related by a Jacobi
transformation.
This observable cannot be reduced to the standard two-body IAV,
because the remaining pair $(n\alpha) = {}^5\mathrm{He}$ is unbound
and does not form a two-body cluster.
In the CFH reference limit ($V_{pA} \to U_{pA}$), the source, after
projection onto the target ground state $\Phi_A$ as in
Sec.~\ref{sec:single}, is
\begin{align}
&\tilde\rho_p^{(\mathrm{ref})}(\mathbf{r}_n, \mathbf{r}_\alpha)
= \int d\mathbf{r}_p\;
\chi_p^{(-)*}(\mathbf{r}_p)
\notag \\
&\qquad \times
(V_{pn} + V_{p\alpha} + U_{pA} - U_p)\,
\chi_a^{(+)}\,\Phi_a(\boldsymbol\zeta', \mathbf{y}') \,,
\label{eq:source_6Li_B}
\end{align}
where the integration variable $\mathbf{r}_p$ is the proton Jacobi
coordinate, namely the proton position relative to the CoM of the
unresolved $n + \alpha + A$ system, $U_{pA}$ is evaluated at the
exact proton-target separation
$\mathbf{r}_{pA} = \mathbf{r}_p
+ \frac{m_n+m_\alpha}{m_n+m_\alpha+m_A}\,\mathbf{r}_{(n\alpha)A}$
[Eq.~(\ref{eq:fragtarget})], built from the source output
variables, and the projectile coordinates
$(\boldsymbol\zeta', \mathbf{y}')$ and the entrance argument of
$\chi_a^{(+)}$
[Eqs.~(\ref{eq:jacobi_B_zeta}), (\ref{eq:jacobi_B_y}), and
(\ref{eq:exit_to_entrance}) with
$i = p,\, j = n,\, k = \alpha$] are determined by the exact
kinematic relations of Sec.~\ref{sec:hamiltonian}.
After the CFH-type Feshbach-Faddeev reduction of the three-body
resolvent, the NEB cross section takes the schematic form
$W_n + W_\alpha + W_{3B}$~\cite{CFH2017}, where $W_n$ describes
neutron absorption by the target (forming compound nucleus $A+n$),
$W_\alpha$ describes $\alpha$ absorption (forming $A+\alpha$), and
$W_{3B}$ represents the correlated $n\alpha$ absorption (forming
$A + n + \alpha$).
The cross term $W_{3B}$ of CFH~\cite{CFH2017} has no two-body
IAV counterpart.
The explicit proton-target coupling adds $V_{pA} - U_{pA}$ to the source.
For a structureless proton this correction represents target
excitations driven by the difference between the microscopic $p$-$A$
interaction and the elastic optical potential.
Unlike the deuteron-target coupling of the pair-detection channel,
$V_{dA} - U_{dA}$, it has no finite-size tidal enhancement from an
extended detected fragment, so its quantitative importance must be
assessed through the reaction calculation rather than inferred from
the operator scale alone.

For ${}^9\mathrm{Be}=\alpha+\alpha+n$, the reaction
${}^9\mathrm{Be}+A\to{}^8\mathrm{Be}+(n+A)^*$ requires an unbound
pair state $b={}^8\mathrm{Be}=(\alpha\alpha)$.  Choosing
$\phi_\alpha$ as a continuum or discretized-bin state selects a
pair energy interval or resonance window.  The source then retains
the three-body projectile wave function, whereas
Ref.~\cite{Villanueva2024} treats ${}^8\mathrm{Be}$ as a preformed
pair.

\section{Discussion}
\label{sec:discussion}

The sum rules and their optical reductions require different assumptions.
The conditions are collected here as (A1)--(A6), followed by the
numerical steps needed for each measurement.  Conditions (A1)--(A3)
define the DWBA sum rules and exact rearrangements within them;
(A4) specifies the optical representation, (A5) its absorption-sign
condition, and (A6) the diagonal approximation for direct propagation
between excited target states.  The diagonal approximation in (A6)
does not require the sign condition in (A5).

\par\smallskip\noindent\textbf{(A1) DWBA truncation of the full $T$-matrix.}
The inclusive cross section is computed from the DWBA transition
amplitudes Eqs.~(\ref{eq:T_A}) and~(\ref{eq:T_B}); residual
distortion in the entrance and exit channels is encoded in the
optical potentials $U_a, U_b, U_i$.
This is the standard sum-rule starting point and fixes
Eqs.~(\ref{eq:master_A}) and~(\ref{eq:master_B}).
\par\smallskip\noindent\textbf{(A2) Spectral identity for the unresolved
subsystem.}
The on-shell sum over $|\Psi^c\rangle$ is replaced by
$-\pi^{-1}\,\mathrm{Im}\,G$ via Sokhotski-Plemelj.
This is exact within (A1) and applies to both partitions.
\par\smallskip\noindent\textbf{(A3) Feshbach $P$-$Q$ rewriting in target Hilbert
space.}
The unresolved-system propagator is split as
$G = G_{PP}+G_{PQ}+G_{QP}+G_{QQ}^{\mathrm{full}}$ with
$P=|\Phi_A\rangle\langle\Phi_A|$, $Q=\mathbf 1-P$.
This is an exact algebraic rearrangement and introduces no
approximation.
The reduced $Q$-space resolvent entering the Feshbach kernel is
denoted $g_Q$.
\par\smallskip\noindent\textbf{(A4) Single-channel optical reduction in the $P$
sector.}
The exact Feshbach polarization kernel from $G_{PP}^{\rm exact}$
is replaced by an energy-independent phenomenological optical
representation in both partitions.
For Partition A this amounts to writing the elastic block of
$G_{kA}^{\rm full}$ as the two-body optical resolvent
$G_k = (E_{k,\alpha}^+ - K_k - U_k)^{-1}$, with dynamic
polarization absorbed into the phenomenological $U_k$
[underlying Eq.~(\ref{eq:NEB_A})].
For Partition B it amounts to replacing
$\Delta\mathcal H_{PP}$ in Eq.~(\ref{eq:GPP_inverse}) by the
CFH-optical kernel $U_j+U_k+\mathcal V_{3B}$ of
Eq.~(\ref{eq:HPP_opt}), absorbing the diagonal one-fragment
Feshbach pieces into phenomenological optical potentials and
retaining the irreducible three-body cross term as $\mathcal
V_{3B}$ [underlying Eq.~(\ref{eq:CFH_NEB})].
In both cases the $E$-derivative of the corresponding exact
polarization kernel is neglected; only the genuine three-body
cross term $\mathcal V_{3B}$ has no Partition-A counterpart.
\par\smallskip\noindent\textbf{(A5) Phenomenological absorption-sign assumption.}
The full kernel $W_j+W_k+W_{3B}$ in
Eq.~(\ref{eq:ImGPP_CFH}) is treated as negative semidefinite
under the standard CFH-optical replacement.
At the microscopic level this is the negative semi-definiteness
of $\mathrm{Im}\,\Delta\mathcal H_{PP}$; at the phenomenological
level it requires the residual three-body $W_{3B}$ to be a small
perturbation around absorptive $W_j+W_k$, an assumption inherited
from the standard CFH optical reduction~\cite{CFH2017} and not introduced
here.
\par\smallskip\noindent\textbf{(A6) Diagonal-intermediate-target-states ansatz.}
Only the outer propagation between target-excited labels in the
direct $g_Q$ term is diagonalized:
$\langle\Phi_{A''}|g_Q|\Phi_{A'}\rangle$ with $A'\neq A''$ is
dropped, giving Eq.~(\ref{eq:gQ_diag}).
The nested dressing inside each diagonal block, defined through
Eq.~(\ref{eq:Qperp_decomp}), is retained and continues to generate
the target-excited optical pieces
$\tilde U_{j,k}^{(A')}$ and $\tilde{\mathcal V}_{3B}^{(A')}$.
The resulting target-excited CFH-like kernels follow from
Eqs.~(\ref{eq:HjkAprime})--(\ref{eq:Qsector_reduction}).
This is used \emph{only} for the direct $g_Q$ part of the
pure-$Q$ contribution $\mathcal I_Q$ of
Eq.~(\ref{eq:target_excited_CFH_B}); the $PQ$ and $QP$ interferences and
the induced $Q\to P\to Q$ part of $G_{QQ}^{\mathrm{full}}$ are not
reduced under this ansatz and remain in their unreduced Feshbach
form.

The optical assumptions (A4) and (A5) concern the reduced absorption
formulas, while (A6) concerns the direct target-excited contribution.
None of these reductions changes the original DWBA sum rules or their
Feshbach decomposition.  Relaxing (A4) requires returning to the energy-dependent
polarization kernels.  The reference interactions $U_{bA}$ and $U_{iA}$
are added and subtracted in the source and cancel before reduction.
The separated terms depend on the reference potentials.  Within a
fixed reduction, only their sum gives the inclusive observable.

For pair detection, the two-body optical propagator can be evaluated
by the methods already used in IAV calculations.  The extra work lies
in the source: Eq.~(\ref{eq:source_6Li_ref}) integrates over both the
pair center-of-mass coordinate $\mathbf r_b$ and the internal coordinate
$\boldsymbol\zeta$, rather than over $\mathbf r_b$ alone.  The
interactions depend on the individual constituent positions, so the
six-dimensional integral does not factorize.  It requires a three-body
projectile wave function and, for an extended pair such as the deuteron
($\zeta_{\mathrm{rms}}\approx3.9$~fm), a converged angular-momentum
expansion.  In the reduced channels where post-prior equivalence has
been established, the prior form avoids the slow convergence caused by
the lack of a natural cutoff in the post-form transition integral.

For ${}^6\mathrm{Li}$, a useful comparison begins with the ordinary
$\alpha+d$ cluster model, $\Phi_a\approx\phi_d f$, and
$V_{bA}\to U_{bA}$.  Replacing the factorized wave function by the full
$\Phi_a$ in Eq.~(\ref{eq:rho_A_ref}) tests projectile correlations.
Adding $V_{bA}-U_{bA}$ then tests the explicit pair-target coupling.
The two steps separate the effects of projectile structure and target
coupling in the calculated spectrum.

For single-particle detection, the CFH reduction~\cite{CFH2017}
requires a three-body wave function acted on by $W_j+W_k+W_{3B}$.
The source also contains $V_{iA}-U_{iA}$.  For structureless $i$,
this target-excitation coupling can be evaluated with the inelastic
form factors used in $(p,p')$ and $(\alpha,\alpha')$ analyses
[Eq.~(\ref{eq:ViA_multipole})]; for composite $i$, the tidal multipoles
of Ref.~\cite{Lei2026nsp} also enter.  Their propagation is treated by
the target-space reduction described above.

The diagonal-target approximation also gives a practical form of
the three-body cross term in Eq.~(\ref{eq:V3B_def}):
\begin{align}
\mathcal V_{3B}
&\approx \sum_{A'\neq 0}\Bigl[
F^{(j)}_{0A'}(\mathbf r_j)\,\tilde G_{jk}^{(A')}\,
F^{(k)}_{A'0}(\mathbf r_k)
\notag\\
&\qquad\quad
+ F^{(k)}_{0A'}(\mathbf r_k)\,\tilde G_{jk}^{(A')}\,
F^{(j)}_{A'0}(\mathbf r_j)\Bigr],
\label{eq:V3B_practical}
\end{align}
with $F^{(j)}_{A'0}(\mathbf r_j)
= \langle\Phi_{A'}|V_{jA}|\Phi_A\rangle$ and
$F^{(j)}_{0A'} = [F^{(j)}_{A'0}]^{*}$ the inelastic transition form
factors of $j+A$, and likewise for $k$.
These are the usual $j+A$ and $k+A$ inelastic form factors, obtained
from rotational or vibrational couplings for a collective target
~\cite{Thompson1988}.  At lowest order,
$\tilde G_{jk}^{(A')}$ is evaluated at
$E_{jkA}-\omega_{A'}$ with the Hamiltonian
$h_j+h_k+K_j+K_k+V_{jk}+U_j+U_k$.  The result is a nonlocal,
energy-dependent second-order kernel.  A first calculation can set
$\mathcal V_{3B}=0$ in the wave equation and evaluate
$\langle\psi_{jk}|W_{3B}|\psi_{jk}\rangle$ perturbatively.  Its
size, and the effect of the off-diagonal target blocks omitted in
(A6), must be tested on the resulting wave function.

Only the action of $G_{PP}^{\mathrm{opt}}$ on the source is needed.
With
$|\psi_{jk}\rangle=G_{PP}^{\mathrm{opt}}|\hat\rho_{jk}\rangle$,
Eq.~(\ref{eq:CFH_NEB}) requires the solution of
$(E_{jkA}^+-\mathcal H_{PP}^{\mathrm{opt}})|\psi_{jk}\rangle
=|\hat\rho_{jk}\rangle$ with outgoing boundary conditions, rather
than construction of the full off-shell propagator.  Expanding the
$jk$ pair in bound and continuum-bin states gives inhomogeneous
coupled equations that can be solved by the Lagrange-mesh $R$-matrix
method~\cite{Lei2020inhomo,Lei2023cdcc}.  This route has the cost and
bin asymptotics of a four-body continuum-discretized
coupled-channels (CDCC) calculation
~\cite{Descouvemont2015}.  Alternatively, a Faddeev-Alt-Grassberger-Sandhas (AGS) treatment
provides exact three-body asymptotics and can include complex,
nonlocal fragment-target optical potentials
~\cite{Deltuva2009,Deltuva2025}.  In either route,
$\mathcal V_{3B}$ can first be added perturbatively.  The homogeneous
equation gives $|\chi_{jk}^{(+)}\rangle$ in
Eq.~(\ref{eq:EBU_B}).

A first application is
${}^6\mathrm{Li}+A\to p+(n+\alpha+A)^*$ above the Coulomb barrier.
It requires a three-body projectile wave function, the $nA$,
$\alpha A$, and $n\alpha$ interactions, a six-dimensional prior-form
source, and a discretized $n\alpha$ continuum
~\cite{Hiyama2003,Descouvemont2003}.  With
$\mathcal V_{3B}=0$, each proton energy and angle requires one
inhomogeneous solve of four-body-CDCC size.

The normal-eigenstate target reduction does not describe
pairing-collective enhancement of two-neutron absorption on an
open-shell target.  For an unresolved neutron
pair in $(t,p)$ or Borromean-projectile reactions, the model-space
sum in Eq.~(\ref{eq:master_B}) includes bound $nn+A$ configurations,
but these are not antisymmetrized physical $A+2$ eigenstates.  The
subsequent reduction is narrower still: its $Q$ space contains
particle-number-conserving states of $H_A$, whereas pairing
vibrations are coherent quasiparticle configurations in the
$A\pm2$ systems~\cite{CFH2017,Potel2013}.  That coherence is
supplied neither by the free $V_{jk}$ nor by the diagonal form of
Eq.~(\ref{eq:V3B_practical}).  A treatment of this physics requires
a quasiparticle target space or explicit two-nucleon transfer
channels.  The $(t,n)$ reaction leaves an unresolved $p+n$ pair and
raises the analogous question for proton-neutron correlations.

In the two-body-projectile limit ($a=d$,
$a+A\to p+(n+A)^*$) Partition B (single-particle detection of $p$,
with the unresolved subsystem reducing to $n+A$) reduces to the
standard two-body IAV sum rule with a structureless detected
particle; this is the same inclusive observable that
Ref.~\cite{Deltuva2025} solves exactly within the Faddeev
framework.
Partition A in the same limit corresponds to detecting the bound
deuteron itself, which is elastic scattering of $d$ off $A$ rather
than an inclusive breakup observable, and is not the relevant
two-body limit for the comparison between IAV and Faddeev.
Ref.~\cite{Deltuva2025} is therefore the relevant benchmark for
the two-body limit of Partition B.

For ${}^6\mathrm{Li}$, the deuteron and proton spectra should be
calculated with the same projectile wave function and consistent
channel interactions and source terms.  Proton detection
leaves the unbound $n\alpha$ pair and requires the three-body
$n+\alpha+A$ propagation.  The inclusive $\alpha$ spectrum, by
comparison, can also be treated in the two-body $\alpha+d$ cluster
model with an unresolved $d+A$ system~\cite{Lei2015,Lei2017}.

For the reference pair-target interaction, the adiabatic folding
potential~\cite{Johnson1970,Harvey1971} removes the diagonal correction
and leaves off-diagonal pair and target-excitation couplings in the
explicit source.  An empirical elastic optical potential is also
admissible, provided polarization effects already included in it are
not counted again~\cite{Feshbach1992,Lei2026nsp}.

\section{Summary and outlook}
\label{sec:summary}

The detected particles determine the unresolved propagator in inclusive
breakup of a three-body projectile.  Pair detection leaves a two-body
propagator and places the projectile's three-body structure in the
pair-projected source.  Comparing this source with its factorized limit
measures, at the amplitude level, what the preformed-cluster
approximation omits.
Factorizing the projectile wave function recovers the two-fragment
detected-cluster result~\cite{Lei2026nsp}.

Single-particle detection requires a three-body propagator.  Its
optical reduction in the target ground state recovers the CFH kernel
$W_j+W_k+W_{3B}$~\cite{CFH2017}, with the absorption-sign condition
(A5).  The detected particle's target interaction adds target-excited
contributions.  The direct part reduces to analogous optical kernels
under (A6); off-diagonal propagation and interference remain in
Feshbach form.  These restrictions belong to the reduced formulas,
not to the original DWBA sum rules.

A first calculation can compare deuteron and proton spectra from
${}^6\mathrm{Li}=\alpha+n+p$ using the same projectile wave function.
The deuteron channel requires the pair-projected source; the proton
channel requires an inhomogeneous three-body solve.  Inelastic form
factors provide a starting point for including $\mathcal V_{3B}$
perturbatively.  Describing pairing-collective enhancement of
two-neutron absorption on open-shell targets, relevant to $(t,p)$
and Borromean projectiles, requires a quasiparticle target space or
explicit two-nucleon transfer channels beyond the target reduction used here.

\begin{acknowledgments}
This work was supported by the National Natural Science Foundation
of China (Grants No.~12475132 and No.~12535009) and the Fundamental
Research Funds for the Central Universities.
AI-based writing assistants were used only for English editing and
for condensing expository text; all mathematical derivations,
physical arguments, and conclusions are the author's.
The author takes full responsibility for the contents of this
paper.
\end{acknowledgments}

\section*{Data Availability}
There are no publicly available research data or software
supporting this paper. Requests for further information or data
should be sent to the author.

\bibliography{references}

\begin{thebibliography}{45}%
\makeatletter
\providecommand \@ifxundefined [1]{%
 \@ifx{#1\undefined}
}%
\providecommand \@ifnum [1]{%
 \ifnum #1\expandafter \@firstoftwo
 \else \expandafter \@secondoftwo
 \fi
}%
\providecommand \@ifx [1]{%
 \ifx #1\expandafter \@firstoftwo
 \else \expandafter \@secondoftwo
 \fi
}%
\providecommand \natexlab [1]{#1}%
\providecommand \enquote  [1]{``#1''}%
\providecommand \bibnamefont  [1]{#1}%
\providecommand \bibfnamefont [1]{#1}%
\providecommand \citenamefont [1]{#1}%
\providecommand \href@noop [0]{\@secondoftwo}%
\providecommand \href [0]{\begingroup \@sanitize@url \@href}%
\providecommand \@href[1]{\@@startlink{#1}\@@href}%
\providecommand \@@href[1]{\endgroup#1\@@endlink}%
\providecommand \@sanitize@url [0]{\catcode `\\12\catcode `\$12\catcode
  `\&12\catcode `\#12\catcode `\^12\catcode `\_12\catcode `\%12\relax}%
\providecommand \@@startlink[1]{}%
\providecommand \@@endlink[0]{}%
\providecommand \url  [0]{\begingroup\@sanitize@url \@url }%
\providecommand \@url [1]{\endgroup\@href {#1}{\urlprefix }}%
\providecommand \urlprefix  [0]{URL }%
\providecommand \Eprint [0]{\href }%
\providecommand \doibase [0]{https://doi.org/}%
\providecommand \selectlanguage [0]{\@gobble}%
\providecommand \bibinfo  [0]{\@secondoftwo}%
\providecommand \bibfield  [0]{\@secondoftwo}%
\providecommand \translation [1]{[#1]}%
\providecommand \BibitemOpen [0]{}%
\providecommand \bibitemStop [0]{}%
\providecommand \bibitemNoStop [0]{.\EOS\space}%
\providecommand \EOS [0]{\spacefactor3000\relax}%
\providecommand \BibitemShut  [1]{\csname bibitem#1\endcsname}%
\let\auto@bib@innerbib\@empty
\bibitem [{\citenamefont {Moro}\ \emph {et~al.}(2025)\citenamefont {Moro},
  \citenamefont {Casal},\ and\ \citenamefont {G{\'o}mez-Ramos}}]{Moro2025}%
  \BibitemOpen
  \bibfield  {author} {\bibinfo {author} {\bibfnamefont {A.~M.}\ \bibnamefont
  {Moro}}, \bibinfo {author} {\bibfnamefont {J.}~\bibnamefont {Casal}},\ and\
  \bibinfo {author} {\bibfnamefont {M.}~\bibnamefont {G{\'o}mez-Ramos}},\
  }\bibfield  {title} {\bibinfo {title} {{The art of modeling nuclear reactions
  with weakly bound nuclei: status and perspectives}},\ }\href
  {https://doi.org/10.1140/epja/s10050-025-01500-0} {\bibfield  {journal}
  {\bibinfo  {journal} {Eur. Phys. J. A}\ }\textbf {\bibinfo {volume} {61}},\
  \bibinfo {pages} {47} (\bibinfo {year} {2025})}\BibitemShut {NoStop}%
\bibitem [{\citenamefont {Budzanowski}\ \emph {et~al.}(1978)\citenamefont
  {Budzanowski}, \citenamefont {Baur}, \citenamefont {Alderliesten},
  \citenamefont {Bojowald}, \citenamefont {Mayer-B{\"o}ricke}, \citenamefont
  {Oelert}, \citenamefont {Turek}, \citenamefont {R{\"o}sel},\ and\
  \citenamefont {Trautmann}}]{Budzanowski1978}%
  \BibitemOpen
  \bibfield  {author} {\bibinfo {author} {\bibfnamefont {A.}~\bibnamefont
  {Budzanowski}}, \bibinfo {author} {\bibfnamefont {G.}~\bibnamefont {Baur}},
  \bibinfo {author} {\bibfnamefont {C.}~\bibnamefont {Alderliesten}}, \bibinfo
  {author} {\bibfnamefont {J.}~\bibnamefont {Bojowald}}, \bibinfo {author}
  {\bibfnamefont {C.}~\bibnamefont {Mayer-B{\"o}ricke}}, \bibinfo {author}
  {\bibfnamefont {W.}~\bibnamefont {Oelert}}, \bibinfo {author} {\bibfnamefont
  {P.}~\bibnamefont {Turek}}, \bibinfo {author} {\bibfnamefont
  {F.}~\bibnamefont {R{\"o}sel}},\ and\ \bibinfo {author} {\bibfnamefont
  {D.}~\bibnamefont {Trautmann}},\ }\bibfield  {title} {\bibinfo {title}
  {{Observation of the $\alpha$-particle breakup process at
  $E_{\alpha,\mathrm{lab}}=172.5$ MeV}},\ }\href
  {https://doi.org/10.1103/PhysRevLett.41.635} {\bibfield  {journal} {\bibinfo
  {journal} {Phys. Rev. Lett.}\ }\textbf {\bibinfo {volume} {41}},\ \bibinfo
  {pages} {635} (\bibinfo {year} {1978})}\BibitemShut {NoStop}%
\bibitem [{\citenamefont {Baur}\ \emph {et~al.}(1980)\citenamefont {Baur},
  \citenamefont {Shyam}, \citenamefont {R{\"o}sel},\ and\ \citenamefont
  {Trautmann}}]{Baur1980}%
  \BibitemOpen
  \bibfield  {author} {\bibinfo {author} {\bibfnamefont {G.}~\bibnamefont
  {Baur}}, \bibinfo {author} {\bibfnamefont {R.}~\bibnamefont {Shyam}},
  \bibinfo {author} {\bibfnamefont {F.}~\bibnamefont {R{\"o}sel}},\ and\
  \bibinfo {author} {\bibfnamefont {D.}~\bibnamefont {Trautmann}},\ }\bibfield
  {title} {\bibinfo {title} {{Importance of the breakup mechanism for composite
  particle scattering}},\ }\href {https://doi.org/10.1103/PhysRevC.21.2668}
  {\bibfield  {journal} {\bibinfo  {journal} {Phys. Rev. C}\ }\textbf {\bibinfo
  {volume} {21}},\ \bibinfo {pages} {2668} (\bibinfo {year}
  {1980})}\BibitemShut {NoStop}%
\bibitem [{\citenamefont {Shyam}\ \emph {et~al.}(1980)\citenamefont {Shyam},
  \citenamefont {Baur}, \citenamefont {R{\"o}sel},\ and\ \citenamefont
  {Trautmann}}]{Shyam1980}%
  \BibitemOpen
  \bibfield  {author} {\bibinfo {author} {\bibfnamefont {R.}~\bibnamefont
  {Shyam}}, \bibinfo {author} {\bibfnamefont {G.}~\bibnamefont {Baur}},
  \bibinfo {author} {\bibfnamefont {F.}~\bibnamefont {R{\"o}sel}},\ and\
  \bibinfo {author} {\bibfnamefont {D.}~\bibnamefont {Trautmann}},\ }\bibfield
  {title} {\bibinfo {title} {{Elastic and inelastic breakup of the ${}^{3}$He
  particle}},\ }\href {https://doi.org/10.1103/PhysRevC.22.1401} {\bibfield
  {journal} {\bibinfo  {journal} {Phys. Rev. C}\ }\textbf {\bibinfo {volume}
  {22}},\ \bibinfo {pages} {1401} (\bibinfo {year} {1980})}\BibitemShut
  {NoStop}%
\bibitem [{\citenamefont {Baur}\ \emph {et~al.}(1983)\citenamefont {Baur},
  \citenamefont {Shyam}, \citenamefont {R{\"o}sel},\ and\ \citenamefont
  {Trautmann}}]{Baur1983}%
  \BibitemOpen
  \bibfield  {author} {\bibinfo {author} {\bibfnamefont {G.}~\bibnamefont
  {Baur}}, \bibinfo {author} {\bibfnamefont {R.}~\bibnamefont {Shyam}},
  \bibinfo {author} {\bibfnamefont {F.}~\bibnamefont {R{\"o}sel}},\ and\
  \bibinfo {author} {\bibfnamefont {D.}~\bibnamefont {Trautmann}},\ }\bibfield
  {title} {\bibinfo {title} {{Calculation of proton-neutron coincidence cross
  sections in 56 MeV deuteron-induced breakup reactions by post form
  distorted-wave Born approximation}},\ }\href
  {https://doi.org/10.1103/PhysRevC.28.946} {\bibfield  {journal} {\bibinfo
  {journal} {Phys. Rev. C}\ }\textbf {\bibinfo {volume} {28}},\ \bibinfo
  {pages} {946} (\bibinfo {year} {1983})}\BibitemShut {NoStop}%
\bibitem [{\citenamefont {Ichimura}\ \emph {et~al.}(1985)\citenamefont
  {Ichimura}, \citenamefont {Austern},\ and\ \citenamefont
  {Vincent}}]{Ichimura1985}%
  \BibitemOpen
  \bibfield  {author} {\bibinfo {author} {\bibfnamefont {M.}~\bibnamefont
  {Ichimura}}, \bibinfo {author} {\bibfnamefont {N.}~\bibnamefont {Austern}},\
  and\ \bibinfo {author} {\bibfnamefont {C.~M.}\ \bibnamefont {Vincent}},\
  }\bibfield  {title} {\bibinfo {title} {{Equivalence of post and prior sum
  rules for inclusive breakup reactions}},\ }\href
  {https://doi.org/10.1103/PhysRevC.32.431} {\bibfield  {journal} {\bibinfo
  {journal} {Phys. Rev. C}\ }\textbf {\bibinfo {volume} {32}},\ \bibinfo
  {pages} {431} (\bibinfo {year} {1985})}\BibitemShut {NoStop}%
\bibitem [{\citenamefont {Austern}\ and\ \citenamefont
  {Vincent}(1981)}]{Austern1981}%
  \BibitemOpen
  \bibfield  {author} {\bibinfo {author} {\bibfnamefont {N.}~\bibnamefont
  {Austern}}\ and\ \bibinfo {author} {\bibfnamefont {C.~M.}\ \bibnamefont
  {Vincent}},\ }\bibfield  {title} {\bibinfo {title} {{Inclusive breakup
  reactions}},\ }\href {https://doi.org/10.1103/PhysRevC.23.1847} {\bibfield
  {journal} {\bibinfo  {journal} {Phys. Rev. C}\ }\textbf {\bibinfo {volume}
  {23}},\ \bibinfo {pages} {1847} (\bibinfo {year} {1981})}\BibitemShut
  {NoStop}%
\bibitem [{\citenamefont {Udagawa}\ and\ \citenamefont
  {Tamura}(1981)}]{Udagawa1981}%
  \BibitemOpen
  \bibfield  {author} {\bibinfo {author} {\bibfnamefont {T.}~\bibnamefont
  {Udagawa}}\ and\ \bibinfo {author} {\bibfnamefont {T.}~\bibnamefont
  {Tamura}},\ }\bibfield  {title} {\bibinfo {title} {{Derivation of
  breakup-fusion cross sections from the optical theorem}},\ }\href
  {https://doi.org/10.1103/PhysRevC.24.1348} {\bibfield  {journal} {\bibinfo
  {journal} {Phys. Rev. C}\ }\textbf {\bibinfo {volume} {24}},\ \bibinfo
  {pages} {1348} (\bibinfo {year} {1981})}\BibitemShut {NoStop}%
\bibitem [{\citenamefont {Li}\ \emph {et~al.}(1984)\citenamefont {Li},
  \citenamefont {Udagawa},\ and\ \citenamefont {Tamura}}]{Li1984}%
  \BibitemOpen
  \bibfield  {author} {\bibinfo {author} {\bibfnamefont {X.-H.}\ \bibnamefont
  {Li}}, \bibinfo {author} {\bibfnamefont {T.}~\bibnamefont {Udagawa}},\ and\
  \bibinfo {author} {\bibfnamefont {T.}~\bibnamefont {Tamura}},\ }\bibfield
  {title} {\bibinfo {title} {{Assessment of approximations made in
  breakup-fusion descriptions}},\ }\href
  {https://doi.org/10.1103/PhysRevC.30.1895} {\bibfield  {journal} {\bibinfo
  {journal} {Phys. Rev. C}\ }\textbf {\bibinfo {volume} {30}},\ \bibinfo
  {pages} {1895} (\bibinfo {year} {1984})}\BibitemShut {NoStop}%
\bibitem [{\citenamefont {Hussein}\ and\ \citenamefont
  {McVoy}(1985)}]{Hussein1985}%
  \BibitemOpen
  \bibfield  {author} {\bibinfo {author} {\bibfnamefont {M.~S.}\ \bibnamefont
  {Hussein}}\ and\ \bibinfo {author} {\bibfnamefont {K.~W.}\ \bibnamefont
  {McVoy}},\ }\bibfield  {title} {\bibinfo {title} {{Inclusive projectile
  fragmentation in the spectator model}},\ }\href
  {https://doi.org/10.1016/0375-9474(85)90364-1} {\bibfield  {journal}
  {\bibinfo  {journal} {Nucl. Phys. A}\ }\textbf {\bibinfo {volume} {445}},\
  \bibinfo {pages} {124} (\bibinfo {year} {1985})}\BibitemShut {NoStop}%
\bibitem [{\citenamefont {G{\'o}mez-Ramos}\ \emph {et~al.}(2021)\citenamefont
  {G{\'o}mez-Ramos}, \citenamefont {G{\'o}mez-Camacho}, \citenamefont {Lei},\
  and\ \citenamefont {Moro}}]{GomezRamos2021}%
  \BibitemOpen
  \bibfield  {author} {\bibinfo {author} {\bibfnamefont {M.}~\bibnamefont
  {G{\'o}mez-Ramos}}, \bibinfo {author} {\bibfnamefont {J.}~\bibnamefont
  {G{\'o}mez-Camacho}}, \bibinfo {author} {\bibfnamefont {J.}~\bibnamefont
  {Lei}},\ and\ \bibinfo {author} {\bibfnamefont {A.~M.}\ \bibnamefont
  {Moro}},\ }\bibfield  {title} {\bibinfo {title} {{The Hussein--McVoy formula
  for inclusive breakup revisited. A Tribute to Mahir Hussein}},\ }\href
  {https://doi.org/10.1140/epja/s10050-021-00376-0} {\bibfield  {journal}
  {\bibinfo  {journal} {Eur. Phys. J. A}\ }\textbf {\bibinfo {volume} {57}},\
  \bibinfo {pages} {57} (\bibinfo {year} {2021})},\ \Eprint
  {https://arxiv.org/abs/2101.09497} {arXiv:2101.09497 [nucl-th]} \BibitemShut
  {NoStop}%
\bibitem [{\citenamefont {Potel}\ \emph {et~al.}(2015)\citenamefont {Potel},
  \citenamefont {Nunes},\ and\ \citenamefont {Thompson}}]{Potel2015}%
  \BibitemOpen
  \bibfield  {author} {\bibinfo {author} {\bibfnamefont {G.}~\bibnamefont
  {Potel}}, \bibinfo {author} {\bibfnamefont {F.~M.}\ \bibnamefont {Nunes}},\
  and\ \bibinfo {author} {\bibfnamefont {I.~J.}\ \bibnamefont {Thompson}},\
  }\bibfield  {title} {\bibinfo {title} {{Inclusive deuteron-induced reactions
  and final neutron states}},\ }in\ \href@noop {} {\emph {\bibinfo {booktitle}
  {{14th International Conference on Nuclear Reaction Mechanisms}}}}\ (\bibinfo
  {year} {2015})\ pp.\ \bibinfo {pages} {155--162},\ \bibinfo {note}
  {{Proceedings contribution}},\ \Eprint {https://arxiv.org/abs/1510.02727}
  {arXiv:1510.02727 [nucl-th]} \BibitemShut {NoStop}%
\bibitem [{\citenamefont {Potel}\ \emph {et~al.}(2017)\citenamefont {Potel},
  \citenamefont {Perdikakis}, \citenamefont {Carlson}, \citenamefont
  {Atkinson}, \citenamefont {Dickhoff}, \citenamefont {Escher}, \citenamefont
  {Hussein}, \citenamefont {Lei}, \citenamefont {Li}, \citenamefont
  {Macchiavelli}, \citenamefont {Moro}, \citenamefont {Nunes}, \citenamefont
  {Pain},\ and\ \citenamefont {Rotureau}}]{Potel2017}%
  \BibitemOpen
  \bibfield  {author} {\bibinfo {author} {\bibfnamefont {G.}~\bibnamefont
  {Potel}}, \bibinfo {author} {\bibfnamefont {G.}~\bibnamefont {Perdikakis}},
  \bibinfo {author} {\bibfnamefont {B.~V.}\ \bibnamefont {Carlson}}, \bibinfo
  {author} {\bibfnamefont {M.~C.}\ \bibnamefont {Atkinson}}, \bibinfo {author}
  {\bibfnamefont {W.~H.}\ \bibnamefont {Dickhoff}}, \bibinfo {author}
  {\bibfnamefont {J.~E.}\ \bibnamefont {Escher}}, \bibinfo {author}
  {\bibfnamefont {M.~S.}\ \bibnamefont {Hussein}}, \bibinfo {author}
  {\bibfnamefont {J.}~\bibnamefont {Lei}}, \bibinfo {author} {\bibfnamefont
  {W.}~\bibnamefont {Li}}, \bibinfo {author} {\bibfnamefont {A.~O.}\
  \bibnamefont {Macchiavelli}}, \bibinfo {author} {\bibfnamefont {A.~M.}\
  \bibnamefont {Moro}}, \bibinfo {author} {\bibfnamefont {F.~M.}\ \bibnamefont
  {Nunes}}, \bibinfo {author} {\bibfnamefont {S.~D.}\ \bibnamefont {Pain}},\
  and\ \bibinfo {author} {\bibfnamefont {J.}~\bibnamefont {Rotureau}},\
  }\bibfield  {title} {\bibinfo {title} {{Toward a complete theory for
  predicting inclusive deuteron breakup away from stability}},\ }\href
  {https://doi.org/10.1140/epja/i2017-12371-9} {\bibfield  {journal} {\bibinfo
  {journal} {Eur. Phys. J. A}\ }\textbf {\bibinfo {volume} {53}},\ \bibinfo
  {pages} {178} (\bibinfo {year} {2017})},\ \Eprint
  {https://arxiv.org/abs/1705.07782} {arXiv:1705.07782 [nucl-th]} \BibitemShut
  {NoStop}%
\bibitem [{\citenamefont {Lei}\ and\ \citenamefont
  {Moro}(2015{\natexlab{a}})}]{Lei2015}%
  \BibitemOpen
  \bibfield  {author} {\bibinfo {author} {\bibfnamefont {J.}~\bibnamefont
  {Lei}}\ and\ \bibinfo {author} {\bibfnamefont {A.~M.}\ \bibnamefont {Moro}},\
  }\bibfield  {title} {\bibinfo {title} {{Reexamining closed-form formulae for
  inclusive breakup: Application to deuteron- and ${}^6$Li-induced
  reactions}},\ }\href {https://doi.org/10.1103/PhysRevC.92.044616} {\bibfield
  {journal} {\bibinfo  {journal} {Phys. Rev. C}\ }\textbf {\bibinfo {volume}
  {92}},\ \bibinfo {pages} {044616} (\bibinfo {year} {2015}{\natexlab{a}})},\
  \Eprint {https://arxiv.org/abs/1510.02602} {arXiv:1510.02602 [nucl-th]}
  \BibitemShut {NoStop}%
\bibitem [{\citenamefont {Lei}\ and\ \citenamefont
  {Moro}(2015{\natexlab{b}})}]{Lei2015b}%
  \BibitemOpen
  \bibfield  {author} {\bibinfo {author} {\bibfnamefont {J.}~\bibnamefont
  {Lei}}\ and\ \bibinfo {author} {\bibfnamefont {A.~M.}\ \bibnamefont {Moro}},\
  }\bibfield  {title} {\bibinfo {title} {{Numerical assessment of post-prior
  equivalence for inclusive breakup reactions}},\ }\href
  {https://doi.org/10.1103/PhysRevC.92.061602} {\bibfield  {journal} {\bibinfo
  {journal} {Phys. Rev. C}\ }\textbf {\bibinfo {volume} {92}},\ \bibinfo
  {pages} {061602} (\bibinfo {year} {2015}{\natexlab{b}})},\ \Eprint
  {https://arxiv.org/abs/1511.03214} {arXiv:1511.03214 [nucl-th]} \BibitemShut
  {NoStop}%
\bibitem [{\citenamefont {Lei}\ and\ \citenamefont {Moro}(2017)}]{Lei2017}%
  \BibitemOpen
  \bibfield  {author} {\bibinfo {author} {\bibfnamefont {J.}~\bibnamefont
  {Lei}}\ and\ \bibinfo {author} {\bibfnamefont {A.~M.}\ \bibnamefont {Moro}},\
  }\bibfield  {title} {\bibinfo {title} {{Comprehensive analysis of large
  $\alpha$ yields observed in ${}^{6}$Li-induced reactions}},\ }\href
  {https://doi.org/10.1103/PhysRevC.95.044605} {\bibfield  {journal} {\bibinfo
  {journal} {Phys. Rev. C}\ }\textbf {\bibinfo {volume} {95}},\ \bibinfo
  {pages} {044605} (\bibinfo {year} {2017})},\ \Eprint
  {https://arxiv.org/abs/1701.00547} {arXiv:1701.00547 [nucl-th]} \BibitemShut
  {NoStop}%
\bibitem [{\citenamefont {Carlson}\ \emph {et~al.}(2016)\citenamefont
  {Carlson}, \citenamefont {Capote},\ and\ \citenamefont {Sin}}]{Carlson2016}%
  \BibitemOpen
  \bibfield  {author} {\bibinfo {author} {\bibfnamefont {B.~V.}\ \bibnamefont
  {Carlson}}, \bibinfo {author} {\bibfnamefont {R.}~\bibnamefont {Capote}},\
  and\ \bibinfo {author} {\bibfnamefont {M.}~\bibnamefont {Sin}},\ }\bibfield
  {title} {\bibinfo {title} {{Inclusive Proton Emission Spectra from Deuteron
  Breakup Reactions}},\ }\href {https://doi.org/10.1007/s00601-016-1054-8}
  {\bibfield  {journal} {\bibinfo  {journal} {Few-Body Syst.}\ }\textbf
  {\bibinfo {volume} {57}},\ \bibinfo {pages} {307} (\bibinfo {year}
  {2016})}\BibitemShut {NoStop}%
\bibitem [{\citenamefont {Lei}\ and\ \citenamefont
  {Moro}(2019{\natexlab{a}})}]{Lei2019}%
  \BibitemOpen
  \bibfield  {author} {\bibinfo {author} {\bibfnamefont {J.}~\bibnamefont
  {Lei}}\ and\ \bibinfo {author} {\bibfnamefont {A.~M.}\ \bibnamefont {Moro}},\
  }\bibfield  {title} {\bibinfo {title} {{Puzzle of Complete Fusion Suppression
  in Weakly Bound Nuclei: A Trojan Horse Effect?}},\ }\href
  {https://doi.org/10.1103/PhysRevLett.122.042503} {\bibfield  {journal}
  {\bibinfo  {journal} {Phys. Rev. Lett.}\ }\textbf {\bibinfo {volume} {122}},\
  \bibinfo {pages} {042503} (\bibinfo {year} {2019}{\natexlab{a}})},\ \Eprint
  {https://arxiv.org/abs/1812.11248} {arXiv:1812.11248 [nucl-th]} \BibitemShut
  {NoStop}%
\bibitem [{\citenamefont {Lei}\ and\ \citenamefont
  {Moro}(2019{\natexlab{b}})}]{Lei2019b}%
  \BibitemOpen
  \bibfield  {author} {\bibinfo {author} {\bibfnamefont {J.}~\bibnamefont
  {Lei}}\ and\ \bibinfo {author} {\bibfnamefont {A.~M.}\ \bibnamefont {Moro}},\
  }\bibfield  {title} {\bibinfo {title} {{Unraveling the reaction mechanisms
  leading to partial fusion of weakly bound nuclei}},\ }\href
  {https://doi.org/10.1103/PhysRevLett.123.232501} {\bibfield  {journal}
  {\bibinfo  {journal} {Phys. Rev. Lett.}\ }\textbf {\bibinfo {volume} {123}},\
  \bibinfo {pages} {232501} (\bibinfo {year} {2019}{\natexlab{b}})},\ \Eprint
  {https://arxiv.org/abs/1910.06625} {arXiv:1910.06625 [nucl-th]} \BibitemShut
  {NoStop}%
\bibitem [{\citenamefont {Moro}\ and\ \citenamefont
  {Lei}(2016)}]{Lei2016review}%
  \BibitemOpen
  \bibfield  {author} {\bibinfo {author} {\bibfnamefont {A.~M.}\ \bibnamefont
  {Moro}}\ and\ \bibinfo {author} {\bibfnamefont {J.}~\bibnamefont {Lei}},\
  }\bibfield  {title} {\bibinfo {title} {{Recent Advances in Nuclear Reaction
  Theories for Weakly Bound Nuclei: Reexamining the Problem of Inclusive
  Breakup}},\ }\href {https://doi.org/10.1007/s00601-016-1085-1} {\bibfield
  {journal} {\bibinfo  {journal} {Few-Body Syst.}\ }\textbf {\bibinfo {volume}
  {57}},\ \bibinfo {pages} {319} (\bibinfo {year} {2016})}\BibitemShut
  {NoStop}%
\bibitem [{\citenamefont {Lei}(2025)}]{Lei2025post}%
  \BibitemOpen
  \bibfield  {author} {\bibinfo {author} {\bibfnamefont {J.}~\bibnamefont
  {Lei}},\ }\bibfield  {title} {\bibinfo {title} {{Numerical assessment of
  convergence in the post-form Ichimura-Austern-Vincent model}},\ }\href
  {https://doi.org/10.1103/l53j-32cp} {\bibfield  {journal} {\bibinfo
  {journal} {Phys. Rev. C}\ }\textbf {\bibinfo {volume} {112}},\ \bibinfo
  {pages} {014609} (\bibinfo {year} {2025})}\BibitemShut {NoStop}%
\bibitem [{\citenamefont {Neoh}\ \emph {et~al.}(2016)\citenamefont {Neoh},
  \citenamefont {Yoshida}, \citenamefont {Minomo},\ and\ \citenamefont
  {Ogata}}]{Neoh2016}%
  \BibitemOpen
  \bibfield  {author} {\bibinfo {author} {\bibfnamefont {Y.~S.}\ \bibnamefont
  {Neoh}}, \bibinfo {author} {\bibfnamefont {K.}~\bibnamefont {Yoshida}},
  \bibinfo {author} {\bibfnamefont {K.}~\bibnamefont {Minomo}},\ and\ \bibinfo
  {author} {\bibfnamefont {K.}~\bibnamefont {Ogata}},\ }\bibfield  {title}
  {\bibinfo {title} {{Microscopic effective reaction theory for
  deuteron-induced reactions}},\ }\href
  {https://doi.org/10.1103/PhysRevC.94.044619} {\bibfield  {journal} {\bibinfo
  {journal} {Phys. Rev. C}\ }\textbf {\bibinfo {volume} {94}},\ \bibinfo
  {pages} {044619} (\bibinfo {year} {2016})}\BibitemShut {NoStop}%
\bibitem [{\citenamefont {Deltuva}(2025)}]{Deltuva2025}%
  \BibitemOpen
  \bibfield  {author} {\bibinfo {author} {\bibfnamefont {A.}~\bibnamefont
  {Deltuva}},\ }\bibfield  {title} {\bibinfo {title} {{Faddeev-type calculation
  of nonelastic breakup in deuteron-nucleus scattering}},\ }\href
  {https://doi.org/10.1016/j.physletb.2025.139825} {\bibfield  {journal}
  {\bibinfo  {journal} {Phys. Lett. B}\ }\textbf {\bibinfo {volume} {868}},\
  \bibinfo {pages} {139825} (\bibinfo {year} {2025})}\BibitemShut {NoStop}%
\bibitem [{\citenamefont {Canto}\ \emph {et~al.}(2006)\citenamefont {Canto},
  \citenamefont {Gomes}, \citenamefont {Donangelo},\ and\ \citenamefont
  {Hussein}}]{Canto2006}%
  \BibitemOpen
  \bibfield  {author} {\bibinfo {author} {\bibfnamefont {L.~F.}\ \bibnamefont
  {Canto}}, \bibinfo {author} {\bibfnamefont {P.~R.~S.}\ \bibnamefont {Gomes}},
  \bibinfo {author} {\bibfnamefont {R.}~\bibnamefont {Donangelo}},\ and\
  \bibinfo {author} {\bibfnamefont {M.~S.}\ \bibnamefont {Hussein}},\
  }\bibfield  {title} {\bibinfo {title} {{Fusion and breakup of weakly bound
  nuclei}},\ }\href {https://doi.org/10.1016/j.physrep.2005.10.006} {\bibfield
  {journal} {\bibinfo  {journal} {Phys. Rep.}\ }\textbf {\bibinfo {volume}
  {424}},\ \bibinfo {pages} {1} (\bibinfo {year} {2006})}\BibitemShut {NoStop}%
\bibitem [{\citenamefont {Keeley}\ \emph {et~al.}(2007)\citenamefont {Keeley},
  \citenamefont {Raabe}, \citenamefont {Alamanos},\ and\ \citenamefont
  {Sida}}]{Keeley2007}%
  \BibitemOpen
  \bibfield  {author} {\bibinfo {author} {\bibfnamefont {N.}~\bibnamefont
  {Keeley}}, \bibinfo {author} {\bibfnamefont {R.}~\bibnamefont {Raabe}},
  \bibinfo {author} {\bibfnamefont {N.}~\bibnamefont {Alamanos}},\ and\
  \bibinfo {author} {\bibfnamefont {J.~L.}\ \bibnamefont {Sida}},\ }\bibfield
  {title} {\bibinfo {title} {{Fusion and direct reactions of halo nuclei at
  energies around the Coulomb barrier}},\ }\href
  {https://doi.org/10.1016/j.ppnp.2007.02.002} {\bibfield  {journal} {\bibinfo
  {journal} {Prog. Part. Nucl. Phys.}\ }\textbf {\bibinfo {volume} {59}},\
  \bibinfo {pages} {579} (\bibinfo {year} {2007})}\BibitemShut {NoStop}%
\bibitem [{\citenamefont {Lehman}\ and\ \citenamefont
  {Rajan}(1982)}]{Lehman1982}%
  \BibitemOpen
  \bibfield  {author} {\bibinfo {author} {\bibfnamefont {D.~R.}\ \bibnamefont
  {Lehman}}\ and\ \bibinfo {author} {\bibfnamefont {M.}~\bibnamefont {Rajan}},\
  }\bibfield  {title} {\bibinfo {title} {{Alpha-deuteron structure of
  ${}^{6}$Li as predicted by three-body models}},\ }\href
  {https://doi.org/10.1103/PhysRevC.25.2743} {\bibfield  {journal} {\bibinfo
  {journal} {Phys. Rev. C}\ }\textbf {\bibinfo {volume} {25}},\ \bibinfo
  {pages} {2743} (\bibinfo {year} {1982})}\BibitemShut {NoStop}%
\bibitem [{\citenamefont {Villanueva}\ \emph {et~al.}(2024)\citenamefont
  {Villanueva}, \citenamefont {Moro}, \citenamefont {Casal},\ and\
  \citenamefont {Lei}}]{Villanueva2024}%
  \BibitemOpen
  \bibfield  {author} {\bibinfo {author} {\bibfnamefont {G.}~\bibnamefont
  {Villanueva}}, \bibinfo {author} {\bibfnamefont {A.~M.}\ \bibnamefont
  {Moro}}, \bibinfo {author} {\bibfnamefont {J.}~\bibnamefont {Casal}},\ and\
  \bibinfo {author} {\bibfnamefont {J.}~\bibnamefont {Lei}},\ }\bibfield
  {title} {\bibinfo {title} {{Neutron-transfer induced breakup of the Borromean
  nucleus ${}^9$Be}},\ }\href {https://doi.org/10.1016/j.physletb.2024.138766}
  {\bibfield  {journal} {\bibinfo  {journal} {Phys. Lett. B}\ }\textbf
  {\bibinfo {volume} {855}},\ \bibinfo {pages} {138766} (\bibinfo {year}
  {2024})}\BibitemShut {NoStop}%
\bibitem [{\citenamefont {Descouvemont}\ \emph {et~al.}(2015)\citenamefont
  {Descouvemont}, \citenamefont {Druet}, \citenamefont {Canto},\ and\
  \citenamefont {Hussein}}]{Descouvemont2015}%
  \BibitemOpen
  \bibfield  {author} {\bibinfo {author} {\bibfnamefont {P.}~\bibnamefont
  {Descouvemont}}, \bibinfo {author} {\bibfnamefont {T.}~\bibnamefont {Druet}},
  \bibinfo {author} {\bibfnamefont {L.~F.}\ \bibnamefont {Canto}},\ and\
  \bibinfo {author} {\bibfnamefont {M.~S.}\ \bibnamefont {Hussein}},\
  }\bibfield  {title} {\bibinfo {title} {{Low-energy
  ${}^9\mathrm{Be}+{}^{208}\mathrm{Pb}$ scattering, breakup, and fusion within
  a four-body model}},\ }\href {https://doi.org/10.1103/PhysRevC.91.024606}
  {\bibfield  {journal} {\bibinfo  {journal} {Phys. Rev. C}\ }\textbf {\bibinfo
  {volume} {91}},\ \bibinfo {pages} {024606} (\bibinfo {year}
  {2015})}\BibitemShut {NoStop}%
\bibitem [{\citenamefont {Carlson}\ \emph {et~al.}(2017)\citenamefont
  {Carlson}, \citenamefont {Frederico},\ and\ \citenamefont
  {Hussein}}]{CFH2017}%
  \BibitemOpen
  \bibfield  {author} {\bibinfo {author} {\bibfnamefont {B.~V.}\ \bibnamefont
  {Carlson}}, \bibinfo {author} {\bibfnamefont {T.}~\bibnamefont {Frederico}},\
  and\ \bibinfo {author} {\bibfnamefont {M.~S.}\ \bibnamefont {Hussein}},\
  }\bibfield  {title} {\bibinfo {title} {{Inclusive breakup of three-fragment
  weakly bound nuclei}},\ }\href
  {https://doi.org/10.1016/j.physletb.2017.01.048} {\bibfield  {journal}
  {\bibinfo  {journal} {Phys. Lett. B}\ }\textbf {\bibinfo {volume} {767}},\
  \bibinfo {pages} {53} (\bibinfo {year} {2017})},\ \Eprint
  {https://arxiv.org/abs/1611.09741} {arXiv:1611.09741 [nucl-th]} \BibitemShut
  {NoStop}%
\bibitem [{\citenamefont {Hussein}\ \emph {et~al.}(2017)\citenamefont
  {Hussein}, \citenamefont {Souza}, \citenamefont {Chimanski}, \citenamefont
  {Carlson},\ and\ \citenamefont {Frederico}}]{Hussein2017}%
  \BibitemOpen
  \bibfield  {author} {\bibinfo {author} {\bibfnamefont {M.~S.}\ \bibnamefont
  {Hussein}}, \bibinfo {author} {\bibfnamefont {L.~A.}\ \bibnamefont {Souza}},
  \bibinfo {author} {\bibfnamefont {E.~V.}\ \bibnamefont {Chimanski}}, \bibinfo
  {author} {\bibfnamefont {B.~V.}\ \bibnamefont {Carlson}},\ and\ \bibinfo
  {author} {\bibfnamefont {T.}~\bibnamefont {Frederico}},\ }\bibfield  {title}
  {\bibinfo {title} {{Inclusive Breakup Theory of Three-Body Halos}},\ }\href
  {https://doi.org/10.1051/epjconf/201716300024} {\bibfield  {journal}
  {\bibinfo  {journal} {EPJ Web Conf.}\ }\textbf {\bibinfo {volume} {163}},\
  \bibinfo {pages} {00024} (\bibinfo {year} {2017})}\BibitemShut {NoStop}%
\bibitem [{\citenamefont {Hussein}\ \emph {et~al.}(2020)\citenamefont
  {Hussein}, \citenamefont {Bertulani}, \citenamefont {Carlson},\ and\
  \citenamefont {Frederico}}]{Hussein2020}%
  \BibitemOpen
  \bibfield  {author} {\bibinfo {author} {\bibfnamefont {M.~S.}\ \bibnamefont
  {Hussein}}, \bibinfo {author} {\bibfnamefont {C.~A.}\ \bibnamefont
  {Bertulani}}, \bibinfo {author} {\bibfnamefont {B.~V.}\ \bibnamefont
  {Carlson}},\ and\ \bibinfo {author} {\bibfnamefont {T.}~\bibnamefont
  {Frederico}},\ }\bibfield  {title} {\bibinfo {title} {{Inclusive breakup
  reaction of a two-cluster projectile on a two-fragment target: A genuine
  four-body problem}},\ }in\ \href
  {https://doi.org/10.1007/978-3-030-32357-8_35} {\emph {\bibinfo {booktitle}
  {{Recent Progress in Few-Body Physics}}}},\ \bibinfo {series} {Springer
  Proceedings in Physics}, Vol.\ \bibinfo {volume} {238}\ (\bibinfo
  {publisher} {Springer International Publishing},\ \bibinfo {year} {2020})\
  pp.\ \bibinfo {pages} {201--208},\ \Eprint {https://arxiv.org/abs/1811.01231}
  {arXiv:1811.01231 [nucl-th]} \BibitemShut {NoStop}%
\bibitem [{\citenamefont {Souza}\ \emph {et~al.}(2021)\citenamefont {Souza},
  \citenamefont {Chimanski},\ and\ \citenamefont {Carlson}}]{Souza2021}%
  \BibitemOpen
  \bibfield  {author} {\bibinfo {author} {\bibfnamefont {L.~A.}\ \bibnamefont
  {Souza}}, \bibinfo {author} {\bibfnamefont {E.~V.}\ \bibnamefont
  {Chimanski}},\ and\ \bibinfo {author} {\bibfnamefont {B.~V.}\ \bibnamefont
  {Carlson}},\ }\bibfield  {title} {\bibinfo {title} {{Inclusive breakup cross
  sections in reactions induced by the nuclides ${}^6$He and ${}^{6,7}$Li in
  the two-body cluster model}},\ }\href
  {https://doi.org/10.1103/PhysRevC.104.034623} {\bibfield  {journal} {\bibinfo
   {journal} {Phys. Rev. C}\ }\textbf {\bibinfo {volume} {104}},\ \bibinfo
  {pages} {034623} (\bibinfo {year} {2021})},\ \Eprint
  {https://arxiv.org/abs/2012.14805} {arXiv:2012.14805 [nucl-th]} \BibitemShut
  {NoStop}%
\bibitem [{\citenamefont {Lei}(2026)}]{Lei2026nsp}%
  \BibitemOpen
  \bibfield  {author} {\bibinfo {author} {\bibfnamefont {J.}~\bibnamefont
  {Lei}},\ }\bibfield  {title} {\bibinfo {title} {{Inclusive breakup reactions
  with nonspectator fragments: Generalization of the Ichimura-Austern-Vincent
  sum rules}},\ }\href {https://doi.org/10.1103/rbph-64rs} {\bibfield
  {journal} {\bibinfo  {journal} {Phys. Rev. C}\ }\textbf {\bibinfo {volume}
  {114}},\ \bibinfo {pages} {014632} (\bibinfo {year} {2026})},\ \Eprint
  {https://arxiv.org/abs/2604.11226} {arXiv:2604.11226 [nucl-th]} \BibitemShut
  {NoStop}%
\bibitem [{\citenamefont {Lei}\ and\ \citenamefont
  {Moro}(2018)}]{Lei2018prior}%
  \BibitemOpen
  \bibfield  {author} {\bibinfo {author} {\bibfnamefont {J.}~\bibnamefont
  {Lei}}\ and\ \bibinfo {author} {\bibfnamefont {A.~M.}\ \bibnamefont {Moro}},\
  }\bibfield  {title} {\bibinfo {title} {{Post-prior equivalence for transfer
  reactions with complex potentials}},\ }\href
  {https://doi.org/10.1103/PhysRevC.97.011601} {\bibfield  {journal} {\bibinfo
  {journal} {Phys. Rev. C}\ }\textbf {\bibinfo {volume} {97}},\ \bibinfo
  {pages} {011601(R)} (\bibinfo {year} {2018})}\BibitemShut {NoStop}%
\bibitem [{\citenamefont {Hiyama}\ \emph {et~al.}(2003)\citenamefont {Hiyama},
  \citenamefont {Kino},\ and\ \citenamefont {Kamimura}}]{Hiyama2003}%
  \BibitemOpen
  \bibfield  {author} {\bibinfo {author} {\bibfnamefont {E.}~\bibnamefont
  {Hiyama}}, \bibinfo {author} {\bibfnamefont {Y.}~\bibnamefont {Kino}},\ and\
  \bibinfo {author} {\bibfnamefont {M.}~\bibnamefont {Kamimura}},\ }\bibfield
  {title} {\bibinfo {title} {{Gaussian expansion method for few-body
  systems}},\ }\href {https://doi.org/10.1016/S0146-6410(03)90015-9} {\bibfield
   {journal} {\bibinfo  {journal} {Prog. Part. Nucl. Phys.}\ }\textbf {\bibinfo
  {volume} {51}},\ \bibinfo {pages} {223} (\bibinfo {year} {2003})}\BibitemShut
  {NoStop}%
\bibitem [{\citenamefont {Descouvemont}\ \emph {et~al.}(2003)\citenamefont
  {Descouvemont}, \citenamefont {Daniel},\ and\ \citenamefont
  {Baye}}]{Descouvemont2003}%
  \BibitemOpen
  \bibfield  {author} {\bibinfo {author} {\bibfnamefont {P.}~\bibnamefont
  {Descouvemont}}, \bibinfo {author} {\bibfnamefont {C.}~\bibnamefont
  {Daniel}},\ and\ \bibinfo {author} {\bibfnamefont {D.}~\bibnamefont {Baye}},\
  }\bibfield  {title} {\bibinfo {title} {{Three-body systems with Lagrange-mesh
  techniques in hyperspherical coordinates}},\ }\href
  {https://doi.org/10.1103/PhysRevC.67.044309} {\bibfield  {journal} {\bibinfo
  {journal} {Phys. Rev. C}\ }\textbf {\bibinfo {volume} {67}},\ \bibinfo
  {pages} {044309} (\bibinfo {year} {2003})}\BibitemShut {NoStop}%
\bibitem [{\citenamefont {Tursunov}\ \emph {et~al.}(2006)\citenamefont
  {Tursunov}, \citenamefont {Baye},\ and\ \citenamefont
  {Descouvemont}}]{Tursunov2006}%
  \BibitemOpen
  \bibfield  {author} {\bibinfo {author} {\bibfnamefont {E.~M.}\ \bibnamefont
  {Tursunov}}, \bibinfo {author} {\bibfnamefont {D.}~\bibnamefont {Baye}},\
  and\ \bibinfo {author} {\bibfnamefont {P.}~\bibnamefont {Descouvemont}},\
  }\bibfield  {title} {\bibinfo {title} {{Analysis of the ${}^{6}$He beta decay
  into the $\alpha+d$ continuum within a three-body model}},\ }\href
  {https://doi.org/10.1103/PhysRevC.73.014303} {\bibfield  {journal} {\bibinfo
  {journal} {Phys. Rev. C}\ }\textbf {\bibinfo {volume} {73}},\ \bibinfo
  {pages} {014303} (\bibinfo {year} {2006})}\BibitemShut {NoStop}%
\bibitem [{\citenamefont {Thompson}(1988)}]{Thompson1988}%
  \BibitemOpen
  \bibfield  {author} {\bibinfo {author} {\bibfnamefont {I.~J.}\ \bibnamefont
  {Thompson}},\ }\bibfield  {title} {\bibinfo {title} {{Coupled reaction
  channels calculations in nuclear physics}},\ }\href
  {https://doi.org/10.1016/0167-7977(88)90005-6} {\bibfield  {journal}
  {\bibinfo  {journal} {Comput. Phys. Rep.}\ }\textbf {\bibinfo {volume} {7}},\
  \bibinfo {pages} {167} (\bibinfo {year} {1988})}\BibitemShut {NoStop}%
\bibitem [{\citenamefont {Lei}\ and\ \citenamefont
  {Descouvemont}(2020)}]{Lei2020inhomo}%
  \BibitemOpen
  \bibfield  {author} {\bibinfo {author} {\bibfnamefont {J.}~\bibnamefont
  {Lei}}\ and\ \bibinfo {author} {\bibfnamefont {P.}~\bibnamefont
  {Descouvemont}},\ }\bibfield  {title} {\bibinfo {title} {{Lagrange-mesh
  $R$-matrix method for inhomogeneous equations}},\ }\href
  {https://doi.org/10.1103/PhysRevC.102.014608} {\bibfield  {journal} {\bibinfo
   {journal} {Phys. Rev. C}\ }\textbf {\bibinfo {volume} {102}},\ \bibinfo
  {pages} {014608} (\bibinfo {year} {2020})}\BibitemShut {NoStop}%
\bibitem [{\citenamefont {Lei}\ and\ \citenamefont {Moro}(2023)}]{Lei2023cdcc}%
  \BibitemOpen
  \bibfield  {author} {\bibinfo {author} {\bibfnamefont {J.}~\bibnamefont
  {Lei}}\ and\ \bibinfo {author} {\bibfnamefont {A.~M.}\ \bibnamefont {Moro}},\
  }\bibfield  {title} {\bibinfo {title} {{Advancing the
  Ichimura-Austern-Vincent model with continuum-discretized coupled-channels
  wave functions for realistic descriptions of two-body projectile breakup}},\
  }\href {https://doi.org/10.1103/PhysRevC.108.034612} {\bibfield  {journal}
  {\bibinfo  {journal} {Phys. Rev. C}\ }\textbf {\bibinfo {volume} {108}},\
  \bibinfo {pages} {034612} (\bibinfo {year} {2023})}\BibitemShut {NoStop}%
\bibitem [{\citenamefont {Deltuva}(2009)}]{Deltuva2009}%
  \BibitemOpen
  \bibfield  {author} {\bibinfo {author} {\bibfnamefont {A.}~\bibnamefont
  {Deltuva}},\ }\bibfield  {title} {\bibinfo {title} {{Three-body direct
  nuclear reactions: Nonlocal optical potential}},\ }\href
  {https://doi.org/10.1103/PhysRevC.79.021602} {\bibfield  {journal} {\bibinfo
  {journal} {Phys. Rev. C}\ }\textbf {\bibinfo {volume} {79}},\ \bibinfo
  {pages} {021602} (\bibinfo {year} {2009})}\BibitemShut {NoStop}%
\bibitem [{\citenamefont {Potel}\ \emph {et~al.}(2013)\citenamefont {Potel},
  \citenamefont {Idini}, \citenamefont {Barranco}, \citenamefont {Vigezzi},\
  and\ \citenamefont {Broglia}}]{Potel2013}%
  \BibitemOpen
  \bibfield  {author} {\bibinfo {author} {\bibfnamefont {G.}~\bibnamefont
  {Potel}}, \bibinfo {author} {\bibfnamefont {A.}~\bibnamefont {Idini}},
  \bibinfo {author} {\bibfnamefont {F.}~\bibnamefont {Barranco}}, \bibinfo
  {author} {\bibfnamefont {E.}~\bibnamefont {Vigezzi}},\ and\ \bibinfo {author}
  {\bibfnamefont {R.~A.}\ \bibnamefont {Broglia}},\ }\bibfield  {title}
  {\bibinfo {title} {{Cooper pair transfer in nuclei}},\ }\href
  {https://doi.org/10.1088/0034-4885/76/10/106301} {\bibfield  {journal}
  {\bibinfo  {journal} {Rep. Prog. Phys.}\ }\textbf {\bibinfo {volume} {76}},\
  \bibinfo {pages} {106301} (\bibinfo {year} {2013})}\BibitemShut {NoStop}%
\bibitem [{\citenamefont {Johnson}\ and\ \citenamefont
  {Soper}(1970)}]{Johnson1970}%
  \BibitemOpen
  \bibfield  {author} {\bibinfo {author} {\bibfnamefont {R.~C.}\ \bibnamefont
  {Johnson}}\ and\ \bibinfo {author} {\bibfnamefont {P.~J.~R.}\ \bibnamefont
  {Soper}},\ }\bibfield  {title} {\bibinfo {title} {{Contribution of Deuteron
  Breakup Channels to Deuteron Stripping and Elastic Scattering}},\ }\href
  {https://doi.org/10.1103/PhysRevC.1.976} {\bibfield  {journal} {\bibinfo
  {journal} {Phys. Rev. C}\ }\textbf {\bibinfo {volume} {1}},\ \bibinfo {pages}
  {976} (\bibinfo {year} {1970})}\BibitemShut {NoStop}%
\bibitem [{\citenamefont {Harvey}\ and\ \citenamefont
  {Johnson}(1971)}]{Harvey1971}%
  \BibitemOpen
  \bibfield  {author} {\bibinfo {author} {\bibfnamefont {J.~D.}\ \bibnamefont
  {Harvey}}\ and\ \bibinfo {author} {\bibfnamefont {R.~C.}\ \bibnamefont
  {Johnson}},\ }\bibfield  {title} {\bibinfo {title} {{Influence of Breakup
  Channels on the Analysis of Deuteron Stripping Reactions}},\ }\href
  {https://doi.org/10.1103/PhysRevC.3.636} {\bibfield  {journal} {\bibinfo
  {journal} {Phys. Rev. C}\ }\textbf {\bibinfo {volume} {3}},\ \bibinfo {pages}
  {636} (\bibinfo {year} {1971})}\BibitemShut {NoStop}%
\bibitem [{\citenamefont {Feshbach}(1992)}]{Feshbach1992}%
  \BibitemOpen
  \bibfield  {author} {\bibinfo {author} {\bibfnamefont {H.}~\bibnamefont
  {Feshbach}},\ }\href@noop {} {\emph {\bibinfo {title} {{Theoretical Nuclear
  Physics: Nuclear Reactions}}}}\ (\bibinfo  {publisher} {Wiley},\ \bibinfo
  {address} {New York},\ \bibinfo {year} {1992})\BibitemShut {NoStop}%
\end{thebibliography}%

\end{document}